\documentclass[fleqn,usenatbib]{mnras}

\usepackage{newtxtext,newtxmath}

\usepackage[T1]{fontenc}

\DeclareRobustCommand{\VAN}[3]{#2}
\let\VANthebibliography\thebibliography
\def\thebibliography{\DeclareRobustCommand{\VAN}[3]{##3}\VANthebibliography}

\usepackage{graphicx}	
\usepackage{amsmath}	
\usepackage{color,soul}
\usepackage{subcaption} 

\newcommand{\Mstar}{$\textrm{M}_{\star}$ }
\newcommand{\Mstarend}{$\textrm{M}_{\star}$}
\newcommand{\SigSFR}{$\Sigma_{\textrm{SFR}}$ }
\newcommand{\SigSFRend}{$\Sigma_{\textrm{SFR}}$}
\newcommand{\Sigstar}{$\Sigma_{\star}$ }
\newcommand{\Sigstarend}{$\Sigma_{\star}$}
\newcommand{\Siggas}{$\Sigma_{\textrm{H}_2}$ }
\newcommand{\Siggasend}{$\Sigma_{\textrm{H}_2}$}
\newcommand{\Siggasmet}{$\Sigma_{\textrm{H}_2, \textrm{met}}$ }

\newcommand{\dfgas}{$\Delta f_{H_2}$ }
\newcommand{\dfgasend}{$\Delta f_{H_2}$}
\newcommand{\dSFE}{$\Delta$SFE }
\newcommand{\dSFEend}{$\Delta$SFE}
\newcommand{\dsigsfr}{$\Delta \Sigma_{SFR}$ }
\newcommand{\dsigsfrend}{$\Delta \Sigma_{SFR}$}
\newcommand{\fgas}{$f_{\textrm{H}_2}$ }
\newcommand{\fgasend}{$f_{\textrm{H}_2}$}
\newcommand{\Mgas}{$\textrm{M}_{\textrm{H}_2}$ }
\newcommand{\Mgasend}{$\textrm{M}_{\textrm{H}_2}$}

\title[ALMaQUEST X - Mergers]{The ALMaQUEST Survey X: What powers merger induced star formation?}

\author[M.D. Thorp et al.]{
Mallory D. Thorp$^{1}$\thanks{E-mail: mallorythorp@uvic.ca}, Sara L. Ellison$^{1}$, Hsi-An Pan$^{2}$, Lihwai Lin$^{3}$, David R. Patton$^{4}$, Asa F. L. Bluck$^{5,6}$, \newauthor Dan Walters$^{1}$, Jillian M. Scudder$^{7}$
\\
$^{1}$Department of Physics \& Astronomy, University of Victoria, Finnerty Road, Victoria, British Columbia, V8P 1A1, Canada\\
$^{2}$Department of Physics, Tamkang University, Tamsui Dist., New Taipei City 251301, Taiwan\\
$^{3}$Institute of Astronomy \& Astrophysics, Academia Sinica, Taipei 10617, Taiwan\\
$^{4}$Department of Physics \& Astronomy, Trent University, 1600 West Bank Drive, Peterborough, ON K9L 0G2, Canada\\
$^{5}$Department of Physics, Florida International University, 11200 SW St., Miami, Fl, USA\\
$^{6}$Hughes Hall College, University of Cambridge, Wollaston Road, Cambridge CB1 2EW, UK\\
$^{7}$Department of Physics \& Astronomy, Oberlin College, Oberlin, OH, 44074, USA\\
}

\date{Accepted XXX. Received YYY; in original form ZZZ}

\pubyear{2021}

\begin{document}
\label{firstpage}
\pagerange{\pageref{firstpage}--\pageref{lastpage}}
\maketitle

\begin{abstract}
Galaxy mergers are known to trigger both extended and central star formation. However, what remains to be understood is whether this triggered star formation is facilitated by enhanced star formation efficiencies, or an abundance of molecular gas fuel. This work presents spatially resolved measurements of CO emission collected with the Atacama Large Millimetre Array (ALMA) for 20 merging galaxies (either pairs or post-mergers) selected from the Mapping Nearby Galaxies at Apache Point Observatory (MaNGA) survey. Eleven additional merging galaxies are selected from the ALMA MaNGA QUEnching and STar formation (ALMaQUEST) survey, resulting in a set of 31 mergers at various stages of interaction and covering a broad range of star formation rates (SFR). We investigate galaxy-to-galaxy variations in the resolved Kennicutt-Schmidt relation, (rKS: \Siggas vs. \SigSFRend), the resolved molecular gas main sequence (rMGMS: \Sigstar vs. \Siggasend), and the resolved star-forming main sequence (rSFMS: \Sigstar vs. \SigSFRend). We quantify offsets from these resolved relations to determine if star formation rate, molecular gas fraction, and/or star formation efficiency (SFE) is enhanced in different regions of an individual galaxy. By comparing offsets in all three parameters we can discern whether gas fraction or SFE powers an enhanced \SigSFRend. We find that merger-induced star formation can be driven by a variety of mechanisms, both within a galaxy and between different mergers, regardless of interaction stage.
\end{abstract}




\section{Introduction}

Evidence has consistently demonstrated that galaxy-galaxy mergers can trigger star formation: from the bluer colour of peculiar galaxies \citep{Larson1978StarGalaxies,BartonGillespie2003Tidally-TriggeredAges,Lambas2012GalaxyInteractions} and excess H$\alpha$ emission noted in interacting pairs \citep{Kennicutt1987THERATES,Knapen2009THESTARBURSTS}, to large single-fibre spectroscopic surveys revealing an excess of star formation as a merger event progresses \citep{Ellison2008GalaxyRelation, Woods2010TRIGGERED0.08-0.38, Scudder2012GalaxyKpc,Patton2013GalaxyGalaxies,Knapen2015InteractingFormation}. Hydrodynamical simulations illustrate how induced star formation likely stems from gas losing angular momentum and fueling centralized star formation, as a result of the non-axisymmetric structures generated by gravitational forces in an interaction \citep{Barnes1991FuelingMergers,Mihos1996GasdynamicsMergers,Iono2004RadialObservations,Hopkins2013StarMedium}. The degree to which star formation is enhanced can vary drastically depending on the properties of an interaction, with orbital parameters \citep{DiMatteo2007StarStudy,DOnghia2010QUASI-RESONANTINTERACTIONS,Moreno2015MappingFormation}, mass ratio between constituents \citep{Cox2006FeedbackMergers,Cox2008TheStarbursts} and gas fraction of the interacting disk \citep{Bournaud2011HYDRODYNAMICSSPHEROIDS,Perez2011ChemicalInteractions,Scudder2015GalaxyInteractions,Fensch2017High-redshiftFormation} all leading to varied star formation strengths. In simulations, star formation tends to peak when the interacting galaxies are either close to pericentric passage or at the moment of coalescence  \citep{Lotz2008GalaxyMergers,Scudder2012GalaxyKpc, Hani2020InteractingStage}. Such results are supported by observational evidence, which has consistently demonstrated that global star formation rate values are greatest for interactions with equal mass ratios and small projected separations \citep{Nikolic2004StarSurvey,Lin2007AEGIS:1,Ellison2008GalaxyRelation,Woods2010TRIGGERED0.08-0.38,Scudder2012GalaxyKpc,Ellison2013GalaxyGalaxies,Bickley2022StarUNIONS}.

Resolved spectroscopic studies have revealed further complexity of merger-induced star formation on a local scale. Integral Field Spectroscopy (IFS) surveys such as the MaNGA Survey (\citealt{Bundy2015OverviewObservatory}), the Calar Alto Legacy Integral Field Area Survey (CALIFA, \citealt{Sanchez2012CALIFASurvey}), and the Sydney-Australian Astronomical-Observatory Multi-object Integral field spectrograph Survey (SAMI, \citealt{Allen2015TheRelease}) have allowed for the collection of resolved spectroscopic data of large samples of galaxies from which robust statistical results can be attained. Investigations of interacting galaxy pairs in IFS surveys have further corroborated the central enhancement of star formation noted in global studies, but often find a diversity of behaviours in the outskirts of the galaxy where star formation can be unaffected, enhanced, or suppressed \citep{Barrera-Ballesteros2015CentralGalaxies,Pan2019SDSS-IVInteractions,Steffen2021SDSS-IVPairs}. Such variation may be linked to the global galaxy properties, with only the higher mass galaxy in a pair showing enhanced star formation in the outskirts \citep{Steffen2021SDSS-IVPairs}. However, evidence also suggests only galaxies at pericentre and coalescence have elevated star formation at large radii \citep{Pan2019SDSS-IVInteractions}.  Significant variations in spatial star formation enhancement are therefore apparent even in a single stage of interaction. 

Whereas pairs of galaxies represent the pre-merger regime, post-merger galaxies allow us to study the late stages of the interaction sequence. On average, post-mergers have central star formation enhancements that moderately fall off with radius, but galaxies with similar central star formation enhancements can have enhanced, normal, or suppressed star formation in the outskirts \citep{Thorp2019SpatiallyMaNGA}. Just as the unique properties of a galaxy and its interaction parameters can lead to a diverse range of global star formation enhancement, there is equal if not greater complexity when star formation is examined on the resolved scale.

A stepping stone towards understanding the variation of star formation seen in merging galaxies is to characterize the gas which fuels star formation on a kpc-scale. The tight correlation between star formation rate (SFR) and molecular gas mass (\Mgasend) which exists on a global scale (often called the Kennicutt-Schmidt relation for its flagship publications \citealt{Schmidt1959TheFormation} and \citealt{Kennicutt1989THEDISKS}) exists on a local scale as well, resulting in the resolved Kennicutt-Schmidt (rKS) \citep{Bigiel2008THESCALES,Leroy2008THEEFFECTIVELY,Schruba2011AGALAXIES}. A high star formation efficiency (SFE$=$SFR$/$\Mgasend) can lead to regions notably above the rKS from a local boost in star formation \citep{Leroy2013MOLECULARGALAXIES,Bolatto2017TheCARMA,Utomo2017THEGALAXIES}. Offsets from the rKS are the primary driver of offsets in the resolved star-forming main sequence (rSFMS, the strong correlation between \SigSFR and \Sigstarend; \citealt{Sanchez2013Mass-metallicityRate,Cano-Diaz2016Spatially-ResolvedSurvey,GonzalezDelgado2016AstrophysicsGalaxies}), demonstrating that although the star formation rate surface density (\SigSFRend) is regulated by molecular gas mass surface density (\Siggasend) as predicted by the rKS, more varied behaviour in star formation stems from changes in SFE \citep{Ellison2020TheEfficiency.}.

However, there are other ways star formation can be augmented, such as a high gas surface density to fuel stellar growth \citep{Bigiel2008THESCALES,Leroy2013MOLECULARGALAXIES,Schruba2011AGALAXIES}. A surplus of molecular gas manifests as an offset from the correlation between \Siggas and the stellar mass surface density (\Sigstarend), otherwise known as the resolved molecular gas main sequence (rMGMS, \citealt{Wong2013CARMAPROPERTIES,Lin2019TheSequence,Ellison2021TheThem}). Environmental effects on the \Sigstar profile can also boost star formation \citep{Usero2015VARIATIONSGALAXIES,Gallagher2018DenseGalaxies,Jimenez-Donaire2019EMPIRE:Galaxies}. Examining how mergers deviate from the rKS and rMGMS will help discern what drives enhancements in star formation on a kpc-scale.

Several attempts have been made to pinpoint the relative importance of the total molecular gas fraction (\fgasend$=$\Mgasend/\Mstarend) and the SFE to driving enhanced star formation in mergers. Early studies measuring total molecular gas mass with single dish telescopes found conflicting results concerning whether total molecular gas fraction or star formation efficiency drives enhanced star formation in mergers \citep{Braine1992AGalaxies, Casasola2004TheSystems,Huchtmeier2008InteractingSurvey}. More recent studies have leaned towards an enhanced gas fraction driving star formation in both interacting pair \citep{Violino2018GalaxyMergers,Pan2018TheProperties} and post-merger galaxies \citep{Ellison2018EnhancedExhaustion}. SFEs for merging galaxies are mostly normal, except for very close pairs and equal-mass systems whose violent interactions lead to enhanced SFEs \citep{Pan2018TheProperties}. What remains unclear is how these global properties drive kpc-scale variations in star formation.

Only recently have observations of resolved molecular gas properties been collected for large samples to complement large optical IFS surveys, allowing for the combined analysis of \Siggas and \SigSFRend. The Extragalactic Database for Galaxy Evolution (EDGE) - CALIFA survey \citep{Bolatto2017TheCARMA} targeted 126 CALIFA galaxies with the Combined Array for Millimeter-wave Astronomy (CARMA) to investigate depletion time gradients within galaxies \citep{Utomo2017THEGALAXIES,Colombo2018TheSequence} and the processes which regulate star formation on a kpc-scale \citep{Barrera-Ballesteros2021EDGE-CALIFAScales}. The ALMA-MaNGA QUEnching and STar formation (ALMaQUEST) survey \citep{Lin2020ALMAQUEST:SURVEY} observed MaNGA galaxies with ALMA to not only confirm that the key scaling relations between star formation rate, molecular gas, and stellar mass that exist on a global scale stem from a kpc-scale relationship, but also that the resolved star-forming main sequence is likely the result of the two other relations \citep{Lin2019TheSequence, Ellison2021TheThem, Baker2022TheSequence}. ALMaQUEST has also revealed that even though the absolute star formation rate in a galaxy is primarily driven by the amount of molecular gas, the scatter in the resolved star formation efficiency is mostly driven by local changes in SFE \citep{Ellison2020The0}. Given that variations in star formation in the outskirts of mergers are by definition scatter from the rSFMS, such a result might imply that SFE may drive star formation for individual regions of galaxy, even if globally gas reservoir is the dominant driver. Further investigation of the interplay between star formation and molecular gas has been done on the scale of molecular clouds with the Physics at High Angular resolution in Nearby Galaxies (PHANGS)-ALMA survey \citep{Leroy2021PHANGS-ALMA:Galaxies}. With resolution on the order of $\sim$100pc, PHANGS-ALMA
found greater scatter in all three scaling relations compared to lower-resolution studies, revealing significant variation in star formation and molecular gas content even within similar morphological environments \citep{Pessa2021AstronomyScale,Querejeta2021StellarGalaxies}.

Although EDGE-CALIFA, the largest of surveys of this kind, does have a small number of mergers that may be responsible for changes in SFR and depletion times \citep{Bolatto2017TheCARMA,Utomo2017THEGALAXIES,Chown2019LinkingGalaxies}, none of the surveys with both IFS and molecular gas data have a sufficient number of mergers to make a dedicated study of interaction induced physics. A handful of detailed case-studies of the resolved star-formation efficiency of pre-merger galaxies have been completed, revealing significant diversity of depletion time on a resolved scale \citep{Tomicic2018Two2276,Bemis2019Kiloparsec-Scale4038/39}. A study dedicated to a diverse sample of merging and post-merger galaxies is required to better understand the most extreme variations in star formation and gas properties.

In the present work we have observed a sample of 31 merging galaxies with a broad range of interaction progressions and star formation rates, with the specific goal of investigating how merger properties effect the mechanisms which drive star formation. Eleven of these mergers are taken from the main ALMaQUEST sample \citep{Lin2020ALMAQUEST:SURVEY}, plus we present observations for 20 additional galaxies with new CO(1-0) data obtained from ALMA. Together, we refer to this sample of 31 galaxies as the ALMaQUEST merger sample.  In this paper, we aim to distinguish whether star formation efficiency or molecular gas fraction drives spatial changes in star formation rate. In Pan et al. (in prep) we will further investigate how the resolved star formation and gas properties vary with merger configuration.

In Section \ref{sec:Data} we describe our methods for selecting a sample of mergers, as well as the MaNGA and ALMA observations utilized in this investigation. Section \ref{sec: analysis} presents our main results, comparing individual resolved scaling relations for our merger sample, as well as comparing offsets from all three scaling relations. We summarize the impact of this work in Section \ref{sec:summary}. Throughout the work we adopt a cosmology in which $\textrm{H}_{\textrm{0}}=70$ km/s/Mpc, $\Omega_M=0.3$, and $\Omega_{\Lambda}=0.7$.

\section{Data}
\label{sec:Data}

\subsection{Merger Sample Selection}
\label{subsec:selection}

\begin{figure*}
	\includegraphics[width=0.99\textwidth]{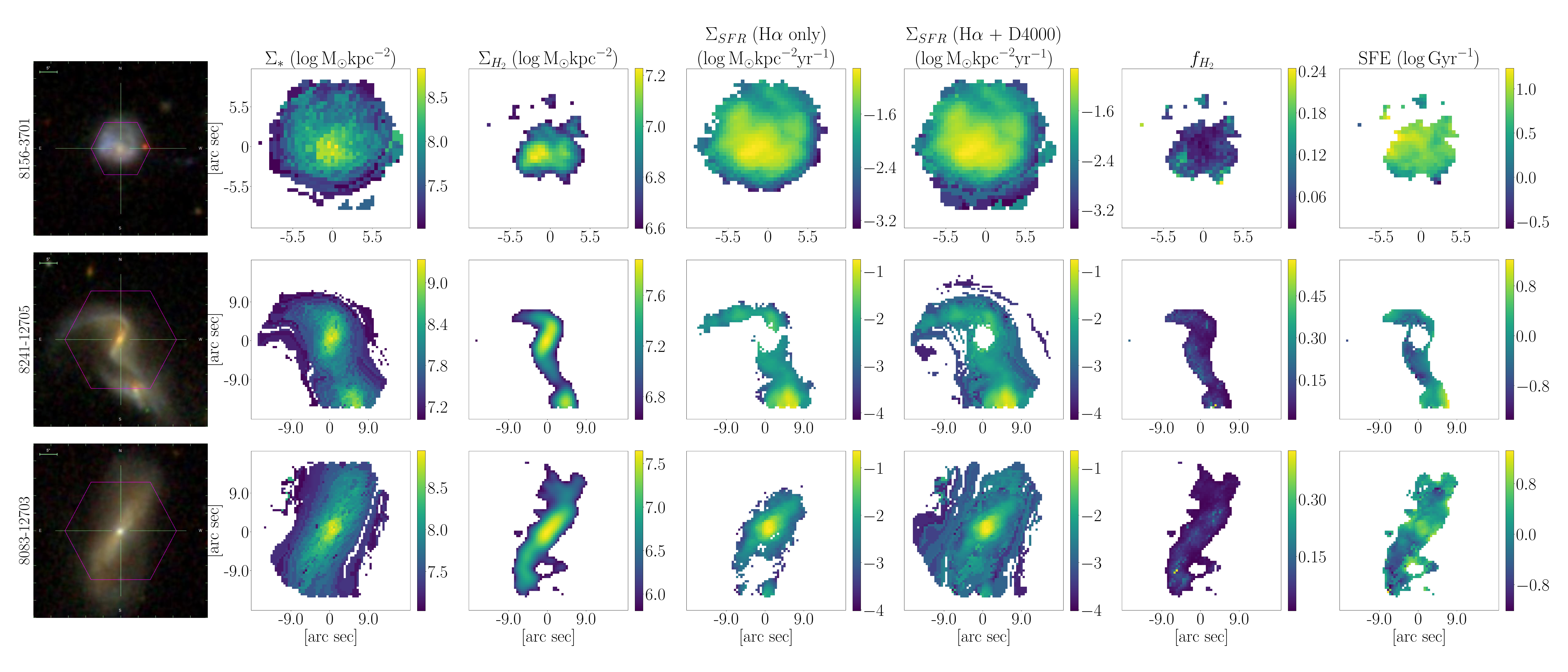}
	\centering
    \caption{Data product maps for three of the galaxies in the ALMaQUEST merger sample (maps for the remaining 28 galaxies are available in Appendix \ref{app:DP}), a post-merger (top), a visual pair (middle), and a spectroscopic pair with the companion out of frame (bottom). From left to right: the SDSS \emph{gri}-image, inclination corrected stellar mass surface density (from {\sc pipe3d}), inclination corrected molecular gas surface density (computed from CO luminosity), inclination corrected star formation rate surface density (computed from H$\alpha$ luminosity), inclination corrected star formation rate surface density (computed from H$\alpha$ luminosity and sSFR-D4000 relation), molecular gas fraction (\Siggasend/\Sigstarend), and star formation efficiency (\SigSFRend/\Siggasend). Note the significant increase in spaxel count when both H$\alpha$ and D4000 \SigSFR are used. The central gap still present in the combined \SigSFR map for 8241-12705 (second row, fifth column) is the result of D4000 exceeding 1.4 in those pixels. By including D4000-\SigSFR measurements we optimize the overlap between \Siggas and \SigSFR for our analysis.}
    \label{fig:ALMaQUEST_Ex}
\end{figure*}

Our sample of merging galaxies is collected from the MaNGA data release 15 (DR15), which was the most recent publicly available release at the time of our observations and for the duration of this project. We visually classify post-merger and interacting galaxies from the $\sim$4800 galaxies in DR15 using the Sloan Digital Sky Survey (SDSS) Sky Server \emph{gri}-images (\emph{r}-band half-light surface brightness limit of 23.0 mag arcsec$^{-2}$; \citealt{Strauss2002SpectroscopicSample}). Post-merger galaxies are distinguished by clear morphological disturbances indicating a recent interaction, such as tidal tails or shells, but with no obvious companion. Galaxies in an interacting pair are identified with similar indicators, with the addition of a clear visual connection to a second disturbed galaxy (such as tidal bridges).

Beyond the visually selected sample, we also identify a group of spectroscopic pairs, where a visible connection between a galaxy and a possible companion is unclear, but spatial and redshift information reveal the two galaxies may be interacting. Spectroscopic pairs are selected using the \cite{Patton2016GalaxySeparations} catalogue, which provides the closest companion for each galaxy in SDSS data release 7 (DR7), with a companion boundary at projected separation $r_p=$ 1 Mpc and a maximum difference in velocity between the two of $\Delta$v $=$ 1000 km/s. We assume any MaNGA galaxies with a projected separation less than 2'' from the \cite{Patton2016GalaxySeparations} galaxy position are the same object. We limit our spectroscopic pairs to those with a mass ratio between 0.1 and 10 (meaning the smaller of the pair is more than 10$\%$ the mass of the larger), and only include galaxies whose companion is within $r_p<$100 kpc and $\Delta$v$<$500 km/s. For simplicity, the rest of this work will not distinguish visually and spectroscopically selected pairs; we will simply refer to ``pairs''. The combination of the pair and post-merger galaxies yields a total sample of 903 galaxies in DR15 which we hence refer to as the parent merger sample, from which we can select a subsample for ALMA observations.

\begin{figure}
	\includegraphics[width=0.9\columnwidth]{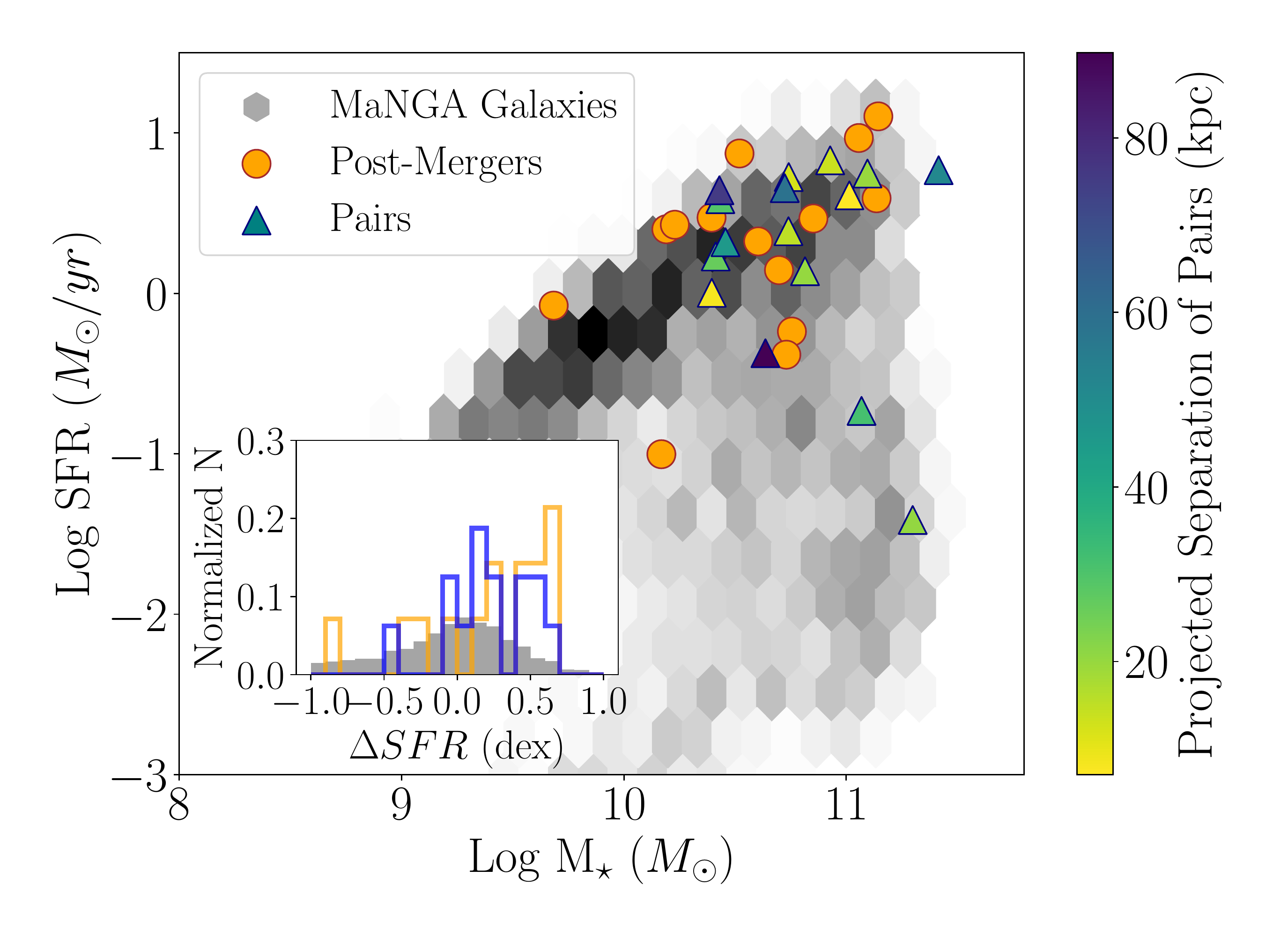}
	\centering
    \caption{Position of our post-merger (orange circles) and pair (triangles) galaxies on the global star-forming main sequence defined by the rest of the MaNGA survey (grey hexbin). Pair galaxies are colour-coded by the projected separation from their closest companion (in kpc). Galaxies that were excluded due to a lack of viable CO spaxels are excluded from this diagram (see Table \ref{tab:properties} and Subsection \ref{subsec:ALMA} for details). An inset histogram shows the distribution in $\Delta$SFR for post-mergers (orange) and pair (blue) galaxies with respect to MaNGA (grey). Although our mergers probe more highly star-forming galaxies than a random population, our sample still covers a broad range of $\Delta$SFR values.}
    \label{fig:Merger_SFMS}
\end{figure}

\begin{table*}
	\centering
	\caption{Summary of global properties of post-merger and pair galaxies observed with ALMA, with post-mergers first (no $r_p$ values), followed by pairs ordered by $r_p$. Key global properties derived from the MaNGA VAC are included (\Mstarend, SFR, $\Delta$SFR, z), as well as merger properties such as $r_p$, $\Delta v$, and mass ratio (the mass of the galaxy divided by that of its companion). Post-mergers have empty columns for merger properties, which require information about a companion galaxy. Pair galaxies that were visually identified only have $r_p$ values, since spectrscopic information is not available for the companion. Also included are spaxel counts for each galaxy that meet our various star-forming and CO S/N cuts, along with the overlap between these regions. Galaxies with less than 10 spaxels of CO+SF(H$\alpha$+D4000) overlap are excluded for plots of individual galaxies.}
	\label{tab:properties}
	\begin{tabular}{lcccccccccccc} 
		\hline
		plate-ifu & $\log \textrm{M}_{\star}$ & $\log$ SFR & $\Delta$ SFR & z & $r_p$ & $\Delta$ v & Mass & SF & SF & CO S/N$>$3 & CO+SF & CO+SF\\
		&  & & & & & & Ratio & (H$\alpha$) & (H$\alpha$+D4000) & & (H$\alpha$) & (H$\alpha$+D4000) \\
		& $\log M_{\odot}$ & $\log M_{\odot}/yr$ & dex &   & kpc & km/s &   & \# Spaxels & \# Spaxels & \# Spaxels & \# Overlap & \# Overlap \\
		\hline
		\hline
        9195-3702 & 11.14 & 1.10 & 0.62 & 0.064 & - & - & - & 0 & 53 & 204 & 0 & 5 \\
        9194-3702 & 11.06 & 0.97 & 0.52 & 0.075 & - & - & - & 0 & 811 & 379 & 0 & 328 \\ 
        8083-9101 & 11.14 & 0.59 & 0.11 & 0.038 & - & - & - & 331 & 1163 & 322 & 60 & 186 \\ 
        8952-12701 & 10.73 & $-$0.38 & $-$0.23 & 0.029 & - & - & - & 21 & 37 & 354 & 18 & 28 \\
        8084-3702 & 10.23 & 0.43 & 0.60 & 0.022 & - & - & - & 240 & 1007 & 552 & 150 & 531 \\
        8156-3701 & 10.52 & 0.87 & 0.66 & 0.053 & - & - & - & 778 & 941 & 279 & 279 & 279 \\
        8081-9101 & 10.60 & 0.32 & 0.23 & 0.028 & - & - & - & 321 & 600 & 436 & 255 & 335 \\
        8615-3703 & 10.19 & 0.40 & 0.45 & 0.018 & - & - & - & 324 & 997 & 538 & 276 & 519 \\
        9512-3704 & 10.70 & 0.14 & $-$0.03 & 0.055 & - & - & - & 168 & 552 & 136 & 33 & 40 \\
        7977-12705 & 10.85 & 0.47 & 0.26 & 0.027 & - & - & - & 981 & 1472 & 538 & 197 & 201 \\ 
        8623-1902 & 10.17 & $-$1.00 & $-$0.84 & 0.025 & - & - & - & 4 & 175 & 257 & 2 & 87 \\
        8616-12702 & 10.76 & $-$0.24 & $-$0.40 & 0.031 & - & - & - & 469 & 1113 & 258 & 78 & 104 \\ 
        9195-3703 & 10.39 & 0.47 & 0.42 & 0.027 & - & - & - & 225 & 833 & 504 & 110 & 460 \\
        8615-1901 & 9.68 & $-$0.08 & 0.51 &0.020 & - & - & - & 461 & 588 & 2 & 2 & 2 \\
        8153-12702 & 9.90 & $-$0.74 & $-$0.45 & 0.038 & 0.46 & - & - & 333 & 1119 & 0 & 0 & 0 \\
        7975-6104 & 11.01 & 0.61 & 0.17 & 0.079 & 7.04 & - & - & 209 & 588 & 391 & 209 & 320 \\
        8241-12705 & 10.40 & 0.00 & $-$0.05 & 0.027 & 8.23 & 80.0 & 3.14 & 847 & 1678 & 592 & 381 & 508 \\
        8082-9102 & 10.74 & 0.73 & 0.53 & 0.037 & 12.24 & - & - & 554 & 1197 & 668 & 422 & 591 \\
        7977-9102 & 10.93 & 0.83 & 0.41 & 0.063 & 13.78 & - & - & 524 & 1129 & 380 & 287 & 354 \\
        9195-9101 & 10.74 & 0.39 & 0.16 & 0.057 & 15.36 & - & - & 678 & 1797 & 722 & 455 & 675 \\
        8616-9101 & 11.10 & 0.75 & 0.28 & 0.092 & 19.89 & 276.0 & 4.19 & 112 & 1044 & 363 & 84 & 249 \\
        8078-12703 & 10.81 & 0.14 & $-$0.02 & 0.028 & 20.20  & 102.0 & - & 1208 & 2597 & 1128 & 444 & 796 \\ 
        7968-12705 & 11.30 & $-$1.41 & $-$1.37 & 0.086 & 20.45 & 13.0 & 0.80 & 0 & 93 & 8 & 0 & 0 \\
        8078-6104 & 10.41 & 0.23 & 0.16 & 0.044 & 26.15 & 109.0 & 0.42 & 197 & 1039 & 433 & 153 & 431 \\
        8085-12701 & 10.43 & 0.59 & 0.61 &0.030 & 29.51 & - & - & 2293 & 2601 & 458 & 456 & 458 \\
        8085-6101 & 11.07 & $-$0.73 & $-$1.09 & 0.052 & 31.48 & - & - & 0 & 0 & 2 & 0 & 0 \\
        8083-12703 & 10.46 & 0.32 & 0.24 & 0.025 & 45.24 & 60.6 & - & 822 & 2201 & 1264 & 641 & 1171 \\
        8082-12704 & 11.42 & 0.77 & 0.05 & 0.132 & 51.28 & - & - & 3 & 1617 & 451 & 3 & 244 \\
        8085-3704 & 10.72 & 0.66 & 0.43 & 0.037 & 58.73 & 37.0 & 0.89 & 319 & 727 & 608 & 299 & 552 \\ 
        8450-6102 & 10.43 & 0.64 & 0.54 & 0.042 & 75.35 & 102.0 & 1.81 & 1162 & 1550 & 536 & 503 & 536 \\
        8728-3701 & 10.64 & $-$0.37 & $-$0.49 & 0.028 & 89.75 & 113.0 & 1.96 & 0 & 31 & 156 & 0 & 0 \\
        \hline
	\end{tabular}
\end{table*}

To select candidate mergers for ALMA observations we first exclude all mergers with declination greater than 20\textdegree, leaving only those with positions that overlap with ALMA's observational range. From the initial parent merger sample only 143 galaxies meet this criterion. We next select mergers for which our target CO line S/N can be achieved in less than 5 hours, to ensure we can observe as many mergers as possible within a competitive proposal (see Section \ref{subsec:ALMA} for more details). What remains is a sample of 6 post-mergers and 14 pairs, which are observed as part of an ALMA Cycle 7 program (2019.1.00260.S, P.I.: Hsi-An Pan, details in Section \ref{subsec:ALMA}). We also make use of eleven galaxies from the ALMaQUEST survey that show clear signs of an interaction within our classification scheme: 8 post-mergers and 3 pairs \citep{Lin2020ALMAQUEST:SURVEY,Ellison2020The0}. The SDSS \emph{gri}-images of three of the galaxies in the final ALMaQUEST merger sample are included in the left panels of Figure \ref{fig:ALMaQUEST_Ex} (continued for all mergers in Appendix \ref{app:DP}). The final sample of 14 post-merger and 17 pair galaxies is summarized in Table \ref{tab:properties}. Included in the table are key global properties from the MaNGA {\sc pipe3d} Value Added Catalogue (VAC) including total stellar mass (\Mstarend), total star formation rate (SFR), and redshift (z). We also include the offset from the global star-forming main sequence $\Delta$SFR, where $\Delta \textrm{SFR}= \log \textrm{SFR}_{\textrm{galaxy}} - \log \textrm{SFR}_{\textrm{control}}$. $\textrm{SFR}_{\textrm{control}}$ is the median SFR value of galaxies within 2$\sigma$ of a fit to the star-forming main sequence (controls are also matched within 0.1 dex in \Mstar and 0.005 in z). Merger properties are listed as well, including $r_p$, $\Delta v$, and mass ratio. Post-merger galaxies have empty $r_p$, $\Delta v$, and mass ratio values since there is no companion with which to compare. Pair galaxies that were visually identified do not have measured $\Delta v$ and mass ratio values, since there is no spectroscopic information from their companion.

The merger sample covers a broad range of \Mstar and SFR. Figure \ref{fig:Merger_SFMS} compares these properties for our post-merger (orange circles) and pair (triangles) galaxies with respect to the rest of MaNGA, shown as grey hexbins. Pairs are colour-coded by the projected separation from their closest companion. Our sample is representative of a typical merger sample, containing galaxies with both greatly elevated and comparatively low SFRs. The inset histogram of Figure \ref{fig:Merger_SFMS} shows the distribution of the offset from the star-forming main sequence $\Delta$SFR of our post-merger (orange) and pair (blue) galaxies with respect to the rest of MaNGA (grey). Although the pair and post-merger samples have on average larger $\Delta$SFR values than non-interacting galaxies, we still cover a broad range of $\Delta$SFR values to reflect the overall variety of behaviours seen in larger galaxy merger studies and to probe any corresponding diversity in the molecular gas properties of those mergers. We note that there may be some selection bias in the sample, given some post-mergers come from the ALMaQUEST starburst sample of \cite{Ellison2020The0}, and the rest still need to have enough gas for detectable CO as described in the following section.

\subsection{ALMA Observations}
\label{subsec:ALMA}

Observations of CO $J=1 \rightarrow 0$ (CO hereafter), rest frame 115 GHz, were completed for 6 post-merger and 14 pair galaxies as part of Cycle 7 ALMA program 2019.1.00260.S (P.I.: Hsi-An Pan). The remaining 11 mergers were already available from the original ALMaQUEST sample \citep{Lin2020ALMAQUEST:SURVEY}. Cycle 7 observations were designed to replicate the methodology used in the original ALMaQUEST survey, the details of which can be found in the survey paper by \cite{Lin2020ALMAQUEST:SURVEY}. We provide a brief summary of those techniques below.

Observations were carried out using the Band 3 receiver, taken in the C43-2 configuration to achieve a synthesized beam full-width half maximum (FWHM) of 2.5'' in order to match the effective resolution of the MaNGA survey \citep{Law2015ObservingSurvey}. The ALMA field of view in our chosen configuration is $\sim$50'' which is sufficient to cover the MaNGA footprint for all of our galaxies. The spectral set up includes one high resolution spectral window targeting the CO emission ($\sim$10 km/s), and three low-resolution spectral windows around the target line to detect the continuum ($\sim$90 km/s). To reach a CO S/N greater than 3 for 50$\%$ of spaxels with H$\alpha$ S/N$>$3, the on-target integration time varied between 0.1-3.3 hours.
The same methodology is used for the main ALMaQUEST sample, ensuring consistent data quality between ALMaQUEST data and that acquired in Cycle 7 \citep{Lin2017SDSS-IVMaNGA,Lin2020ALMAQUEST:SURVEY}.

Data were calibrated using the ALMA data reduction software Common Astronomy Software Applications (CASA, \citealt{McMullin2007CASAApplications}) using version 5.4 for all but 3 galaxies (which were observed in earlier cycles and thus used CASA version 4.5), along with the standard ALMA reduction pipeline. In ALMA's Band 3 the systematic flux uncertainty inherent with calibration is roughly 5-10$\%$ \citep{2019acpg.rept.....D}. Continua were subtracted in the visibility domain for a handful of galaxies both in the original ALMaQUEST and the Cycle 7 galaxies. Once the continuum was subtracted, the task CLEAN was used to clean data down to 1$\sigma$ and produce spectral line data cubes with a Briggs weighting (robust parameter=0.5), resulting in a native effective beam size ranging from 1.6''-2.8'' depending on the target. The data were then re-imaged with a user-specified spaxel size (0.5'') and a restoring beamsize (2.5'' $\times$ 2.5'') to match the MaNGA image grid (prior) and spatial resolution (latter). The final cubes have channel widths of 11 km/s and $\sigma_{\textrm{rms}}\sim$0.2-2 mJy/beam. We applied the CASA task IMMOMENTS to these datacubes to determine moment 0 (integrated intensity) maps. IMMOMENTS can also generate moment 1 (intensity-weighted velocity) and moment 2 (intensity-weighted velocity dispersion) maps, but those are not relevant to the analysis of this work and are saved for future projects. Moment 0 maps are constructed by integrating the CO emission from a set velocity range without any clipping in signal. Table \ref{tab:properties} lists the number of spaxels in each galaxy where the CO line emission from the moment 0 map has S/N$>3$. Four of the merger galaxies observed as part of Cycle 7 (8615-1901, 8153-12702, 7968-12705, 8085-6101, all four early-type galaxies) have less than ten spaxels with CO S/N$>$3, and are thus excluded from any further analysis that looks at individual galaxies (rather than spaxels).

The CO luminosity per spaxel determined from the moment 0 map is converted to molecular gas surface density \Siggas ($\textrm{M}_{\odot}/\textrm{kpc}^2$) using a constant conversion factor of $\alpha_{\textrm{CO}}$=4.35 ($\textrm{M}_{\odot} (\textrm{K } \textrm{km/s } \textrm{pc}^2)^{-1}$) \citep{Bolatto2013THEFACTOR}. All \Siggas values are then inclination corrected using the b/a axial ratios from the NASA Sloan Atlas (NSA) catalogue, themselves determined from single Sérsic fits. Three example \Siggas maps are provided in Figure \ref{fig:ALMaQUEST_Ex}, with the rest of the \Siggas maps from the sample available in Appendix \ref{app:DP}. There has been extensive research into the viability of a constant conversion factor, and many suggest using a metallicity dependent \citep{Accurso2017DerivingRelations,Sun2020DynamicalGalaxies} or metallicity and line intensity dependent \citep{Narayanan2012ALaw} conversion instead. The requirements needed to determine accurate metallicity measurements would limit our spaxel count significantly, given the sparse overlap between high S/N star-forming spaxels and CO measurements (see Table \ref{tab:properties}), as we will discuss further in Subsection \ref{subsec:SFR-D4000}. To minimize the loss of spaxels we choose to use a constant conversion factor. In Appendix \ref{app:alpha_CO} we characterize the difference in \Siggas measurements when a metallicity dependent $\alpha_{\textrm{CO}}$ is used (for spaxels where that is possible), and confirm that key results from this work cannot result from inaccuracies of conversion factor.

\subsection{MaNGA Data Products}
\label{subsec:MANGA}

The work presented here primarily uses MaNGA data products from the {\sc pipe3d} spectral fitting pipeline, described in detail in \cite{Sanchez2016Pipe3DFIT3D, Sanchez2016Pipe3DDataproducts}. Along with global values provided by the {\sc pipe3d} VAC mentioned previously, we also make extensive use of the {\sc pipe3d} stellar mass surface density (\Sigstarend) and emission line fluxes. We correct all emission line fluxes for dust using a Milky Way dust extinction curve \citep{Cardelli1989THEEXTINCTION}, assuming an intrinsic H$\alpha$/H$\beta$ ratio of 2.85. Star formation rate surface densities (\SigSFRend) are determined from the dust corrected H$\alpha$ luminosity using the \cite{Kennicutt1994PastGalaxies} relation and assuming a Salpeter intitial mass function \citep{Salpeter1955THEEVOLUTION}. Both \Sigstar and \SigSFR are inclination corrected using the b/a axial ratios (derived from single Sérsic fits) provided in the NSA catalogue. The same b/a ratio is used to compute an inclination corrected galactocentric radius from the V-band centre of the MaNGA map. We also use the 4000\AA \textrm{ } break strength (D4000) provided by {\sc pipe3d}, as it serves a crucial role in expanding our star formation rate measurements described in the next section.

\vfill
\subsubsection{Star Formation Rates - D4000 vs H$\alpha$}
\label{subsec:SFR-D4000}

The work presented here requires that both SFRs and molecular gas surface densities are measured in a given spaxel. Ideally we would use SFRs determined from H$\alpha$ and only consider the \SigSFR values of star-forming spaxels defined by the \cite{Kauffmann2003TheAGN} designation on a
Baldwin, Phillips \& Terlevich diagram (BPT; \citealt{Baldwin1981CLASSIFICATIONOBJECTS}). Along with a \cite{Kauffmann2003TheAGN} star-forming cut, we also impose a S/N$>3$ cut for the flux of each diagnostic emission line, as well as requiring an H$\alpha$ equivalent width (EW)$>$6\AA{ }limit to ensure H$\alpha$ flux stems from a young stellar population. Of the 13,046 spaxels with CO detections S/N$>3$, only 44.5\% pass all of these star-forming criteria. Table \ref{tab:properties} provides for each galaxy the total count of spaxels which pass the star-forming criteria cut for H$\alpha$, which have CO S/N$>3$, and which pass both criteria. 

\begin{figure}
	\includegraphics[width=0.95\columnwidth]{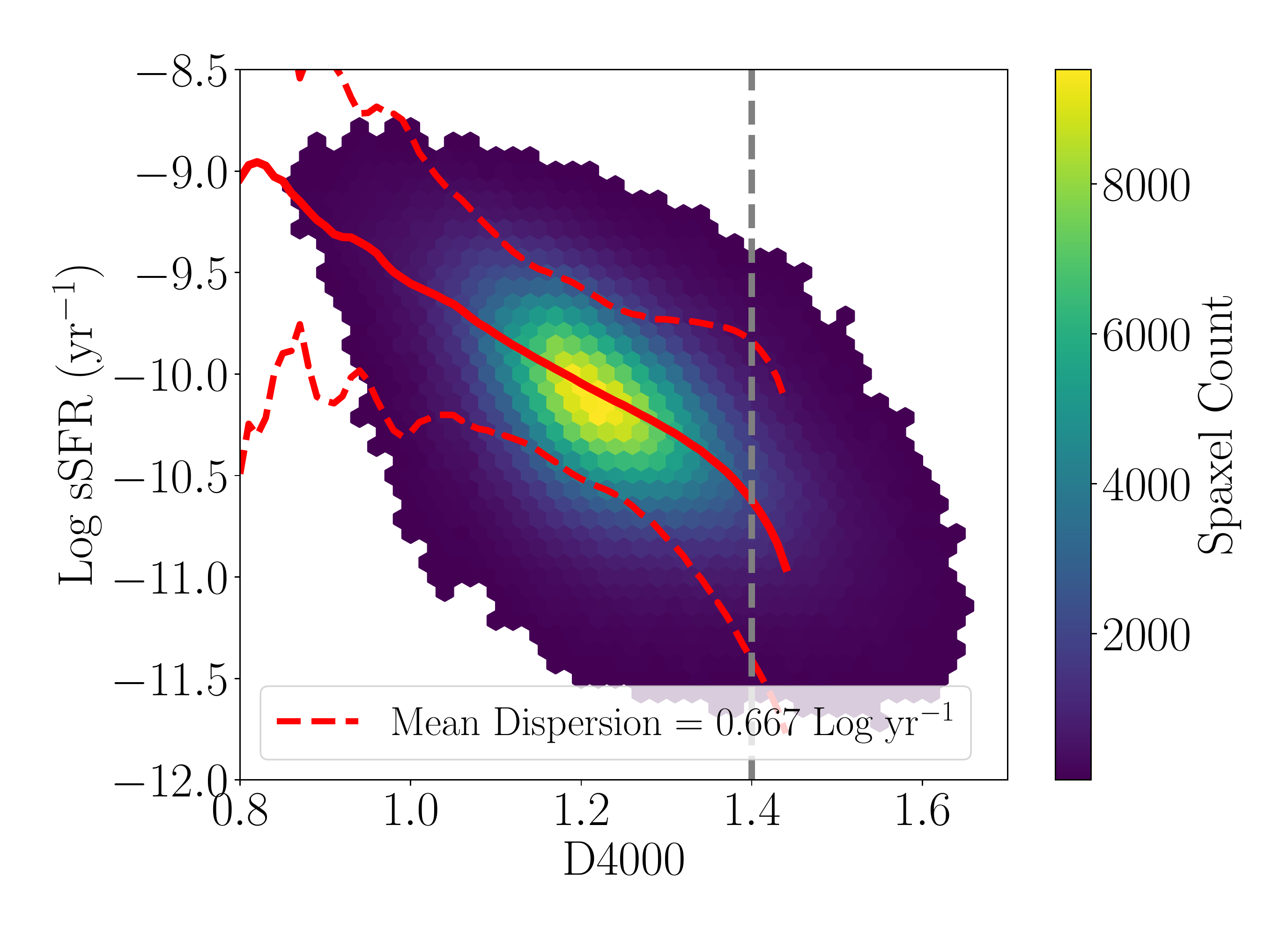}
	\centering
    \caption{The sSFR-D4000 distribution for all star-forming spaxels in MaNGA (shown as a density hexbin). The median sSFR value in each D4000 bin is shown as red line out to D4000=1.45; this value is used to approximate \SigSFR based on D4000 for spaxels which do not meet are star-forming criteria. The standard deviation of sSFR within each bin is shown as a red dashed line. The median sSFR value steeply changes at high D4000, leading to vary different sSFR values in relatively close D4000 bins. We therefore only use the D4000-sSFR calibration when D4000$<$1.4 (vertical dashed line).}
    \label{fig:sSFR-D4000}
\end{figure}

\begin{figure*}
	\includegraphics[width=0.95\textwidth]{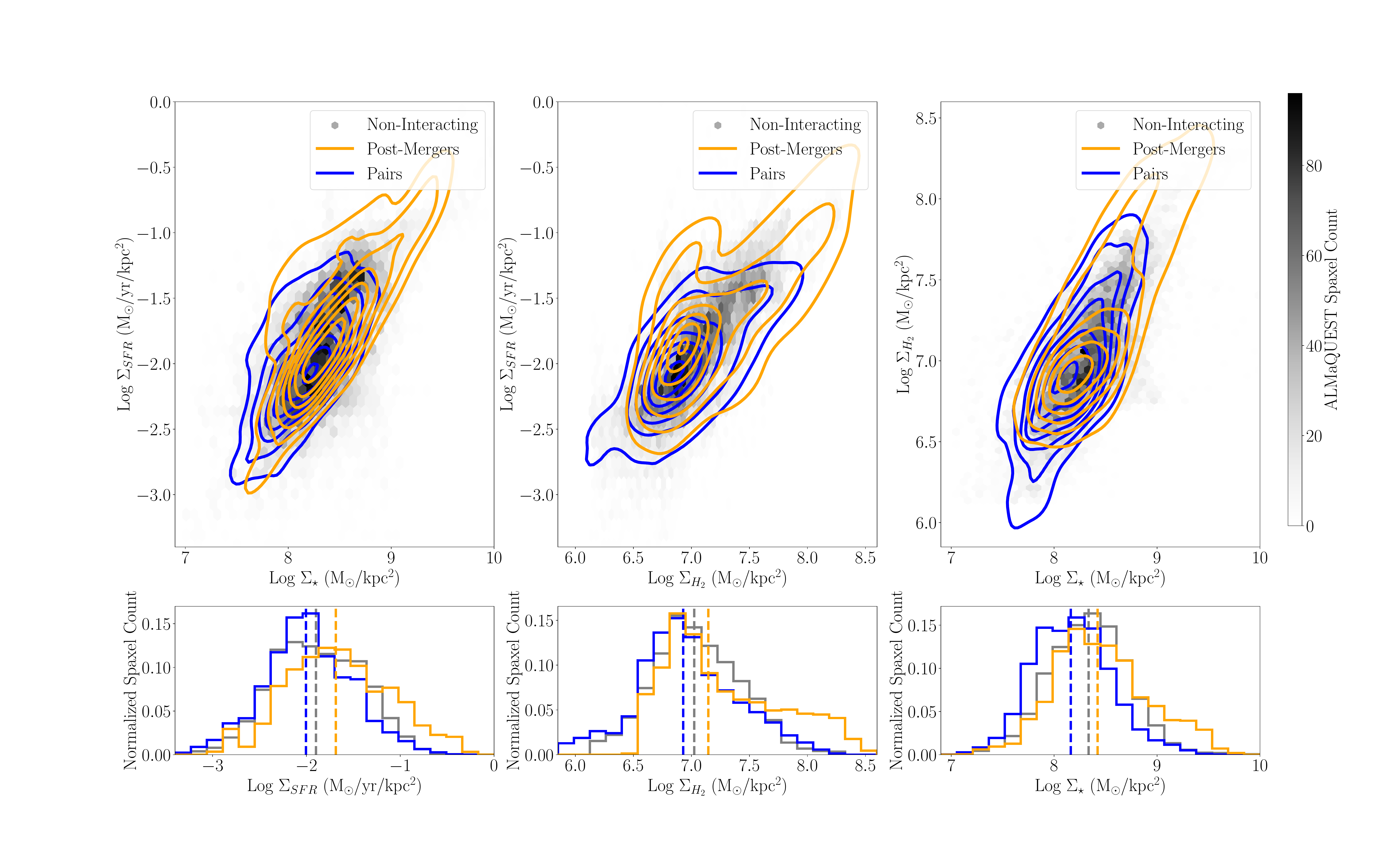}
	\centering
    \caption{Resolved scaling relations for non-interacting ALMaQUEST galaxies (grey histogram), pairs (blue contours), and post-mergers (orange contours). The rSFMS is shown on the left, the rKS is shown in the middle, and the rMGMS is show on the right. All measurements required a CO S/N$>3$, along with the quality criteria to measure \SigSFR with either H$\alpha$ or D4000 methods as described in Subsection \ref{subsec:SFR-D4000}. This results in 22782 isolated spaxels, 7990 pair spaxels, and 4012 post-merger spaxels. Below each histogram distributions of \SigSFRend, \Siggasend, and \Sigstar are included, separated into non-interacting, post-merger, and pair populations. The median value of each distribution is shown as a horizontal dashed line. Note that the non-interacting ALMaQUEST sample is biased towards higher \SigSFR values resulting from the population of starbursts in the original ALMaQUEST sample (see Subsection \ref{subsec: offsets} for clarification on how this differs from offset parameter controls). A KS-test reveals that all three populations are distinct from each other in \SigSFRend, \Siggasend, and \Sigstarend. The pair sample has a small offset to lower \SigSFRend, \Siggasend, and \Sigstarend, which is driven by two pair galaxies with uniquely low \Sigstar and \Siggas values (8078-6104 and 8083-12703). But overall all three populations cover a similar range of surface densities.}
    \label{fig:All_scaling_relations}
\end{figure*}

\begin{figure*}
	\includegraphics[width=0.95\textwidth]{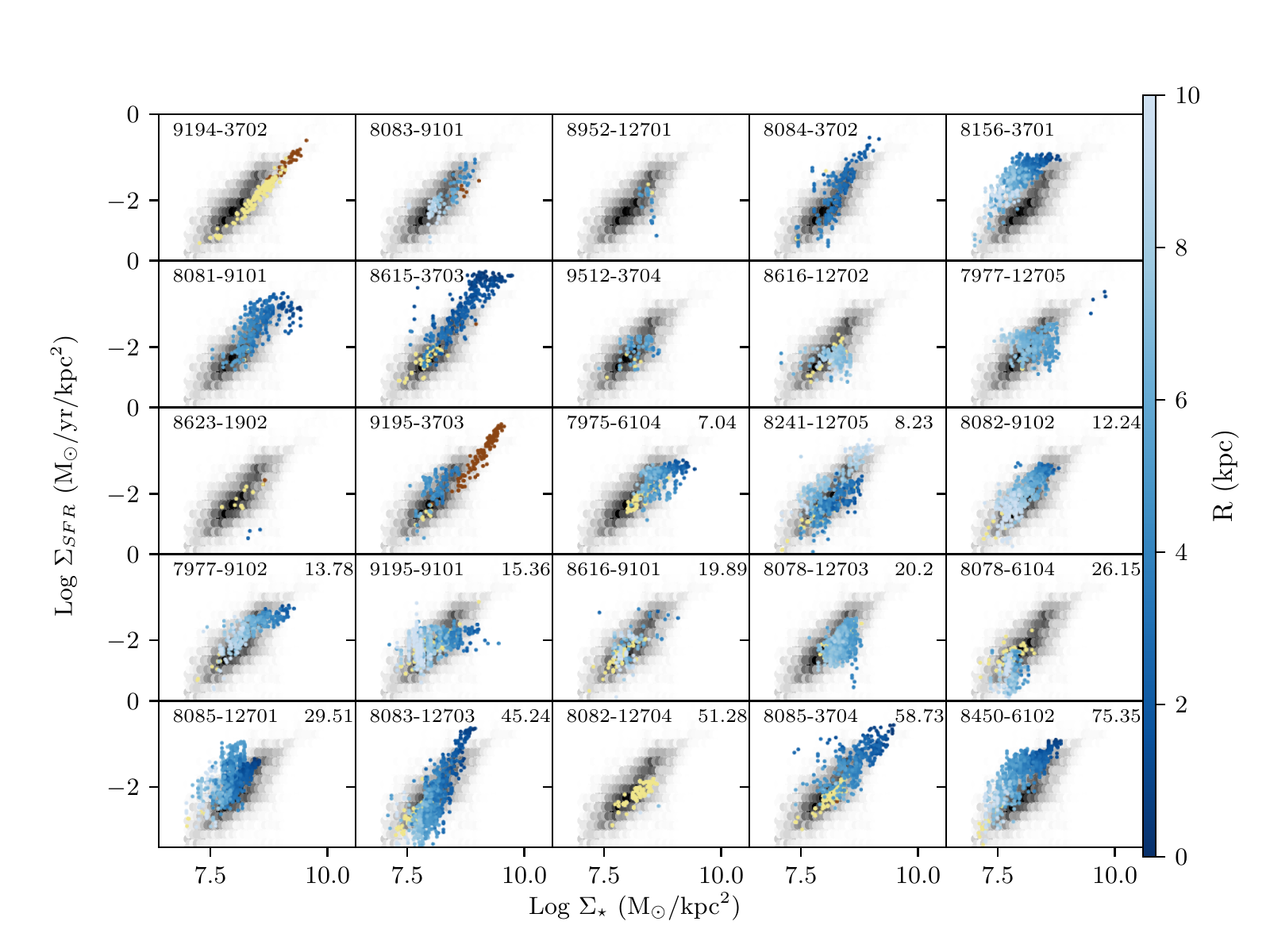}
	\centering
    \caption{Resolved star-forming main sequence for all merger galaxies, with the MaNGA plate-ifu of the galaxy in the left corner and $r_p$ (kpc) in the right corner (left blank for post-mergers). Star-forming spaxels are colour-coded by radius, and spaxels with \SigSFR determined by D4000 are yellow (low H$\alpha$ S/N) or brown (AGN contamination). The non-interacting galaxies in ALMaQUEST are shown as a grey histogram, for reference (including both H$\alpha$-\SigSFR and D4000-\SigSFRend). Note that there is a diversity of star formation behaviour in the merger sample. Some post-mergers and pairs are significantly offset from the rest of ALMaQUEST. Those which exhibit suppressed star-formation (such as 8241-12705, 7975-6104, and 8078-6104) can be derived from both H$\alpha$- and D4000- \SigSFR values, confirming the validity of our combined \SigSFR method.}
    \label{fig:SFMS_resolved}
\end{figure*}

\begin{figure*}
	\includegraphics[width=0.95\textwidth]{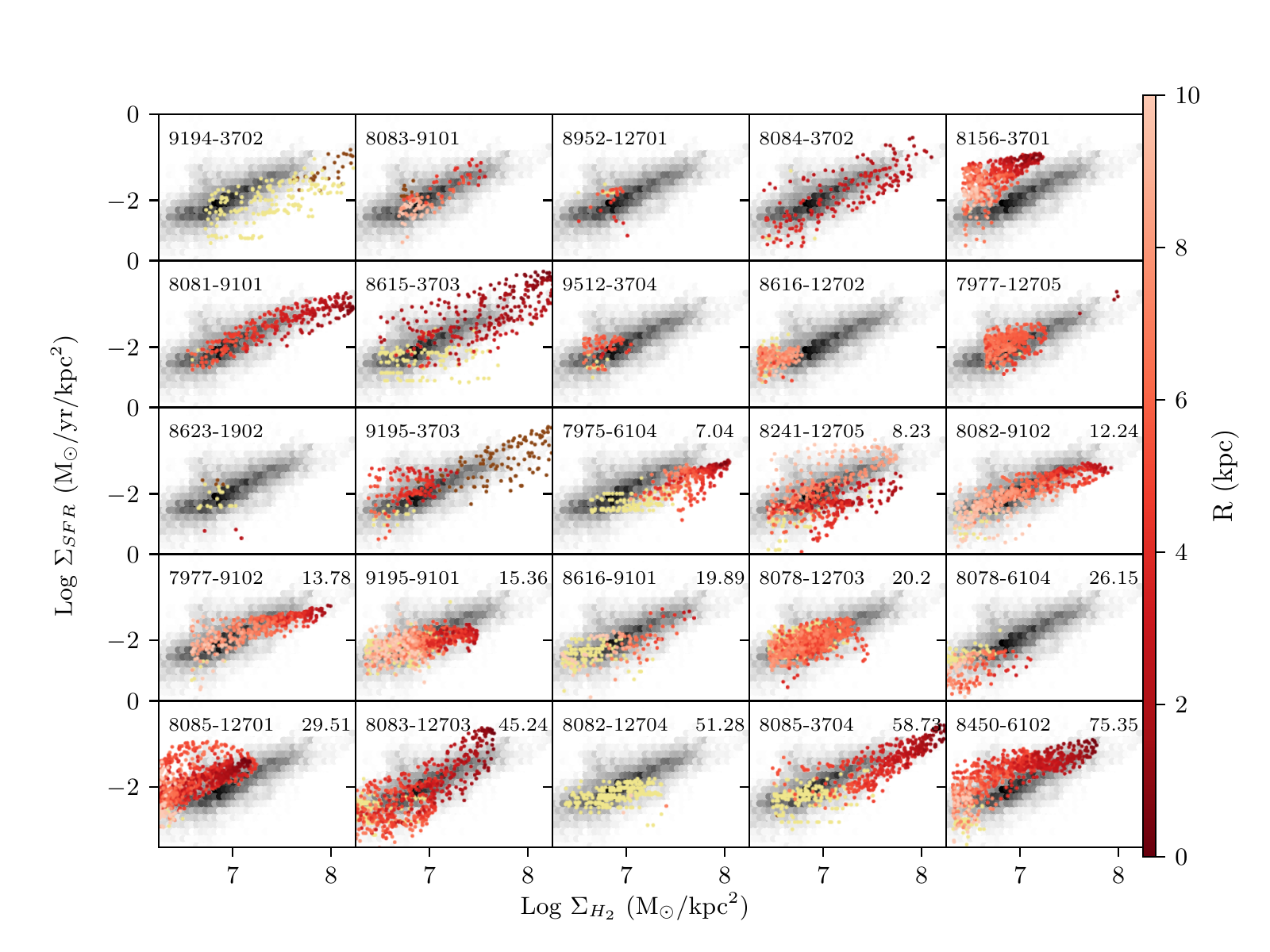}
	\centering
    \caption{Resolved Kennicutt-Schmidt relation for all merger galaxies, with the MaNGA plate-ifu of the galaxy in the left corner and $r_p$ (kpc) in the right corner (left blank for post-mergers). Star-forming spaxels are red and colour-coded by radius, and spaxels with \SigSFR determined by D4000 are yellow (low H$\alpha$ S/N) or brown (AGN contamination). The non-interacting galaxies in ALMaQUEST are shown as a grey histogram, for reference, (including both H$\alpha$-\SigSFR and D4000-\SigSFRend). By examining mergers on an individual basis we see many mergers that are offset, both above and below, the non-interacting rKS.}
    \label{fig:KS_resolved}
\end{figure*}

\begin{figure*}
	\includegraphics[width=0.95\textwidth]{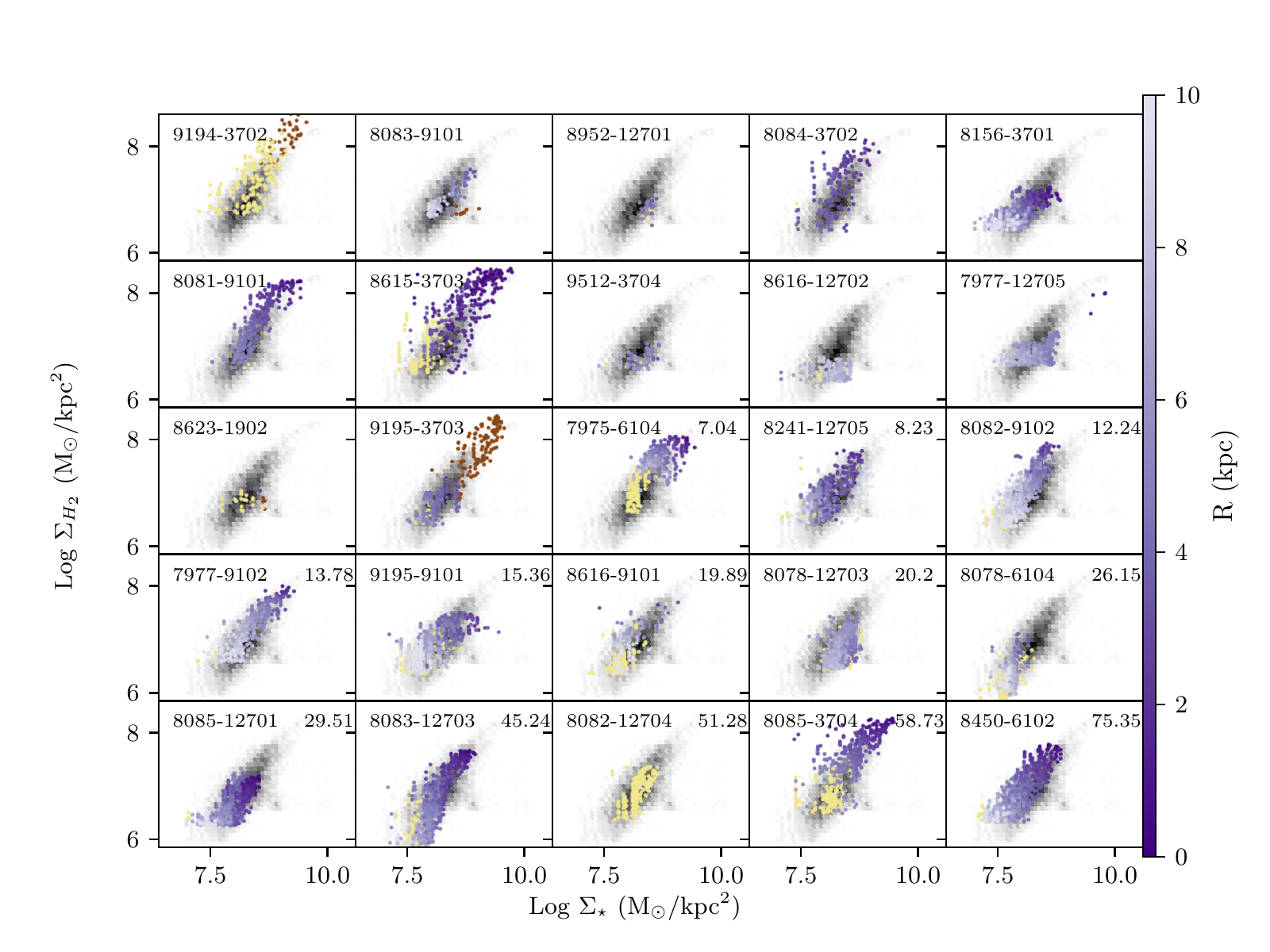}
	\centering
    \caption{Resolved molecular gas main sequence for all merger galaxies, with the MaNGA plate-ifu of the galaxy in the left corner and $r_p$ (kpc) in the right corner (left blank for post-mergers). Star-forming spaxels are colour-coded purple by radius, and spaxels with \SigSFR determined by D4000 are yellow (low H$\alpha$ S/N) or brown (AGN contamination). The non-interacting galaxies in ALMaQUEST are shown as a grey histogram, for reference. Many merging galaxies have tails at the high mass end of the molecular gas main sequence, at times for AGN spaxels as well (supporting a paradigm where mergers fuel AGN).}
    \label{fig:MGMS_resolved}
\end{figure*}

To maximize the number of spaxels with both CO and SFR measurements, we elect to approximate \SigSFR for spaxels that are not star-forming (based on our BPT and EW criteria) using the relationship between sSFR and D4000 found for both global \citep{Brinchmann2004TheUniverse} and local \citep{Spindler2018SDSS-IVEnvironment, Wang2019OnGalaxies,Bluck2020ArePhenomena} scales. We adopt an empirical approach similar to that in \cite{Bluck2020ArePhenomena}, who tested the validity of this approximation in the MaNGA {\sc pipe3d} data products. For bins of D4000 we compute a median sSFR value based on all star-forming spaxels (using our previous criteria). For any spaxel that does not meet our star-forming criteria, either due to low signal-to-noise or AGN contamination based on the BPT cut, we assign the median sSFR from the closest D4000 bin, and convert that to \SigSFR by multiplying sSFR and \Sigstarend. Figure \ref{fig:sSFR-D4000} displays the complete sSFR-D4000 distribution for star-forming spaxels in MaNGA along with a red line showing the median sSFR value, up to D4000=1.45, used to approximate \SigSFR when a spaxel is not star forming.

Spaxels with D4000>1.45 are generally quenched and therefore the sSFR-D4000 relation is no longer viable \citep{Bluck2020ArePhenomena}. As can be seen from Figure \ref{fig:sSFR-D4000}, the D4000-sSFR relation turns steeply downwards at D4000$>$1.4.  We therefore only estimate \SigSFR from D4000 when D4000$<$1.4, which represents a slightly stricter cut than that used in \cite{Bluck2020ArePhenomena}. Table \ref{tab:properties} provides the increased spaxel count when both H$\alpha$- and D4000-\SigSFR values are employed. Of the 13,046 spaxels with CO S/N>3, now 78.8\% of spaxels have measurable \SigSFR, almost doubled from the 44.5\% of our previous criteria. 9195-3702 and 8450-6102, despite having a considerable number of spaxels with CO S/N>3, still have less than 10 spaxels of overlap between good CO and \SigSFR measurements. These two galaxies are thus excluded from individual galaxy analysis later in this work, along with the other four galaxies previously mentioned. Thus we have 25 galaxies which can be studied on an individual basis, and we limit our studies to these 25 for the rest of the work. Multiple tests are performed to check how using D4000 approximated \SigSFR might impact our results, details of which are included in Appendix \ref{app:sSFR-d4000}. Adopting this method of measuring \SigSFR sacrifices accuracy for completeness in our analysis; to assess how this will impact our results we repeat key parts of our analysis using only H$\alpha-$\SigSFR (for galaxies where that is possible). Unless specified otherwise, for the rest of this work \SigSFR and any derived products, use the combined H$\alpha$+D4000 \SigSFR values for both the merger and isolated sample. 

Figure \ref{fig:ALMaQUEST_Ex} summarizes the various data products from MaNGA and ALMA used within this work for three galaxies from our sample: a post-merger galaxy (top), an interacting pair (middle), and a spectroscopic pair (bottom). We display maps of H$\alpha$-\SigSFRend, as well as our combined H$\alpha$+D4000-\SigSFR map. The middle galaxy is a clear case where we are able to recover a significant number of spaxels using the combined \SigSFR values. In particular, a significant fraction of spaxels in the outskirts of the galaxy are recovered to maximize the spatial coverage of our analysis. Though the middle galaxy demonstrates how the combined H$\alpha$+D4000-\SigSFR method recovers lower \SigSFR values, the method also recovers AGN dominated spaxels. Thus the main limitation of our analysis becomes the CO signal-to-noise, and the overlap between CO and \SigSFR.

\subsubsection{Spaxel Offsets from Resolved Relations}
\label{subsec: offsets}

Following \cite{Ellison2020The0}, we compute offsets from the rKS, rSFMS, and rMGMS to quantify how an individual spaxel may deviate from the average spaxel behaviour. All offsets are computed as the log difference between a spaxel value and the median value of a set of control spaxels that defines the average behaviour, i.e. $\Delta X = \log X - <\log X_{\textrm{control}}>_{\textrm{median}}$. For example, we define an offset from the SFMS as \dsigsfrend, with positive \dsigsfr values corresponding to enhanced star formation compared to the control sample. The sample of control spaxels is collected from all DR15 MaNGA spaxels from non-merging galaxies (as classified in Subsection \ref{subsec:selection}) with b/a$>$0.34 (excluding galaxies with inclination greater than 70\textdegree) and which are star-forming based on our cuts for the H$\alpha$-\SigSFR measurements described in Section \ref{subsec:SFR-D4000}. All MaNGA \SigSFR values are included, not just ALMaQUEST, such that the requirement for CO detections in ALMaQUEST spaxels does not bias our control spaxel set to higher \SigSFR. Note that we exclude D4000-\SigSFR values from the control sample, given we want to know the difference from the star-forming population that defines the resolved star-forming main sequence. The majority of spaxels on the rSFMS for which we can measure accurate D4000-\SigSFR tend to not meet our star-forming criteria due to low S/N in BPT diagnostic emission lines, rather than being classified as AGN or composite spaxels. Including D4000-\SigSFR in the control would lead to slightly lower \dsigsfr values as a result of the contribution from low S/N spaxels. We have tested whether the inclusion of D4000-\SigSFR in our control would alter any major conclusions of this work, and find our results remain unchanged. From the set of star-forming controls we select a subset that is matched within 0.1 dex \Sigstarend, 0.1 $R/R_{e}$ (where R is the inclination corrected galactocentric radius, and the effective radius $R_{e}$ is the \emph{r}-band half-light radius from the NSA catalogue), and 0.1 dex \Mstar of the merger spaxel. The median of this control set defines our ``regular'' star-forming behaviour. Similar offsets can be determined for the other scaling relations.

An offset from the resolved KS is referred to as \dSFEend; a value above the KS would have a larger star formation rate given the molecular gas in a spaxel, i.e. an enhancement in the efficiency at which gas is converted to stars. Rather than rely on S/N and BPT cuts to define star-forming spaxels, as was achieved with \dsigsfrend, we instead make use of the \dsigsfr value to construct the star-forming rKS. The control sample is also limited to galaxies in ALMaQUEST, rather than all of DR15, given \Siggas is required to compute \dSFEend. Thus the control sample is selected from non-interacting ALMaQUEST galaxies with b/a$>$0.34, with the additional cut of -0.5< \dsigsfrend <0.5 to select star-forming spaxels. A subset of the control is found by matching to the merger spaxel within 0.1 dex \Siggasend, 0.1 $R/R_{e}$, and 0.1 dex \Mstarend. 

Using the same control spaxel sample as for \dSFE, we determine offsets from the rMGMS, referred to as \dfgas. The control spaxels are again taken from non-interacting ALMaQUEST galaxies and matched within 0.1 dex \Siggasend, 0.1 $R/R_{e}$, and 0.1 dex \Mstarend. Unlike the other two offset metrics, \dfgas can be computed for any spaxels with CO S/N>3. Since there is no dependence on whether \SigSFR is measurable, or if measured \SigSFR values overlap with good CO measurements, maps of \dfgas tend to be more complete than the other two offsets. Maps of offset parameters for the merger sample are available in Appendix \ref{app:maps}.

One key difference between \cite{Ellison2020The0}, who first introduced these offset methods, and the methods used here is we do not match control spaxels by metallicity. Limiting our analysis to spaxels with valid metallicity measurements drastically diminishes our total spaxel count, as described in detail in Section \ref{subsec:ALMA}, limiting the spatial coverage of some galaxies and removing three from viable examination entirely. Nonetheless, we have repeated the analysis described in this section with a metallicity control (similar to that used in \citealt{Ellison2020The0}), and find little change in our key results for those galaxies on which this check can be performed (i.e., those with large CO+SFR(H$\alpha$) overlap).

\vfill
\section{Analysis}
\label{sec: analysis}

\subsection{Resolved Scaling Relations}
\label{subsec: Relations}

The stellar mass surface density (\Sigstarend), SFR surface density (\SigSFRend), and molecular gas surface density (\Siggasend) are all interconnected in three well established resolved relations: the resolved Kennicutt-Schmidt relation and the resolved molecular gas main sequence, which together drive the resolved star-forming main sequence\citep{Lin2019TheSequence, Ellison2021TheThem, Baker2022TheSequence}. We investigate these three scaling relations both for merger populations, and for individual galaxies in our merger sample.


Figure \ref{fig:All_scaling_relations} compares the resolved scaling relations for post-merger (orange) and pair (blue) galaxies with respect to the relatively isolated galaxies in the rest of ALMaQUEST (grey). Histograms of the \SigSFRend, \Siggasend, and \Sigstar are provided for context as well. Note that the grey histogram does not represent the control spaxel population used to compute various offset parameters as described in Subsection \ref{subsec: offsets}, rather all spaxels where \SigSFR can be measured for ALMaQUEST. Given the selection of the original ALMaQUEST sample, including a sample of starburst galaxies, the non-interacting spaxels may be biased to high \SigSFRend. This is likely why the median \SigSFR for non-interacting galaxies (shown as a dashed line in the \SigSFR histogram) is slightly larger than the median \SigSFR of the pair sample. We perform Kolmagorov-Smirnov (KS) tests between all three samples to ascertain whether the pair or post-merger \SigSFRend/\Siggasend/\Sigstarend values could be drawn randomly from the non-interacting spaxel sample. We find the probability of this hypothesis to be approximately zero for all three properties ($P_{KS}\approx0$). The pair sample has a tail towards lower \Siggasend, which seems to be driven by two individual galaxies with uniquely low \Sigstar and \Siggas values (8078-6104 and 8083-12703). The post-merger sample also has a tail towards large \SigSFRend, \Siggasend, and \Sigstar values, which manifests at the upper end of each scaling relation. Along with the clear lack of post-merger spaxels with \Siggas$<6.5$log $\textrm{M}_{\odot}/\textrm{kpc}^2$, it is clear that our post-merger sample probes regions of heightened SFR and molecular gas properties.

There is significant diversity on a galaxy-per-galaxy basis to consider, as has been seen in other studies (\citealt{Vulcani2019GASP.Galaxies,Ellison2021TheThem,Pessa2021AstronomyScale}, Brown et al. in prep). Figure \ref{fig:SFMS_resolved} shows the rSFMS for each post-merger and pair galaxy. The star-forming spaxels are colour-coded by radius, while the D4000-\SigSFR are a plain yellow if they did not meet our S/N cuts or brown if they are AGN based on the \cite{Kauffmann2003TheAGN} criteria. The non-interacting ALMaQUEST spaxels are shown as a grey histogram for comparison (including both H$\alpha$-\SigSFR and D4000-\SigSFRend). By examining galaxies as individuals we see clear divergence from the spaxels in non-interacting galaxies, as well as a diversity of behaviour within post-merger and pair classifications. Many mergers show a large population of spaxels above the star-forming main sequence, as is expected if mergers trigger star formation, such as 8156-3701 (1st row, 5th column), 8615-3703 (2nd row, 2nd column), and 8085-12701 (5th row, 1st column). We note that three post-mergers (8084-3702, 8156-3701, 8615-3703) and one pair (8450-6102) were selected from the starburst sample as part of the original ALMaQUEST (ALMA program 2018.1.00541.S, PI: Ellison), which could bias our merger sample towards strongly star-forming galaxies. Yet many spaxels and at times entire mergers are also below the star-forming main sequence, such as 9194-3702 (1st row, 1st column), 7975-6104 (3rd row, 3rd column), and 8078-12703 (4th row, 4th column). In particular, both galaxies with concentrations of ``AGN'' spaxels as defined by the BPT diagram tend to lie below the non-interacting rSFMS (see 9194-3702 (1st row, 1st column) and 9195-3703 (3rd row, 2nd column)), as has been observed previously with global quantities \citep{Ellison2016TheGalaxies,Leslie2016QuenchingSequence,Mcpartland2019DissectingUniverse}. This range of enhanced and suppressed star formation rates drives much of the overall scatter of our sample on the global star-forming main sequence (refer back to Figure \ref{fig:Merger_SFMS}).

Mergers present diverse behaviour in the rKS as well, as shown in in Figure \ref{fig:KS_resolved}. Many post-mergers and pairs have overall greater scatter in the rKS than the non-interacting sample (see 8615-3703 (2nd row, 2nd column), 8085-12701 (5th row, 1st column), 8083-12703 (5th row, 2nd column)), looking ``puffier''; a similar effect as was seen in \cite{Ellison2021TheThem}. The increased scatter could represent smaller regions of enhanced SFE in a merger that, on average, has a low or normal global SFE. Note that although D4000-\SigSFR values tend to be lower on the rKS, many still overlap with H$\alpha$-\SigSFR spaxels. Spaxels which are classified as ``AGN'' by a BPT diagram (brown in the figure) tend to exist at larger \Siggas and \SigSFR values, reflecting their central location in the galaxy and the overall radial dependence of the rKS (smaller radii predominantly fill the upper end of the rKS). However, AGN spaxels at large \Siggas could also support a scenario where the infall of molecular gas fuels a central AGN. The two largest AGN spaxel populations belonging to post-mergers appears to support such a scenario, though two galaxies alone are not enough to determine if one scenario is more likely than another. Such a query can be further investigated by looking for enhancements in the molecular gas main sequence at high \Siggas values.

Figure \ref{fig:MGMS_resolved} appears to confirm that variations in the rMGMS are the least drastic of the three relations, as was found for non-interacting galaxies by \cite{Ellison2021TheThem}. Interestingly, the AGN spaxels (shown in brown) are not offset to below the rMGMS, as would be implied by the comparatively low gas fractions found for AGN spaxels with MaNGA \citep{Sanchez2018SDSSGalaxies} and EDGE-CALIFA \citep{Ellison2021TheQuenching}. Rather the AGN spaxels in the merger sample are comparable to star-forming isolated spaxels, or even enhanced in the case of 9194-3702 (1st row, 1st column) and 9195-3703 (3rd row, 2nd column). What seems to be an inconsistency may  stem from the unique nature of the merger stage. Rather than capture central gas depletion triggered by an AGN, as in the isolated galaxies of \cite{Ellison2021TheQuenching}, mergers may have more recently funnelled molecular gas to the galaxy's centre which has yet to be consumed by the AGN. In the next section we will investigate what drives a merger galaxy to be offset from the resolved scaling relations by comparing spaxel offsets from scaling relations.

\subsection{Efficiency vs. Fuel Driven Enhanced Star Formation}
\label{subsec: which mechanism?}

\begin{figure*}
	\includegraphics[width=0.95\textwidth]{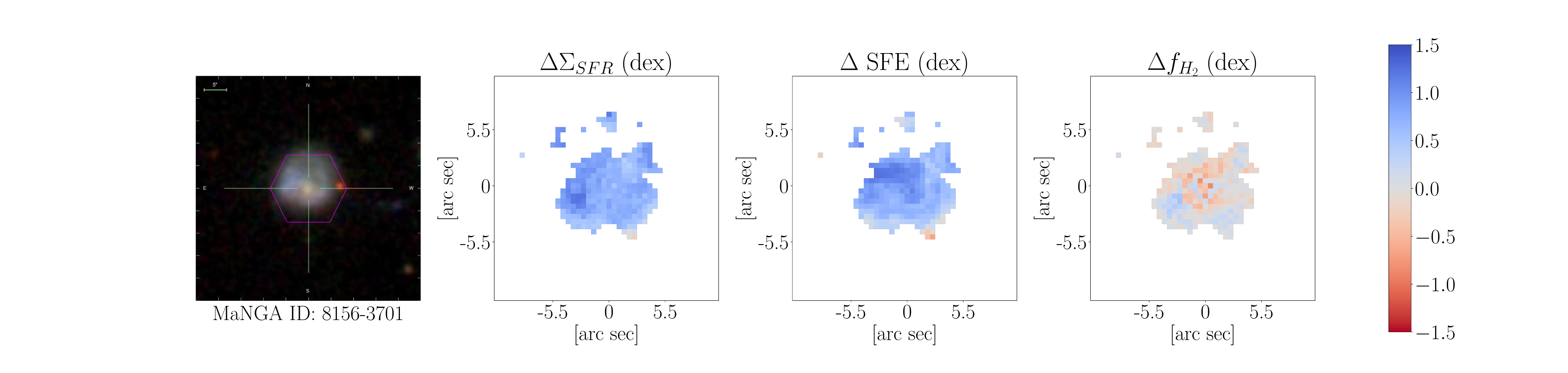}
	\centering
    \caption{Maps of the offset parameters for an example post-merger galaxy. 1st column: the SDSS \emph{gri}-image of the galaxy, for reference. 2nd column: \dsigsfr distribution, with blue representing an enhancement in star formation. 3rd column: \dSFE distribution, with blue representing enhanced star formation efficiency. 4th column: \dfgas distribution, with blue representing enhanced gas fraction in a spaxel. The correlation between a uniform enhancement in \SigSFR and SFE, along with the patchy deficits of molecular gas, imply that the star formation in this galaxy is driven by an enhanced efficiency at which gas is converted to stars.}
    \label{fig:offset_maps}
\end{figure*}

We can use maps of spaxel offsets from resolved scaling relations (\dsigsfrend, \dSFEend, and \dfgasend, as described in Subsection \ref{subsec: offsets}), to discern whether changes in SFE or \fgas drive enhancements in star formation rate. Figure \ref{fig:offset_maps}  shows the offset maps for an example post-merger galaxy, maps for the entire sample are available in Appendix \ref{app:maps}, for which we can attempt to discern whether star formation is driven by enhanced SFE or enhanced \fgasend. Star formation in this post-merger is generally enhanced (\dsigsfrend$>0$), as is expected given the average spatial enhancement of post-mergers (see \citealt{Thorp2019SpatiallyMaNGA}). That enhancement corresponds to a global enhancement in star formation efficiency (positive \dSFEend). Interestingly, this post-merger mostly has a deficit in molecular gas (negative \dfgasend), with some smaller regions of surplus gas. Figure \ref{fig:offset_maps} thus implies that the boost of star formation in this post-merger in particular is predominantly driven by an enhanced efficiency at which gas is converted to stars, not an enhanced amount of gas to fuel that star formation.

\begin{figure*}
	\includegraphics[width=0.95\textwidth]{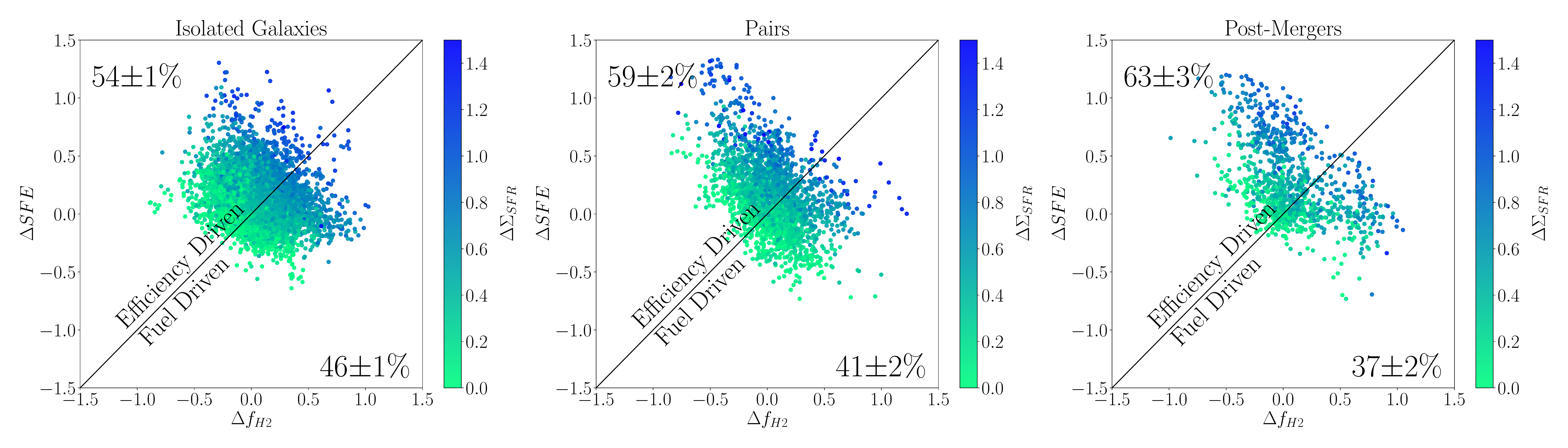}
	\centering
    \caption{\dSFE versus \dfgas for all spaxels that have \dsigsfrend$>$0, separated into spaxels which belong to isolated galaxies (left), pair galaxies (middle), and post-merger galaxies (right). Spaxels are colour-coded by \dsigsfrend. The line of equality is shown to delineate the efficiency driven regime (where \dSFEend$>$\dfgasend) and the fuel driven regime (where \dSFEend$<$\dfgasend). The percentage of isolated/pair/post-merger spaxels in each regime is overplotted on the figure, with the uncertainty in the percentage being propogated error of the number of spaxels above/below the line divided by the total number of spaxels. Based on these percentages there is a slight bias towards efficiency driven star formation in post-mergers and pairs (though more so in the former), but other than that the amount of efficiency and fuelled spaxels with enhanced star formation is relatively even.}
    \label{fig:SFMS_mechanism}
\end{figure*}

From the offset maps alone we can infer some interesting behaviour. Many pair galaxies, for example, have large \dfgas values along spiral arms (see 9195-9101, 7977-9102, and 8241-12705 in Appendix \ref{app:maps}). Some pair galaxies have central enhancements in \Siggas that correspond to a central burst of star formation, such as 9195-3703 and 8078-6104. These two galaxies support a scenario where inflow of molecular gas fuels merger triggered central star formation. Post-merger galaxies 8615-3703 and 8084-3702 have an excess of molecular gas across the galaxy's surface, rather than a central concentration. Yet that is not a universal scenario; within our sample, 8156-3701 has one of the strongest enhancements in SFR, but has suppressed \Siggas across the galaxy.

Although the distribution of these offset parameters can reveal interesting results for individual galaxy behaviour, it is difficult to extract general trends for merging and post-merger galaxies from visual examination alone. In particular it is difficult to parse whether enhanced SFE or enhanced \Siggas drives any merger-triggered star formation. To discern which mechanism is likely more influential over star formation enhancements, we implement an analysis which includes all three offset parameters in a single diagram as pioneered by \cite{Ellison2020The0} using ALMaQUEST and \cite{Moreno2021SpatiallyInteractions} using simulations.

Figure \ref{fig:SFMS_mechanism} shows the offsets in star formation efficiency versus the offsets in molecular gas, for all spaxels with enhanced star formation (\dsigsfrend$>0$). We separate the galaxies into three categories: the non-interacting set from ALMaQUEST, pair galaxies, and post-merger galaxies. For these diagrams we plot the line of equality which is used to distinguish an ``efficiency driven'' and ``fuel driven'' regime, i.e. where one offset is greater than the other. If a spaxel lies above this line of equality, then \dSFE$>$\dfgasend; if \dsigsfrend$>$0 this would imply enhanced efficiency is prompting the enhanced star formation, more so than the gas fraction. All three galaxy populations have a relatively equal percentage of spaxels in the efficiency driven and fuel driven regime, implying that enhanced star formation is equally driven by enhanced fuel and SFE when all spaxels are examined together. There is a slight bias towards efficiency driven star formation for the post-merger spaxels, with 63\% of spaxels in the efficiency driven regime. We remind the reader that the original ALMaQUEST sample includes 11 galaxies that were deliberately selected to be starbursts \citep{Ellison2020The0}. Three of these starbursts are in our post-merger sample, one is in the pairs sample, and 12 are in our isolated sample. Since \cite{Ellison2020The0} showed that starbursts tend to be dominated by high SFEs, the isolated sample in Figure \ref{fig:SFMS_mechanism} is not statistically representative of a normal galaxy distribution.

\begin{figure}
	\includegraphics[width=0.85\columnwidth]{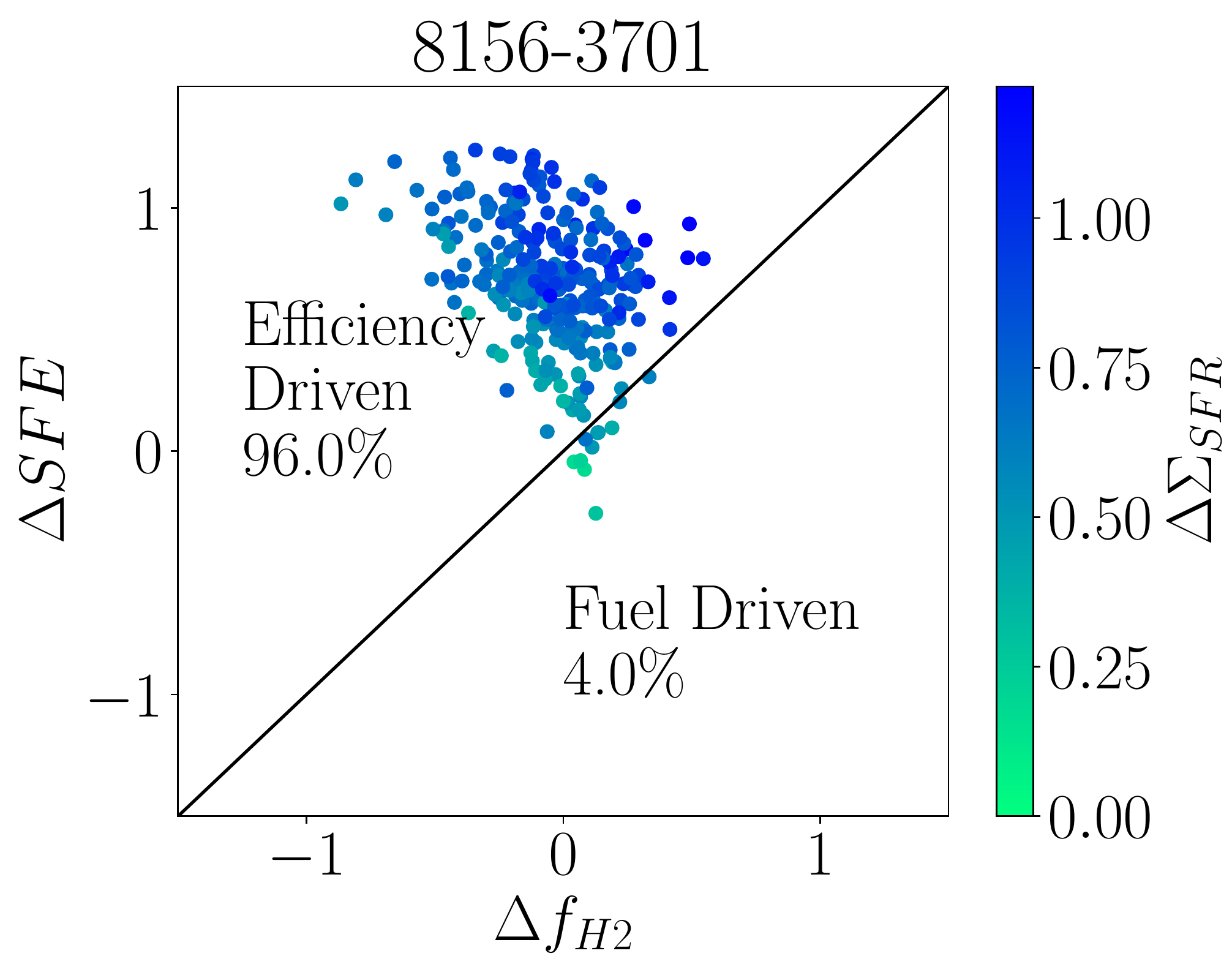}
	\centering
    \caption{\dSFE versus \dfgas for all spaxels with \dsigsfrend$>$0 in an example post-merger galaxy (MaNGA plate-ifu = 8156-3701). In this galaxy almost all spaxels are in the efficiency driven regime, as we might expect from examining the distribution of offset parameters in Figure \ref{fig:offset_maps}.}
    \label{fig:Example_Mechanism}
\end{figure}

Figure \ref{fig:SFMS_mechanism} might lead us to believe that enhancements in both SFE and \fgas are equally important for driving SFR enhancements in both post-mergers and pairs. However, in the previous subsection, we found that ensemble spaxel distributions showed considerable diversity when plotted on a galaxy-by-galaxy basis. We now investigate whether the same is true for distributions of offset properties. Figure \ref{fig:Example_Mechanism} replicates the combined offset diagram from Figure \ref{fig:SFMS_mechanism}, but for a single post-merger galaxy (the same one that is shown in Figure \ref{fig:offset_maps}). 96\% of spaxels with \dsigsfrend$>0$ in this galaxy are in the efficiency driven regime, confirming that the star formation in this galaxy is driven by enhanced SFE as is evident from visually inspecting Figure \ref{fig:offset_maps}. Figure \ref{fig:Example_Mechanism} highlights that whilst the ensemble of spaxels across all galaxies might have a relatively even split between those whose star formation is enhanced by SFE or gas fraction (Figure \ref{fig:SFMS_mechanism}), individual galaxies may be strongly driven by one process or the other.

\begin{figure*}
	\includegraphics[width=0.95\textwidth]{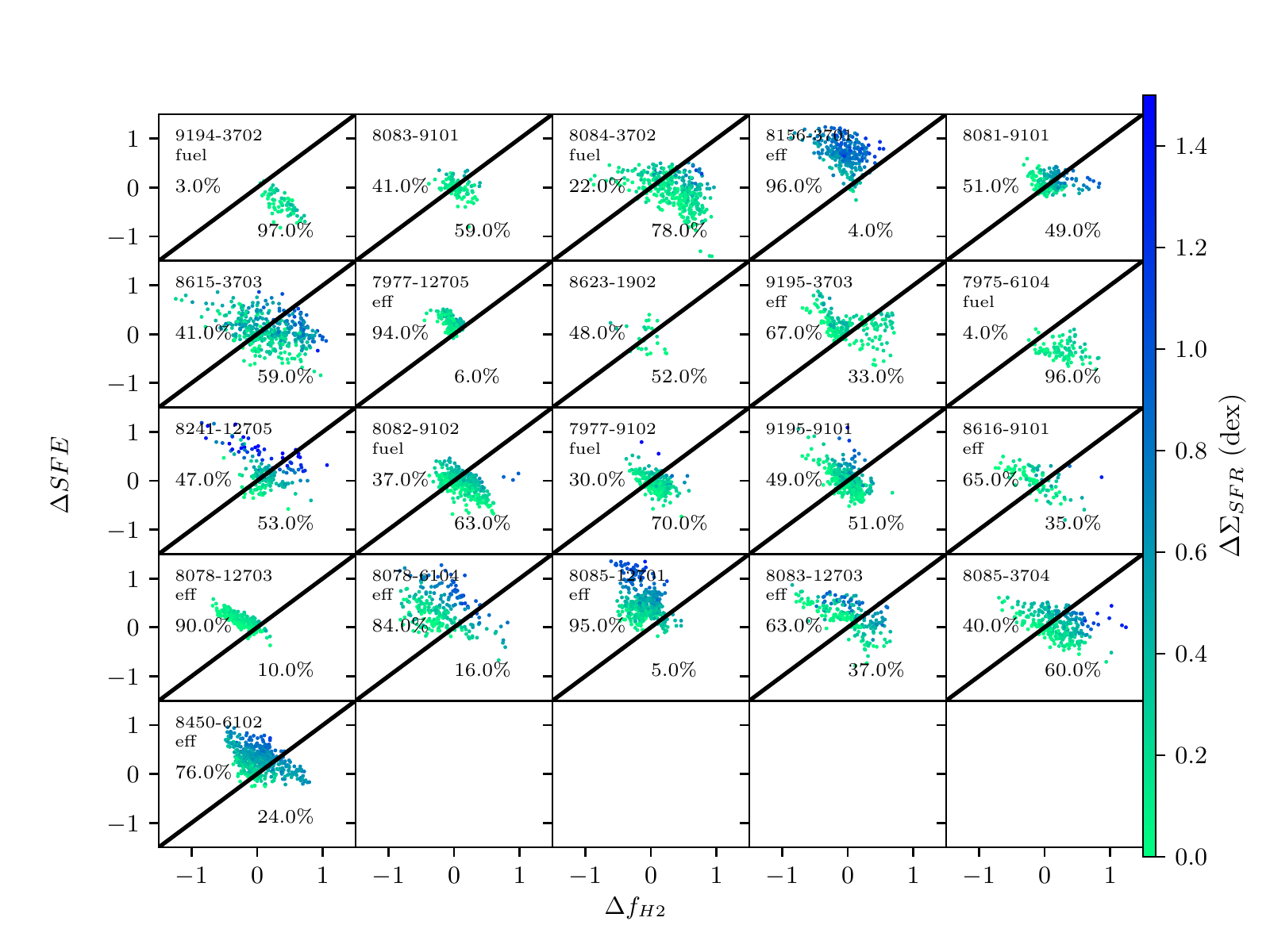}
	\centering
    \caption{\dSFE versus \dfgas for all spaxels with \dsigsfrend$>$0 in each post-merger galaxies with more than 20 spaxels with \dsigsfrend$>$0, the plate-ifu number of each galaxy is provided in the left corner of every plot. Spaxels are colour-coded by \dsigsfrend. The percentage of spaxels above and below the line of equality are provided as well, which help discern whether the star formation in a galaxy is predominantly driven by enhanced SFE or enhanced \fgasend. We label galaxies which are predominantly efficiency driven ($<40\%$ below the line) and fuel driven ($>60\%$ above the line) for reference. For some galaxies there is clearly a dominant mechanism driving star formation, but many galaxies have contributions from both efficiency and fuel driven SFR enhancements.}
    \label{fig:delta_plots_all}
\end{figure*}

Offset diagrams like that in Figure \ref{fig:Example_Mechanism} are provided for all mergers which have more than 20 spaxels where \dsigsfrend$>0$ in Figure \ref{fig:delta_plots_all}, including the percentage of spaxels in the fuel/efficiency driven regimes. There are cases of both post-merger and pair galaxies that are indisputably driven by a single mechanism, such as 9194-3702 (fuel driven) or 8085-12701 (efficiency driven), but for the majority the dominant star formation driver is less clear. In order to classify the dominance of fuel or efficiency in driving star formation enhancements on a galaxy-by-galaxy basis, we quantify the percentage of spaxels in the fuel/efficiency driven regime of our offset diagrams. If more than 60$\%$ of a galaxy's \dsigsfrend$>0$ spaxels are in the fuel-driven regime (i.e., \dfgasend$>$\dSFEend), then we classify the galaxy as ``fuel-driven''. If less than 40$\%$ of a galaxy's \dsigsfrend$>0$ spaxels are in the fuel-driven regime, then we classify the galaxy as ``efficiency driven''. If the percentage of spaxels in the fuel-driven regime is between 40$\%$ and 60$\%$, we can assume that enhanced amounts of gas and enhanced SFE are approximately equally contributing to the enhanced star formation. Note that this method of classification gives equal weight to all spaxels, regardless of the strength of star-formation in each spaxel. By doing so our analysis focuses on what mechanism drives star-formation for most regions in a galaxy, not necessarily the mechanism which drives ``the most'' star-formation.

\begin{table}
	\centering
	\caption{The fraction of \dsigsfrend$>0$ spaxels in the fuel driven regime (i.e., \dfgasend$>$\dSFEend) for each merger with more than 20 \dsigsfrend$>0$ spaxels. The fraction of fuel driven spaxels is computed using the combined H$\alpha$+D4000 \SigSFRend, as well as just the H$\alpha$ star formation rates (with appropriate cuts). Note some galaxies do not have fractions computed for H$\alpha$, because they do not have more than 20 spaxels that meet our star-forming cuts. Fractions greater than 0.6, what we consider a fuel driven merger, are coloured blue. Fractions less than 0.4, what we consider an efficiency driven merger, are coloured red. Those between 0.4 and 0.6 are left black, since both mechanisms contribute relatively equally.}
	\label{tab:mechanism}
	\begin{tabular}{ccc} 
		\hline
		plate-ifu & Fraction of & Fraction of\\
		 & Fuel Driven Spaxels & Fuel Driven Spaxels\\
		 &  &  (H$\alpha$ only)\\
		\hline
		\hline
		9195-3702 & - & - \\
        9194-3702 & \textcolor{blue}{0.97} & - \\
        8083-9101 & 0.59 & 0.50 \\
        8952-12701 & - & - \\
        8084-3702 & \textcolor{blue}{0.78} & \textcolor{blue}{0.75} \\
        8156-3701 & \textcolor{red}{0.04} & \textcolor{red}{0.04} \\ 
        8081-9101 & 0.50 & 0.53 \\
        8615-3703 & \textcolor{blue}{0.60} & \textcolor{blue}{0.64} \\
        9512-3704 & - & - \\ 
        7977-12705 & \textcolor{red}{0.06} & \textcolor{red}{0.06} \\
        8623-1902 & 0.52 & - \\ 
        8616-12702 & - & - \\
        9195-3703 & \textcolor{red}{0.33} & \textcolor{red}{0.05} \\ 
        \hline
        7975-6104 & \textcolor{blue}{0.97} & \textcolor{blue}{0.97} \\ 
        8241-12705 & 0.52 & 0.52 \\
        8082-9102 & \textcolor{blue}{0.64} & \textcolor{blue}{0.62} \\
        7977-9102 & \textcolor{blue}{0.69} & \textcolor{blue}{0.68} \\ 
        9195-9101 & 0.51 & 0.55 \\
        8616-9101 & \textcolor{red}{0.34} & \textcolor{blue}{0.60} \\ 
        8078-12703 & \textcolor{red}{0.10} & \textcolor{red}{0.14} \\ 
        8078-6104 & \textcolor{red}{0.16} & \textcolor{red}{0.21} \\
        8085-12701 & \textcolor{red}{0.05} & \textcolor{red}{0.05} \\
        8083-12703 & \textcolor{red}{0.38} & \textcolor{red}{0.39} \\ 
        8082-12704 & - & - \\ 
        8085-3704 & \textcolor{blue}{0.60} & \textcolor{blue}{0.67} \\ 
        8450-6102 & \textcolor{red}{0.24} & \textcolor{red}{0.24} \\
		\hline
	\end{tabular}
\end{table}

The fraction of fuel driven spaxels for every merger is listed in Table \ref{tab:mechanism}. We calculate this fraction for both our combined \SigSFR values and for the H$\alpha$-\SigSFR only, although some mergers do not have enough H$\alpha$-\SigSFR spaxels to meet our criteria. For galaxies which have enough H$\alpha$-\SigSFR spaxels to measure the fuel-driven fraction, we find adding D4000-\SigSFR to the analysis does not change the classification of the galaxy as fuel or efficiency driven (except for 8616-9101). Interestingly, those galaxies we are only able to analyze with the inclusion of D4000-\SigSFR tend to have large fractions of fuel driven spaxels, i.e. the star formation is driven by an excess of fuel. All four recovered galaxies miss our H$\alpha$-\SigSFR criteria based on low signal-to-noise, implying that the star formation in these galaxies is truly low (and thus a small or negative \dSFE is to be expected). Further discussion on how D4000-\SigSFR impacts the fraction of fuel driven spaxels is included in Appendix \ref{app:sSFR-d4000}, where we find the inclusion of D4000-\SigSFR does not drastically change our results. For both \SigSFR measurement methods there is a relatively similar number of fuel and efficiency driven galaxies, with slightly fewer galaxies driven by neither mechanism. We also consider whether variations in the fraction of fuel driven spaxels could be driven by inaccuracies resulting from using a constant $\alpha_{\textrm{CO}}$ conversion factor. We investigate the error in \Siggas from using a constant conversion factor, and its impact on the fraction of fuel driven spaxels, in Appendix \ref{app:alpha_CO}, and find our results are robust to variations in $\alpha_{\textrm{CO}}$.

\begin{figure}
	\includegraphics[width=0.9\columnwidth]{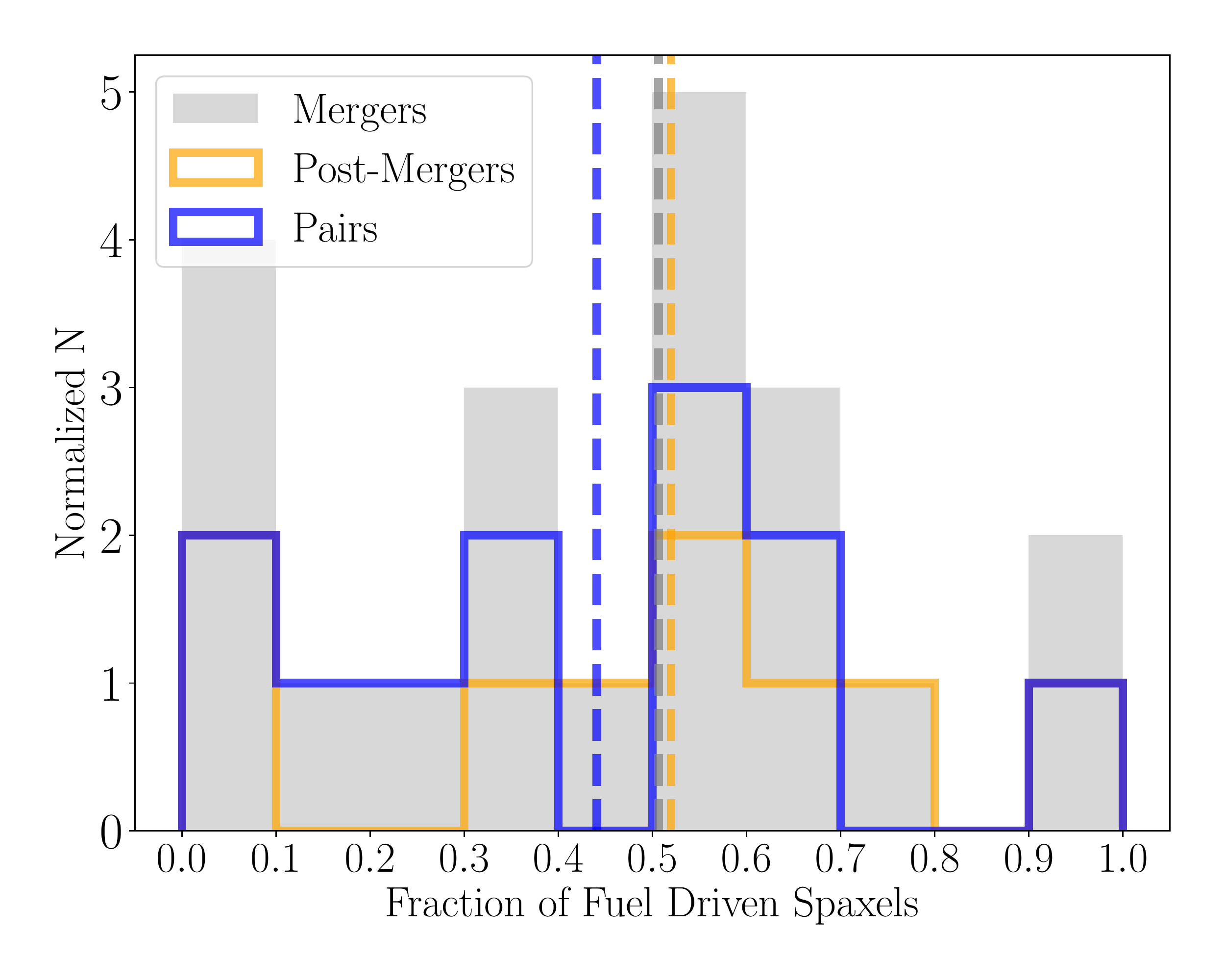}
	\centering
    \caption{Distributions of the fraction of fuel driven spaxels for the merger sample (grey), along with the pair (blue) and post-merger (orange) sub-samples. The median of each distribution is shown with a vertical dashed line: 0.51 for the entire merger sample, 0.44 for pairs, and 0.52 for post-mergers. The median value of each sub-smaple being so similar, and so close to 0.5, confirms that there are an equal number of fuel and efficiency driven mergers.}
    \label{fig:FF_hist}
\end{figure}

Figure \ref{fig:FF_hist} displays the distribution of the fraction of fuel driven spaxels, both for the merger sample as a whole (grey) and the pair (blue) and post-merger (orange) sub-samples. Both the post-merger and pair sample show relatively equal distribution in fuel fraction implying a diversity of star formation drivers throughout the interaction process. There is an intriguing excess of extremely low fuel fractions (below 0.1), compared to the more uniform distribution above a fuel fraction of 0.5, implying that ``efficiency driven'' mergers have \dSFEend$>$\dfgas in most spaxels. However, given only 3 post-mergers are below a fuel fraction of 0.4, this could be the result of small number statistics.

Both observations \citep{Ellison2013GalaxyGalaxies, Patton2013GalaxyGalaxies,Knapen2015InteractingFormation} and simulations \citep{Cox2008TheStarbursts,Torrey2012THEGALAXIES,Moreno2019InteractingContent,Patton2020InteractingContext} have demonstrated that merger induced star formation peaks at coalescence. Despite the diversity demonstrated in Table 2, it is interesting to note that 2 of the 3 post-mergers with a global $\Delta$SFR $>0.5$ are fuel driven enhancements (9194-3702, 8084-3702). We might wonder, then, if there is some correlation between the fraction of fuel driven spaxels and the interaction stage, with the largest fraction of fuel driven spaxels occurring at coalescence when the star formation is strongest. The higher percentage of efficiency driven galaxies in the pair sample (50\% as opposed to 33\% in the post-mergers) would agree with a scenario where efficiency driven star-formation becomes less important towards coalescence. Such a correlation would fit with the hypothesis that merger-induced star formation stems from non-axisymmetric structures causing a loss of angular momentum for gas, which then funnels into the centre of the galaxy towards the end of an interaction.

\begin{figure}
	\includegraphics[width=0.95\columnwidth]{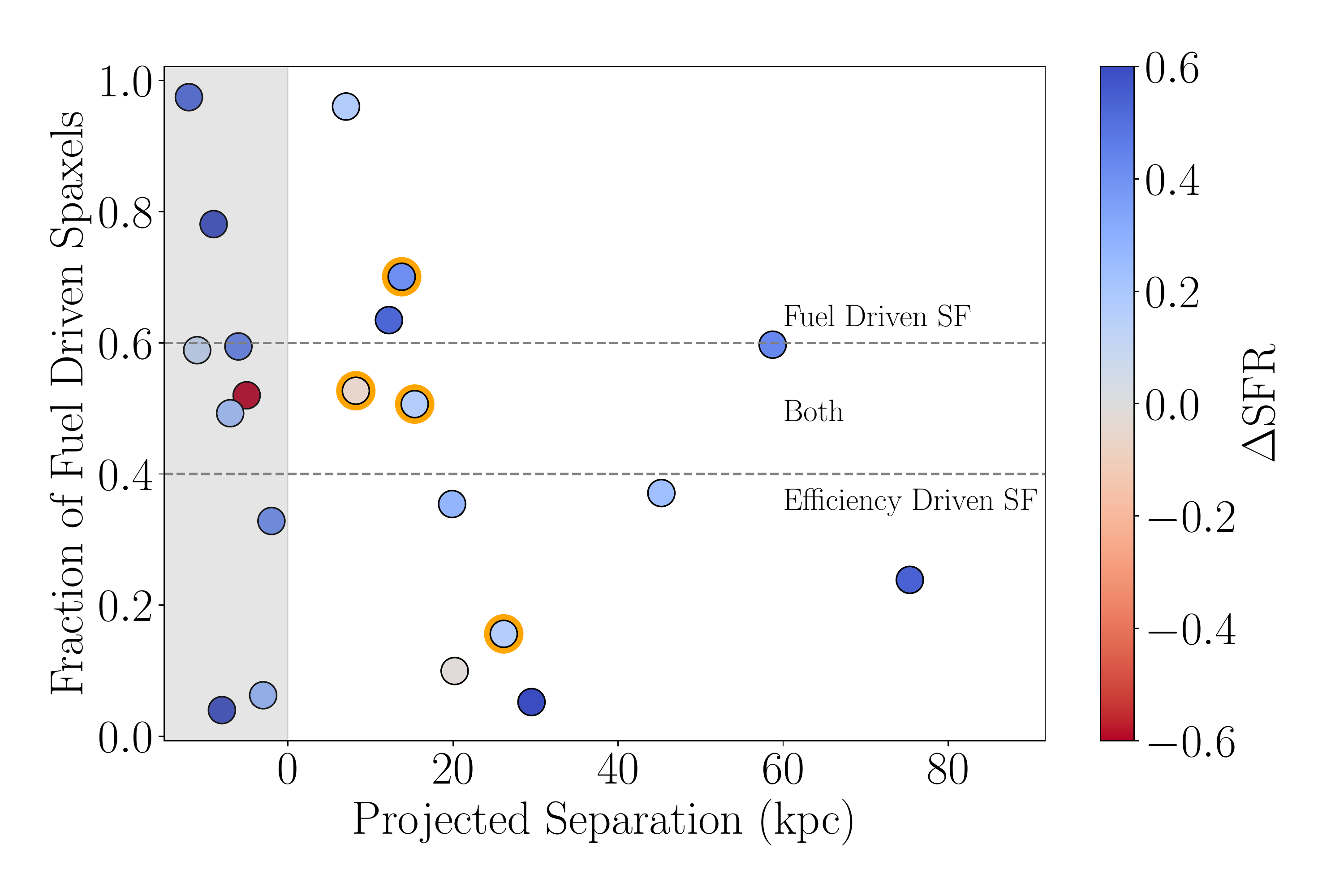}
	\centering
    \caption{Fraction of fuel driven spaxels as a function of projected separation in kpc, with post-mergers in the grey region (since projected separation cannot be measured without a companion). All points are colour-coded by the galaxy's global offset from the star-forming main sequence, $\Delta$SFR (bluer points have globally enhanced SFR, red points have globally suppressed SFR, compared with the global SFMS). Note that galaxies with negative $\Delta$SFR can still have regions of enhanced star-formation, which we use to compute a fuel fraction. Pairs that have tidal arms connecting the two galaxies are highlighted with orange. Neither projected separation, nor this secondary morphology characteristic, show any trend with the fraction of fuel driven spaxels.}
    \label{fig:Fraction_mechanism}
\end{figure}

Figure \ref{fig:Fraction_mechanism} is a visual representation of Table \ref{tab:mechanism}, showing the fraction of fuel driven spaxels as a function of projected separation between pairs, with each point colour-coded by global offset from the star-forming main sequence. The post-mergers are shown with a projected separation as arbitrary values below zero, in the grey bar. Figure \ref{fig:Fraction_mechanism} reveals no clear correlation with the fraction of fuel driven spaxels and the projected separation (Pearson's correlation coefficient=$-$0.355). Thus there is no evidence from our sample that the merger stage is strongly linked to whether enhanced star formation is driven by elevated fuel or efficiency. It is important to note that projected separation is not a perfect representation of merger stage due to both uncertainty on whether a galaxy has already completed one or more pericentric passages, as well as projection effects. Morphological signs can serve as an additional indicator for interaction stage given galaxies are most disturbed at first passage and right before coalescence \citep{Lotz2004AClassification,Lotz2008GalaxyMergers,Hambleton2011AdvancedGalaxies}. Unfortunately most of our mergers have some degree of visual disturbance, making it difficult to directly link morphology to a distinct interaction stage (outside of pre- and post-coalescence). We attempt to account for this by highlighting pair galaxies which are connected to their companion by a tidal arm on Figure \ref{fig:Fraction_mechanism}, as a more concrete indicator that first passage has already occurred. Even with this additional category, no trend with merger stage and fraction of fuel driven spaxels arises, further affirming that interaction stage plays a secondary role influencing how star formation is enhanced. 

Figure \ref{fig:Fraction_mechanism} encapsulates one of the main conclusions of this paper: interacting galaxies show a broad diversity of the relative contributions of fuel and efficiency, with no obvious dependence on the merger stage we adopted. This sample suggests that the properties of the galaxies themselves (e.g. gas fractions, morphologies) as well as the details of the interaction (orbit and mass ratio) play a larger role in how star formation is enhanced. We have checked to see if the fraction of fuel driven spaxels correlates with \Mstarend, \Mgasend, \fgasend, SFR, and $\Delta$SFR, and found only \Mgas and \fgas have a Pearson's correlation coefficient greater than 0.6 (note that \fgas and \Mgas are themselves correlated, with a Pearson's correlation coefficient of 0.69). However, a correlation between the fraction of spaxels driven by enhanced amounts of molecular gas and large gas fractions is not a particularly noteworthy result. More likely it is the combined properties of the galaxy and its companion that drive the diversity of star formation properties observed. Recent simulations have found that mass ratio and orbital geometry can both impact how star formation is powered in mergers \citep{Moreno2021SpatiallyInteractions}. It is possible that, with a larger sample, more pronounced correlations with merger stage or integrated galaxy property would emerge, but for this sample such trends are not obvious.

\begin{figure*}
	\includegraphics[width=0.9\textwidth]{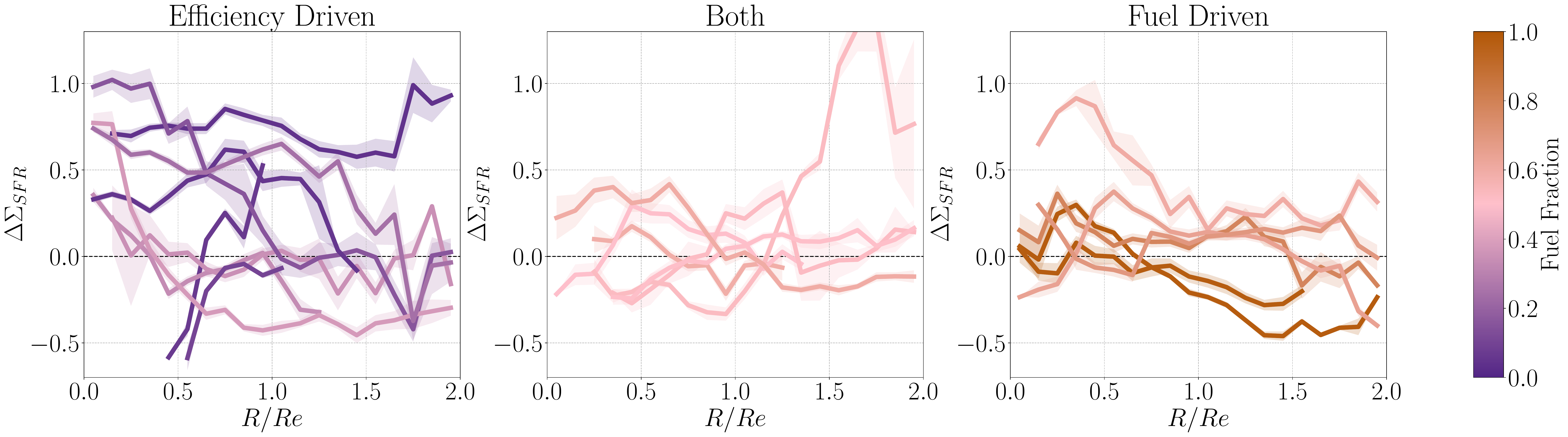}
	\centering
    \caption{Radial profiles of the median \dsigsfr in bins of effective radius for post-merger and pair galaxies included in our analyses, separated into groups of ``Efficiency Driven'', ``Both'', and ``Fuel Driven'' based on the fuel fraction value (the second column of Table \ref{tab:mechanism}). Each profile is also colour-coded by the fuel fraction. Radial profiles are plotted as a function of the effective radius, with the solid line representing the median in a bin of 0.1 $R_e$, and the faded width of the line representing the error on the mean. Efficiency driven mergers tend to have the strongest and most extended enhancements in star formation. Whereas fuel driven mergers are more likely to have centrally concentrated bursts of star formation, likely due to an inflow of gas.}
    \label{fig:profiles}
\end{figure*}

\subsection{What drives resolved changes in star formation?}
\label{sec:Discussion}

We have shown that, for both pair and post-merger galaxies, either enhanced SFE or enhanced \fgas can drive the subsequent star formation enhancement that results from an interaction. The merger sample is split evenly between galaxies whose star formation driven by SFE, by \fgasend, and those equally driven by both. An underlying issue yet to be addressed is how the mechanism which powers star formation relates to the spatial variations in star formation.

Figure \ref{fig:profiles} displays radial profiles of \dsigsfr for the merger sample separated into distinct categories of efficiency driven, fuel driven, or driven by both based on the fraction of fuel driven spaxels in Table \ref{tab:mechanism} (profiles are additionally colour-coded by the exact fuel fraction computed). By creating radial profiles of offsets in \SigSFR we can investigate whether the fuel fraction we have computed can help explain the diversity of merger radial profiles seen with previous MaNGA studies (i.e., \citealt{Thorp2019SpatiallyMaNGA,Pan2019SDSS-IVInteractions,Steffen2021SDSS-IVPairs}). A more detailed analysis of merger radial profiles for different ALMaQUEST data products will be included in Pan et al. (in prep); what is discussed here is simply to assess how our global fuel fraction might relate to variations in \SigSFR already noted by previous works.

The efficiency driven and fuel driven radial profiles in Figure \ref{fig:profiles} have clearly distinct behaviour. Fuel driven mergers tend to have greater \dsigsfr at small radii, though that does not always lead to a suppression of star formation in the outskirts. Efficiency driven mergers have both greater peaks in \dsigsfr (5 out of 9 surpass 0.5 dex enhancement at one point) and a greater diversity in profile behaviour. Profiles for the efficiency driven mergers can both increase \dsigsfr with radius, decrease \dsigsfr with radius, or have consistent \dsigsfr out to 2$R_e$. Thus it is likely a diversity of fuel fraction values could explain some of the diversity of post-merger profiles noted in \cite{Thorp2019SpatiallyMaNGA}. Interestingly, the fuel driven mergers show similar profiles to the starburst \dsigsfr profiles found in \cite{Ellison2020The0}, despite non-interacting starbursts being primarily driven by enhanced SFE. It is possible that mergers provide one route to a starburst by funnelling gas to a centralized starburst, and the resulting SFR enhancement profile is indistinguishable from secular starbursts. Efficiency driven mergers, on the other hand, are more likely to have stronger and continuous SFR enhancements that distinguish them from isolated galaxies of a similar $\Delta$SFR. In the case of the later, the strongest bursts of star formation cannot be driven by enhanced gas reservoirs alone.

There are still issues with applying a single ``fuel driven'' prescription to what is clearly quite variable on the resolved scale. Simulations of the resolved gas properties of mergers have demonstrated that mergers with overall suppressed star formation can still have appreciable amounts of centralized cold-dense gas \citep{Moreno2021SpatiallyInteractions}. We can see this in our handful of fuel-driven mergers that have suppressed star formation beyond 1$R_e$, despite centralized boosts in star-formation. It is possible that, though a merger is primarily fuel or efficiency driven, there are regions that deviate from this norm. After all, the mergers which are driven by ``both'' are lacking any trend in radial profile unlike the other two categories. It is clearly worthwhile to explore how the dominant mechanism might vary within a galaxy.

We attempt to pinpoint spatial variations in the dominant star-forming mechanism by examining the difference between \dSFE and \dfgasend. The difference between two offset parameters provides a rough approximation of which offset is dominant in an individual spaxel by simply quantifying which is larger. The difference between \dSFE and \dfgas is a bit like a fuel fraction for each spaxel, though we note that the number attributed to each has no physical meaning beyond comparing offset values. We can thus construct maps of \dSFEend$-$\dfgas to demonstrate which mechanism is more important within a certain region of a galaxy. We only consider this for spaxels where \dsigsfrend, \dSFEend, and \dfgas are positive, to avoid interpreting the contribution of suppressed SFE or \fgasend.

\begin{figure}
	\includegraphics[width=0.99\columnwidth]{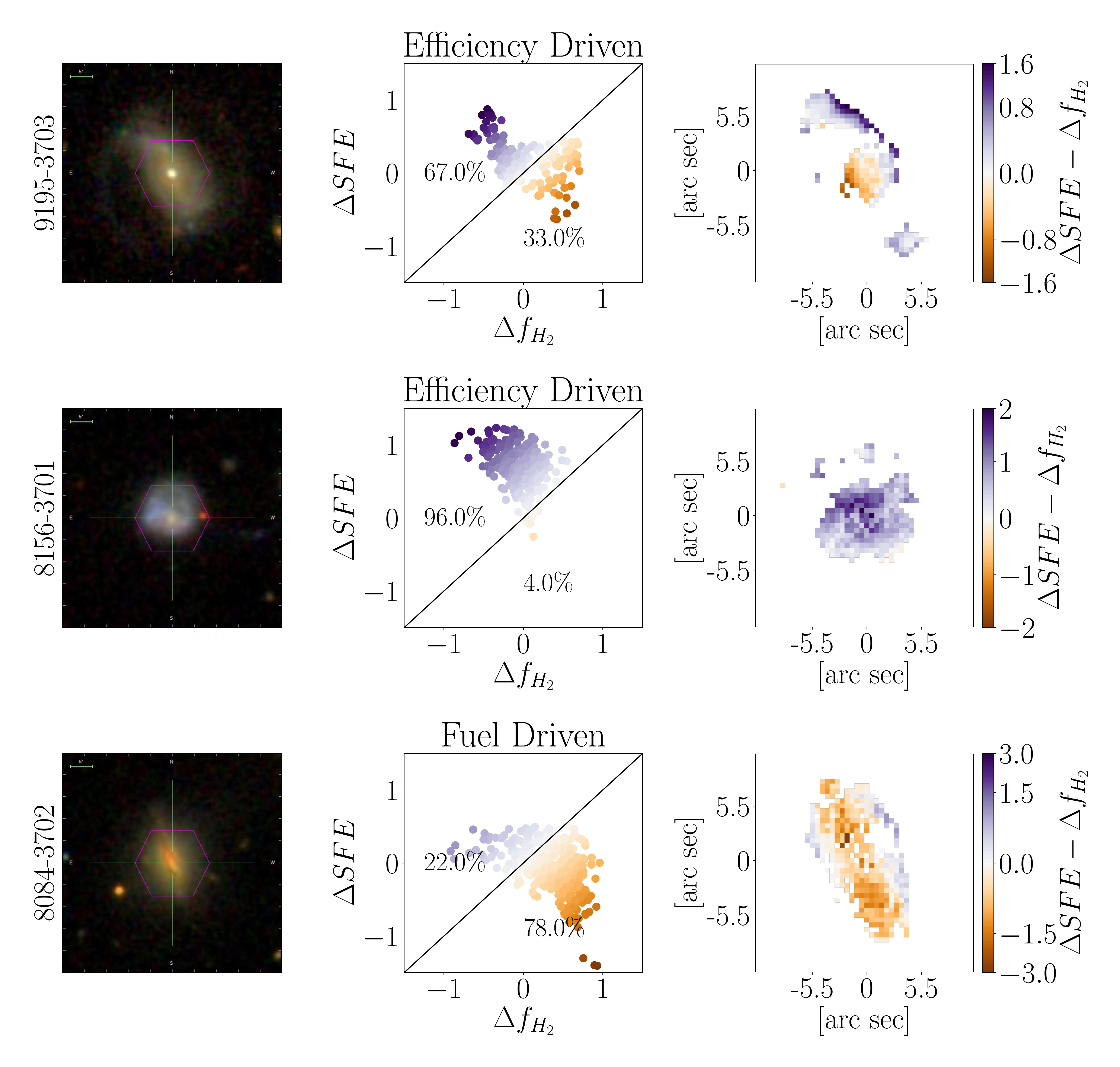}
	\centering
    \caption{1st column: \emph{gri}-image of the merger galaxy. 2nd column: \dSFE versus \dfgas for all spaxels where \dsigsfrend$>$0, colour-coded by \dSFEend$-$\dfgasend. 3rd column: spatially resolved map of \dSFEend$-$\dfgasend for pixels where all three offset parameters are positive. The colour bars change slightly for each merger, which have different spans of \dSFEend$-$\dfgasend, but zero is always white. Note that \dSFEend-\dfgas scales with distance from the line of equality shown in the 2nd column. Within each galaxy there is distinct variability in the importance of \dSFE and \dfgasend, particularly between the centre and outskirts of the galaxy.}
    \label{fig:Sub_method}
\end{figure}

Figure \ref{fig:Sub_method} displays maps of \dSFEend$-$\dfgas for three example galaxies, in conjunction with the \dSFE vs. \dfgas diagram colour-coded by \dSFEend$-$\dfgasend. Note how the magnitude of \dSFEend$-$\dfgasend scales with distance from the line of equality, where the contribution from SFE and \fgas is equal. Regions where \dSFE is greater are shown in purple, and regions where \dfgas is greater are shown in brown. We specifically chose three galaxies that are predominantly efficiency or fuel driven, to demonstrate how this clear-cut classification may break down on a spaxel-per-spaxel basis.

Figure \ref{fig:Sub_method} demonstrates that, even within a given galaxy, the dominant driver of enhanced star formation can vary spatially. The first row of Figure \ref{fig:Sub_method} shows a post-merger galaxy with a global fuel fraction of 0.33, indicating that SFE is the dominant driver of enhanced star formation.  However, a more complicated picture emerges when we look at the spatially resolved map of \dSFEend$-$\dfgas in the right column. Although the extended spiral arms are indeed dominated by enhanced SFE, the core is dominated by enhanced gas fractions. Similar results were found for the Antennae merging galaxies, which had central enhancements in dense gas along with lowered SFE \citep{Bemis2019Kiloparsec-Scale4038/39}. By examining maps of \dSFEend$-$\dfgas for all mergers (included with offset parameter maps in Appendix \ref{app:maps}), we find that most of our merger sample follows a similar trend of \dfgas dominating the central regions of the galaxy.

To quantify whether \dfgas is more important in the centre of a merging galaxy, we compute a gradient in \dSFEend$-$\dfgas for each merger, subtracting the median \dSFEend$-$\dfgas within 1$R_e$ from the median where $R>1R_e$. All but two mergers have a positive gradient (i.e., \dfgas is more important in the centres than the outskirts), confirming that similar to previous case studies \dfgas plays a key role in triggering central bursts of star formation. Positive gradients are found both within efficiency driven and fuel driven mergers, with no correlation to the global fraction of fuel driven spaxels (Pearson's correlation coefficient=$-$0.19). Efficiency driven mergers have mostly efficiency driven spaxels (positive values in \dSFEend$-$\dfgas maps) with fuel driven central starbursts; but given there are more spaxels at large radii than small, those efficiency driven spaxels dominate the percentage of star-forming regions overall. The two mergers with negative gradients, 8156-3701 and 8081-9101, are not unique in their global galaxy properties, but have distinct \dsigsfr profiles. 8156-3701 (a post-merger with fuel fraction=0.04) has \dsigsfr$>0.5$ out to two effective radii, powered by predominantly $\Delta$SFE. In contrast, 8081-9101 (a post-merger fuel fraction=0.5) has brief suppressions in star formation at the centre and edge of the galaxy, but enhancements $\sim$0.25 dex for the majority of the galaxy. Though no substantial result can be derived from two galaxies alone, it is interesting that mergers without central fuel enhancements display two of the most chaotic \dsigsfr profiles. However, on average, our sample confirms that central bursts of star formation in mergers require centralized enhancements in gas fraction.

Clearly the consistency of positive gradients in \dSFEend$-$\dfgas does not negate further variations between galaxies, as can be seen in the examples from Figure \ref{fig:Sub_method} alone. The middle row of Figure \ref{fig:Sub_method} shows a post-merger with a negative gradient in \dSFEend$-$\dfgas. Though star formation in this merger is clearly driven by enhanced SFE (it has a fuel fraction of 0.04), \fgas plays a slightly more important role in the outskirts than in the centre (though \dSFE is always greater). But even galaxies with a positive gradient, as expected, can have local variations from the norm. The last row of Figure \ref{fig:Sub_method} shows a fuel driven post-merger (fuel fraction 0.78) with a positive \dSFEend$-$\dfgas gradient. However, regions where \dSFEend$-$\dfgas$<0$ persist out to large radii as well, following a central extended structure within the galaxy. In this case fuel-driven star formation is not limited to the centre of the galaxy (though \dSFE becomes more important at large radii as well). Such variations are similar to those found in studies of isolated galaxies, where the slope and offset of the rSFMS, rKS, and rMGMS can vary significantly depending on the local galaxy environment (particularly for spiral arms, \citealt{Pessa2021AstronomyScale}). In summary, we find that there is not only considerable diversity from galaxy-to-galaxy in terms of the relative importance of fuel and efficiency, but even within a given galaxy, different regions show considerable variation.

\section{Summary \& Conclusion}
\label{sec:summary}

We have presented MaNGA+ALMA CO(1-0) observations of 31 merging galaxies from the ALMaQUEST merger sample in order to study whether interaction-induced star formation is driven primarily by enhancements in SFE or gas fraction.  The sample includes 14 post-merger galaxies and 17 pairs with projected separations up to 90 kpc. We compare the rSFMS, rKS, and rMGMS of the merger galaxies to relatively isolated galaxies in the ALMaQUEST survey. Although the ensemble of spaxels in the post-mergers, pairs and isolated galaxies have similar scaling relations (Figure \ref{fig:All_scaling_relations}), significant variation can be seen when examining the resolved relations for individual mergers (see Figures \ref{fig:SFMS_resolved}, \ref{fig:KS_resolved}, and \ref{fig:MGMS_resolved}).

In order to better understand the spatial variations in each merger, we construct maps of offsets from these resolved relations (\dsigsfrend, \dSFEend, and \dfgasend). By directly comparing \dSFE and \dfgas for all spaxels with enhanced star formation, we can identify whether the star formation is driven by an enhanced gas reservoir, or an enhanced efficiency at which gas is converted into stars. We find that when all spaxels in all galaxies are considered together, \fgas and SFE contribute equally to enhanced star formation (Figure \ref{fig:SFMS_mechanism}).

However, the approximately equal importance of fuel and efficiency across the entire sample is misleading..  When we examine galaxies on an individual basis, a different picture emerges. Some mergers are clearly dominated by either efficiency or fuel driven star formation, with all points lying in one regime of Figure \ref{fig:delta_plots_all}. About a third of the merger sample is predominantly efficiency driven, a third is fuel driven, and a third is driven equally by both. We investigate whether the dominant star formation mechanism might be correlated with global galaxy properties (SFR, \Mstarend, \Mgasend, \fgasend, $\Delta$SFR), and find only an unsurprising correlation with fuel driven star formation and large gas fractions.

The driving star formation mechanism does not depend on the stage of an interaction either. The percentage of fuel driven post-mergers is somewhat more than the percentage of fuel driven pair galaxies. But fuel fraction does not correlate with other indicators of interaction stage, such as projected separation and tidal connections (Figure \ref{fig:Fraction_mechanism}). The progression of the interaction does not lead to clear evolution in the fuel fraction, which implies the unique properties of the progenitors are more influential on which mechanisms powers enhancements in star-formation.

In section \ref{sec:Discussion} we investigate how the dominant star-forming mechanism leads to variations on a spatial scale. Radial profiles of \dsigsfr for the merger sample are shown in Figure \ref{fig:profiles}, separated into three categories based on fuel fraction. We find that ``fuel driven'' mergers (with fuel fractions > 0.6) have relatively distinct \dsigsfr profiles compared to  ``efficiency driven'' mergers (fuel fraction < 0.4), which often have stronger and more extended \dsigsfr enhancements. Thus a range of fuel fractions in mergers could explain the diversity of star formation offset profiles seen in previous merger studies \citep{Thorp2019SpatiallyMaNGA,Pan2019SDSS-IVInteractions,Steffen2021SDSS-IVPairs}. 

The range of \dsigsfr profiles in these three categories implies more internal variation within a galaxy than a single fuel fraction can capture. Figure \ref{fig:Sub_method} demonstrates this variation in the dominant mechanism within a galaxy by presenting 2D maps of \dSFEend$-$\dfgasend, indicating which offset parameter is largest in each spaxel. Even galaxies which are driven predominantly by a single mechanism on a global scale can exhibit internal deviations from the dominant mechanism. For example, both efficiency and fuel driven mergers tend to have \dfgasend$>$\dSFE in their centre, confirming that enhanced amounts of molecular gas are crucial to merger-induced central starbursts, even if the rest of the galaxy has efficiency driven star formation.

The work presented here is the first to investigate the resolved molecular gas and star formation properties for a relatively large set of mergers spanning a wide range of sSFRs and stages of interaction, with the explicit goal of understanding how gas and star formation evolve with an interaction. Our work adds to the growing evidence that, despite following overall scaling relationships, galaxies are diverse in their details. Moreover, we have shown that in addition to diversity on a global scale (i.e. some mergers have their star formation driven by fuel, others by efficiency), there is significant internal variation on kpc-scales.

\section*{Acknowledgements}

We acknowledge and respect the Lekwungen
peoples on whose traditional territories the University of Victoria stands and where the majority of this work was conducted. We strive to honour the Songhees, Esquimalt, and WSÁNEĆ peoples who were the first astronomers of this land and whose continued stewardship is crucial to its preservation.

MDT thanks Jorge Moreno, Shoshannah Byrne-Mamahit, Salvatore Quai, Robert Bickley, Scott Wilkinson and Joanna Woo for discussion throughout the creation of this work, as well as William Baker and Fangting Yuan for key editorial remarks. SLE and DRP gratefully acknowledges NSERC of Canada for Discovery Grants which helped to fund this research. HAP acknowledges support by the Ministry of Science and Technology of Taiwan under grant 110-2112-M-032-020-MY3. SLE, MDT, LL and HAP also gratefully acknowledge grant MOST 107-2119-M-001-024 and 108-2628-M-001-001-MY3 for travel funding that  facilitated  both  the  ALMA  data  reduction  and  analysis of ALMaQUEST survey.

This paper makes use of the following ALMA data: ADS/JAO.ALMA\#2015.1.01225.S, ADS/JAO.ALMA\#2017.1.01093.S, ADS/JAO.ALMA\#2018.1.00558.S, ADS/JAO.ALMA\#2018.1.00541.S, ADS/JAO.ALMA\#2019.1.00260.S. ALMA is a partnership of ESO (representing its member states), NSF (USA) and NINS (Japan), together with NRC (Canada), MOST and ASIAA (Tai- wan), and KASI (Republic of Korea), in cooperation with the Republic of Chile. The Joint ALMA Observatory is operated by ESO, AUI/NRAO and NAOJ. The National Radio Astronomy Observatory is a facility of the National Science Foundation operated under cooperative agreement by Associated Universities, Inc.

Funding for the Sloan Digital Sky Survey IV has been provided by the Alfred P. Sloan Foundation, the U.S. Department of Energy Office of Science, and the Participating Institutions. SDSS-IV acknowledges support and resources from
the Center for High-Performance Computing at the University of Utah. The SDSS web site is www.sdss.org. SDSSIV is managed by the Astrophysical Research Consortium
for the Participating Institutions of the SDSS Collaboration
including the Brazilian Participation Group, the Carnegie
Institution for Science, Carnegie Mellon University, the
Chilean Participation Group, the French Participation Group,
Harvard-Smithsonian Center for Astrophysics, Instituto de
Astrofísica de Canarias, The Johns Hopkins University, Kavli
Institute for the Physics and Mathematics of the Universe
(IPMU) / University of Tokyo, Lawrence Berkeley National Laboratory, Leibniz Institut für Astrophysik Potsdam
(AIP), Max-Planck-Institut für Astronomie (MPIA Heidelberg), Max-Planck-Institut für Astrophysik (MPA Garching),
Max-Planck-Institut für Extraterrestrische Physik (MPE), National Astronomical Observatory of China, New Mexico State
University, New York University, University of Notre Dame,
Observatário Nacional / MCTI, The Ohio State University,
Pennsylvania State University, Shanghai Astronomical Observatory, United Kingdom Participation Group, Universidad
Nacional Autónoma de México, University of Arizona, University of Colorado Boulder, University of Oxford, University of Portsmouth, University of Utah, University of Virginia,
University of Washington, University of Wisconsin, Vanderbilt University, and Yale University.

\section*{Data Availability}

The MaNGA data cubes used in this work are publicly available at https://www.sdss.org/dr15/. The ALMA data used in this work are publicly available after the standard one year proprietary period via the ALMA archive: http://almascience.nrao.edu/aq/.



\bibliographystyle{mnras}
\bibliography{references.bib}

\begin{thebibliography}{}
\makeatletter
\relax
\def\mn@urlcharsother{\let\do\@makeother \do\$\do\&\do\#\do\^\do\_\do\%\do\~}
\def\mn@doi{\begingroup\mn@urlcharsother \@ifnextchar [ {\mn@doi@}
  {\mn@doi@[]}}
\def\mn@doi@[#1]#2{\def\@tempa{#1}\ifx\@tempa\@empty \href
  {http://dx.doi.org/#2} {doi:#2}\else \href {http://dx.doi.org/#2} {#1}\fi
  \endgroup}
\def\mn@eprint#1#2{\mn@eprint@#1:#2::\@nil}
\def\mn@eprint@arXiv#1{\href {http://arxiv.org/abs/#1} {{\tt arXiv:#1}}}
\def\mn@eprint@dblp#1{\href {http://dblp.uni-trier.de/rec/bibtex/#1.xml}
  {dblp:#1}}
\def\mn@eprint@#1:#2:#3:#4\@nil{\def\@tempa {#1}\def\@tempb {#2}\def\@tempc
  {#3}\ifx \@tempc \@empty \let \@tempc \@tempb \let \@tempb \@tempa \fi \ifx
  \@tempb \@empty \def\@tempb {arXiv}\fi \@ifundefined
  {mn@eprint@\@tempb}{\@tempb:\@tempc}{\expandafter \expandafter \csname
  mn@eprint@\@tempb\endcsname \expandafter{\@tempc}}}

\bibitem[\protect\citeauthoryear{Accurso et~al.,}{Accurso
  et~al.}{2017}]{Accurso2017DerivingRelations}
Accurso G.,  et~al., 2017, Monthly Notices of the Royal Astronomical Society,
  470, 4750

\bibitem[\protect\citeauthoryear{Allen et~al.,}{Allen
  et~al.}{2015}]{Allen2015TheRelease}
Allen J.~T.,  et~al., 2015, \mn@doi [Monthly Notices of the Royal Astronomical
  Society] {10.1093/mnras/stu2057}, 446, 1567

\bibitem[\protect\citeauthoryear{Baker, Maiolino, Bluck, Lin, Ellison,
  Belfiore, Pan  \& Thorp}{Baker et~al.}{2022}]{Baker2022TheSequence}
Baker W.~M.,  Maiolino R.,  Bluck A. F.~L.,  Lin L.,  Ellison S.~L.,  Belfiore
  F.,  Pan H.-A.,   Thorp M.,  2022, \mn@doi [Monthly Notices of the Royal
  Astronomical Society] {10.1093/mnras/stab3672}, 510, 3622

\bibitem[\protect\citeauthoryear{Baldwin, Phillips  \& Terlevich}{Baldwin
  et~al.}{1981}]{Baldwin1981CLASSIFICATIONOBJECTS}
Baldwin J.~A.,  Phillips M.~M.,   Terlevich R.,  1981, Publications of the
  Astronomical Society of the Pacific, 93, 5

\bibitem[\protect\citeauthoryear{Barnes \& Hernquist}{Barnes \&
  Hernquist}{1991}]{Barnes1991FuelingMergers}
Barnes J.~E.,  Hernquist L.~E.,  1991, \mn@doi [The Astrophysical Journal]
  {10.1086/185978}, 370, L65

\bibitem[\protect\citeauthoryear{Barrera-Ballesteros
  et~al.,}{Barrera-Ballesteros
  et~al.}{2015}]{Barrera-Ballesteros2015CentralGalaxies}
Barrera-Ballesteros J.~K.,  et~al., 2015, \mn@doi [Astronomy {\&} Astrophysics]
  {10.1051/0004-6361/201425397}, 579, A45

\bibitem[\protect\citeauthoryear{Barrera-Ballesteros
  et~al.,}{Barrera-Ballesteros
  et~al.}{2021}]{Barrera-Ballesteros2021EDGE-CALIFAScales}
Barrera-Ballesteros J.~K.,  et~al., 2021, \mn@doi [Monthly Notices of the Royal
  Astronomical Society] {10.1093/mnras/stab755}, 18, 1

\bibitem[\protect\citeauthoryear{Barton~Gillespie, Geller  \&
  Kenyon}{Barton~Gillespie
  et~al.}{2003}]{BartonGillespie2003Tidally-TriggeredAges}
Barton~Gillespie E.~B.,  Geller M.~J.,   Kenyon S.~J.,  2003, \mn@doi [The
  Astrophysical Journal] {10.1086/344724}, 582, 668

\bibitem[\protect\citeauthoryear{Bemis \& Wilson}{Bemis \&
  Wilson}{2019}]{Bemis2019Kiloparsec-Scale4038/39}
Bemis A.,  Wilson C.~D.,  2019, \mn@doi [The Astronomical Journal]
  {10.3847/1538-3881/ab041d}, 157, 18

\bibitem[\protect\citeauthoryear{Bickley, Ellison, Patton, Bottrell, Gwyn  \&
  Hudson}{Bickley et~al.}{2022}]{Bickley2022StarUNIONS}
Bickley R.~W.,  Ellison S.~L.,  Patton D.~R.,  Bottrell C.,  Gwyn S.,   Hudson
  M.~J.,  2022, \mn@doi [Monthly Notices of the Royal Astronomical Society]
  {10.1093/mnras/stac1500}, 514, 3294

\bibitem[\protect\citeauthoryear{Bigiel, Leroy, Walter, Brinks, De~Blok, Madore
   \& Thornley}{Bigiel et~al.}{2008}]{Bigiel2008THESCALES}
Bigiel F.,  Leroy A.,  Walter F.,  Brinks E.,  De~Blok W. J.~G.,  Madore B.,
  Thornley M.~D.,  2008, \mn@doi [The Astronomical Journal]
  {10.1088/0004-6256/136/6/2846}, 136, 2846

\bibitem[\protect\citeauthoryear{Bluck, Maiolino, S{\'{a}}nchez, Ellison,
  Thorp, Piotrowska, Teimoorinia  \& Bundy}{Bluck
  et~al.}{2020}]{Bluck2020ArePhenomena}
Bluck A.,  Maiolino R.,  S{\'{a}}nchez S.,  Ellison S.,  Thorp M.,  Piotrowska
  J.,  Teimoorinia H.,   Bundy K.,  2020, \mn@doi [Monthly Notices of the Royal
  Astronomical Society] {10.1093/mnras/stz3264}, 492, 96

\bibitem[\protect\citeauthoryear{Bolatto, Wolfire  \& Leroy}{Bolatto
  et~al.}{2013}]{Bolatto2013THEFACTOR}
Bolatto A.~D.,  Wolfire M.,   Leroy A.~K.,  2013, \mn@doi [Annual Review of
  Astronomy and Astrophysics] {10.1146/annurev-astro-082812-140944}, 51, 207

\bibitem[\protect\citeauthoryear{Bolatto et~al.,}{Bolatto
  et~al.}{2017}]{Bolatto2017TheCARMA}
Bolatto A.~D.,  et~al., 2017, \mn@doi [The Astrophysical Journal]
  {10.3847/1538-4357/aa86aa}, 846, 159

\bibitem[\protect\citeauthoryear{Bournaud et~al.,}{Bournaud
  et~al.}{2011}]{Bournaud2011HYDRODYNAMICSSPHEROIDS}
Bournaud F.,  et~al., 2011, \mn@doi [The Astrophysical Journal]
  {10.1088/0004-637X/730/1/4}, 730, 4

\bibitem[\protect\citeauthoryear{Braine \& Combes}{Braine \&
  Combes}{1992}]{Braine1992AGalaxies}
Braine J.,  Combes F.,  1992, \mn@doi [Astronomy {\&} Astrophysics]
  {1992A{\&}A...264..433B}, 269, 7

\bibitem[\protect\citeauthoryear{Brinchmann, Charlot, White, Tremonti,
  Kauffmann, Heckman  \& Brinkmann}{Brinchmann
  et~al.}{2004}]{Brinchmann2004TheUniverse}
Brinchmann J.,  Charlot S.,  White S. D.~M.,  Tremonti C.,  Kauffmann G.,
  Heckman T.,   Brinkmann J.,  2004, \mn@doi [Monthly Notices of the Royal
  Astronomical Society] {10.1111/j.1365-2966.2004.07881.x}, 351, 1151

\bibitem[\protect\citeauthoryear{Bundy et~al.,}{Bundy
  et~al.}{2015}]{Bundy2015OverviewObservatory}
Bundy K.,  et~al., 2015, \mn@doi [The Astrophysical Journal]
  {10.1088/0004-637X/798/1/7}, 798, 7

\bibitem[\protect\citeauthoryear{Cano-D{\'{i}}az et~al.,}{Cano-D{\'{i}}az
  et~al.}{2016}]{Cano-Diaz2016Spatially-ResolvedSurvey}
Cano-D{\'{i}}az M.,  et~al., 2016, \mn@doi [The Astrophysical Journal]
  {10.3847/2041-8205/821/2/L26}, 821, L26

\bibitem[\protect\citeauthoryear{Cardelli, Clayton  \& Mathis}{Cardelli
  et~al.}{1989}]{Cardelli1989THEEXTINCTION}
Cardelli J.~A.,  Clayton G.~C.,   Mathis J.~S.,  1989, \mn@doi [The
  Astrophysical Journal] {10.1086/167900}, 345, 245

\bibitem[\protect\citeauthoryear{Casasola, Bettoni  \& Galletta}{Casasola
  et~al.}{2004}]{Casasola2004TheSystems}
Casasola V.,  Bettoni D.,   Galletta G.,  2004, \mn@doi [Astronomy {\&}
  Astrophysics] {10.1051/0004-6361:20040283}, 422, 941

\bibitem[\protect\citeauthoryear{Chown et~al.,}{Chown
  et~al.}{2019}]{Chown2019LinkingGalaxies}
Chown R.,  et~al., 2019, \mn@doi [Monthly Notices of the Royal Astronomical
  Society] {10.1093/mnras/stz349}, 484, 5192

\bibitem[\protect\citeauthoryear{Colombo et~al.,}{Colombo
  et~al.}{2018}]{Colombo2018TheSequence}
Colombo D.,  et~al., 2018, \mn@doi [Monthly Notices of the Royal Astronomical
  Society] {10.1093/MNRAS/STX3233}, 475, 1791

\bibitem[\protect\citeauthoryear{Cox, Jonsson, Primack  \& Somerville}{Cox
  et~al.}{2006}]{Cox2006FeedbackMergers}
Cox T.~J.,  Jonsson P.,  Primack J.~R.,   Somerville R.~S.,  2006, \mn@doi
  [Monthly Notices of the Royal Astronomical Society]
  {10.1111/j.1365-2966.2006.11107.x}, 373, 1013

\bibitem[\protect\citeauthoryear{Cox, Jonsson, Somerville, Primack  \&
  Dekel}{Cox et~al.}{2008}]{Cox2008TheStarbursts}
Cox T.~J.,  Jonsson P.,  Somerville R.~S.,  Primack J.~R.,   Dekel A.,  2008,
  \mn@doi [Monthly Notices of the Royal Astronomical Society]
  {10.1111/j.1365-2966.2007.12730.x}, 384, 386

\bibitem[\protect\citeauthoryear{D'Onghia, Vogelsberger, Faucher-Giguere  \&
  Hernquist}{D'Onghia et~al.}{2010}]{DOnghia2010QUASI-RESONANTINTERACTIONS}
D'Onghia E.,  Vogelsberger M.,  Faucher-Giguere C.-A.,   Hernquist L.,  2010,
  \mn@doi [The Astrophysical Journal] {10.1088/0004-637X/725/1/353}, 725, 353

\bibitem[\protect\citeauthoryear{Di~Matteo, Combes, Melchior  \&
  Semelin}{Di~Matteo et~al.}{2007}]{DiMatteo2007StarStudy}
Di~Matteo P.,  Combes F.,  Melchior A.~L.,   Semelin B.,  2007, \mn@doi
  [Astronomy {\&} Astrophysics] {10.1051/0004-6361:20066959}, 468, 61

\bibitem[\protect\citeauthoryear{Diaz~Trigo, Carpenter, Maude, Miura  \&
  Plunkett}{Diaz~Trigo et~al.}{2019}]{2019acpg.rept.....D}
Diaz~Trigo M.,  Carpenter J.,  Maude L.,  Miura R.,   Plunkett A.,  2019, {ALMA
  Cycle 7 Proposer's Guide}, 2019, \mn@doi{10.5281/zenodo.4511962}

\bibitem[\protect\citeauthoryear{Ellison, Patton, Simard  \&
  McConnachie}{Ellison et~al.}{2008}]{Ellison2008GalaxyRelation}
Ellison S.~L.,  Patton D.~R.,  Simard L.,   McConnachie A.~W.,  2008, \mn@doi
  [The Astronomical Journal] {10.1088/0004-6256/135/5/1877}, 135, 1877

\bibitem[\protect\citeauthoryear{Ellison, Mendel, Patton  \& Scudder}{Ellison
  et~al.}{2013}]{Ellison2013GalaxyGalaxies}
Ellison S.~L.,  Mendel J.~T.,  Patton D.~R.,   Scudder J.~M.,  2013, \mn@doi
  [Monthly Notices of the Royal Astronomical Society] {10.1093/mnras/stt1562},
  435, 3627

\bibitem[\protect\citeauthoryear{Ellison, Teimoorinia, Rosario  \&
  Mendel}{Ellison et~al.}{2016}]{Ellison2016TheGalaxies}
Ellison S.~L.,  Teimoorinia H.,  Rosario D.~J.,   Mendel J.~T.,  2016, \mn@doi
  [MNRAS] {10.1093/mnrasl/slw012}, 458, 34

\bibitem[\protect\citeauthoryear{Ellison, Catinella  \& Cortese}{Ellison
  et~al.}{2018}]{Ellison2018EnhancedExhaustion}
Ellison S.~L.,  Catinella B.,   Cortese L.,  2018, \mn@doi [Monthly Notices of
  the Royal Astronomical Society] {10.1093/mnras/sty1247}, 478, 3447

\bibitem[\protect\citeauthoryear{Ellison, Thorp, Pan, Lin, Scudder, Bluck,
  S{\'{a}}nchez  \& Sargent}{Ellison et~al.}{2020a}]{Ellison2020The0}
Ellison S.~L.,  Thorp M.~D.,  Pan H.-A.,  Lin L.,  Scudder J.~M.,  Bluck A.
  F.~L.,  S{\'{a}}nchez S.~F.,   Sargent M.,  2020a, \mn@doi [Monthly Notices
  of the Royal Astronomical Society] {10.1093/mnras/staa001}, 492, 6027

\bibitem[\protect\citeauthoryear{Ellison et~al.,}{Ellison
  et~al.}{2020b}]{Ellison2020TheEfficiency.}
Ellison S.,  et~al., 2020b, \mn@doi [Monthly Notices of the Royal Astronomical
  Society] {10.1093/mnrasl/slz179}, 493, L39

\bibitem[\protect\citeauthoryear{Ellison, Lin, Thorp, Pan, Scudder,
  S{\'{a}}nchez, Bluck  \& Maiolino}{Ellison
  et~al.}{2021a}]{Ellison2021TheThem}
Ellison S.~L.,  Lin L.,  Thorp M.~D.,  Pan H.-A.,  Scudder J.~M.,
  S{\'{a}}nchez S.~F.,  Bluck A. F.~L.,   Maiolino R.,  2021a, \mn@doi [Monthly
  Notices of the Royal Astronomical Society] {10.1093/mnras/staa3822}, 501,
  4777

\bibitem[\protect\citeauthoryear{Ellison et~al.,}{Ellison
  et~al.}{2021b}]{Ellison2021TheQuenching}
Ellison S.~L.,  et~al., 2021b, MNRAS, 505, L46

\bibitem[\protect\citeauthoryear{Fensch et~al.,}{Fensch
  et~al.}{2017}]{Fensch2017High-redshiftFormation}
Fensch J.,  et~al., 2017, \mn@doi [Monthly Notices of the Royal Astronomical
  Society] {10.1093/mnras/stw2920}, 465, 1934

\bibitem[\protect\citeauthoryear{Gallagher et~al.,}{Gallagher
  et~al.}{2018}]{Gallagher2018DenseGalaxies}
Gallagher M.~J.,  et~al., 2018, \mn@doi [The Astrophysical Journal]
  {10.3847/1538-4357/aabad8}, 858, 90

\bibitem[\protect\citeauthoryear{Gonz{\'{a}}lez~Delgado
  et~al.,}{Gonz{\'{a}}lez~Delgado
  et~al.}{2016}]{GonzalezDelgado2016AstrophysicsGalaxies}
Gonz{\'{a}}lez~Delgado R.~M.,  et~al., 2016, \mn@doi [Astronomy {\&}
  Astrophysics] {10.1051/0004-6361/201628174}, 590

\bibitem[\protect\citeauthoryear{Hambleton, Gibson, Brook, Stinson, Conselice,
  Bailin, Couchman  \& Wadsley}{Hambleton
  et~al.}{2011}]{Hambleton2011AdvancedGalaxies}
Hambleton K.~M.,  Gibson B.~K.,  Brook C.~B.,  Stinson G.~S.,  Conselice C.~J.,
   Bailin J.,  Couchman H.,   Wadsley J.,  2011, \mn@doi [Monthly Notices of
  the Royal Astronomical Society] {10.1111/j.1365-2966.2011.19532.x}, 418, 801

\bibitem[\protect\citeauthoryear{Hani, Gosain, Ellison, Patton  \& Torrey}{Hani
  et~al.}{2020}]{Hani2020InteractingStage}
Hani M.~H.,  Gosain H.,  Ellison S.~L.,  Patton D.~R.,   Torrey P.,  2020,
  \mn@doi [Monthly Notices of the Royal Astronomical Society]
  {10.1093/mnras/staa459}, 493, 3716

\bibitem[\protect\citeauthoryear{Hopkins, Cox, Hernquist, Narayanan, Hayward
  \& Murray}{Hopkins et~al.}{2013}]{Hopkins2013StarMedium}
Hopkins P.~F.,  Cox T.~J.,  Hernquist L.,  Narayanan D.,  Hayward C.~C.,
  Murray N.,  2013, \mn@doi [Monthly Notices of the Royal Astronomical Society]
  {10.1093/mnras/stt017}, 430, 1901

\bibitem[\protect\citeauthoryear{Huchtmeier, Petrosian, Krishna, McLean  \&
  Kunth}{Huchtmeier et~al.}{2008}]{Huchtmeier2008InteractingSurvey}
Huchtmeier W.~K.,  Petrosian A.,  Krishna G.,  McLean B.,   Kunth D.,  2008,
  \mn@doi [Astronomy {\&} Astrophysics] {10.1051/0004-6361:200810734}, 492, 367

\bibitem[\protect\citeauthoryear{Iono, Yun  \& Mihos}{Iono
  et~al.}{2004}]{Iono2004RadialObservations}
Iono D.,  Yun M.~S.,   Mihos J.~C.,  2004, \mn@doi [The Astrophysical Journal]
  {10.1086/424797}, 616, 199

\bibitem[\protect\citeauthoryear{Jim{\'{e}}nez-Donaire
  et~al.,}{Jim{\'{e}}nez-Donaire
  et~al.}{2019}]{Jimenez-Donaire2019EMPIRE:Galaxies}
Jim{\'{e}}nez-Donaire M.~J.,  et~al., 2019, \mn@doi [The Astrophysical Journal]
  {10.3847/1538-4357/ab2b95}, 880, 127

\bibitem[\protect\citeauthoryear{Kauffmann et~al.,}{Kauffmann
  et~al.}{2003}]{Kauffmann2003TheAGN}
Kauffmann G.,  et~al., 2003, \mn@doi [Monthly Notices of the Royal Astronomical
  Society] {10.1111/j.1365-2966.2003.07154.x}, 346, 1055

\bibitem[\protect\citeauthoryear{Kennicutt}{Kennicutt}{1989}]{Kennicutt1989THEDISKS}
Kennicutt R.~C.,  1989, \mn@doi [The Astrophysical Journal] {10.1086/167834},
  344, 685

\bibitem[\protect\citeauthoryear{Kennicutt, Keel, Van Der~Hulst, Hummel  \&
  Roettiger}{Kennicutt et~al.}{1987}]{Kennicutt1987THERATES}
Kennicutt R.~C.,  Keel W.~C.,  Van Der~Hulst J.~M.,  Hummel E.,   Roettiger
  K.~A.,  1987, \mn@doi [The Astronomical Journal] {10.1086/114384}, 93, 1011

\bibitem[\protect\citeauthoryear{Kennicutt, Tamblyn  \& Congdon}{Kennicutt
  et~al.}{1994}]{Kennicutt1994PastGalaxies}
Kennicutt R.~C.,  Tamblyn P.,   Congdon C.~W.,  1994, \mn@doi [Astrophysical
  Journal] {10.1086/174790}, 435, 22

\bibitem[\protect\citeauthoryear{Knapen \& James}{Knapen \&
  James}{2009}]{Knapen2009THESTARBURSTS}
Knapen J.~H.,  James P.~A.,  2009, \mn@doi [The Astrophysical Journal]
  {10.1088/0004-637X/698/2/1437}, 698, 1437

\bibitem[\protect\citeauthoryear{Knapen, Cisternas  \& Querejeta}{Knapen
  et~al.}{2015}]{Knapen2015InteractingFormation}
Knapen J.~H.,  Cisternas M.,   Querejeta M.,  2015, \mn@doi [Monthly Notices of
  the Royal Astronomical Society] {10.1093/mnras/stv2135}, 454, 1742

\bibitem[\protect\citeauthoryear{Lambas, Alonso, Mesa  \& O’Mill}{Lambas
  et~al.}{2012}]{Lambas2012GalaxyInteractions}
Lambas D.~G.,  Alonso S.,  Mesa V.,   O’Mill A.~L.,  2012, \mn@doi [Astronomy
  {\&} Astrophysics] {10.1051/0004-6361/201117900}, 539, A45

\bibitem[\protect\citeauthoryear{Larson \& Tinsley}{Larson \&
  Tinsley}{1978}]{Larson1978StarGalaxies}
Larson R.~B.,  Tinsley B.~M.,  1978, \mn@doi [The Astrophysical Journal]
  {10.1086/155753}, 219, 46

\bibitem[\protect\citeauthoryear{Law et~al.,}{Law
  et~al.}{2015}]{Law2015ObservingSurvey}
Law D.~R.,  et~al., 2015, \mn@doi [The Astronomical Journal]
  {10.1088/0004-6256/150/1/19}, 150, 19

\bibitem[\protect\citeauthoryear{Leroy, Walter, Brinks, Bigiel, De~Blok, Madore
   \& Thornley}{Leroy et~al.}{2008}]{Leroy2008THEEFFECTIVELY}
Leroy A.~K.,  Walter F.,  Brinks E.,  Bigiel F.,  De~Blok W. J.~G.,  Madore B.,
    Thornley M.~D.,  2008, \mn@doi [The Astronomical Journal]
  {10.1088/0004-6256/136/6/2782}, 136, 2782

\bibitem[\protect\citeauthoryear{Leroy et~al.,}{Leroy
  et~al.}{2013}]{Leroy2013MOLECULARGALAXIES}
Leroy A.~K.,  et~al., 2013, \mn@doi [The Astronomical Journal]
  {10.1088/0004-6256/146/2/19}, 146, 19

\bibitem[\protect\citeauthoryear{Leroy et~al.,}{Leroy
  et~al.}{2021}]{Leroy2021PHANGS-ALMA:Galaxies}
Leroy A.~K.,  et~al., 2021, \mn@doi [The Astrophysical Journal Supplement
  Series] {10.3847/1538-4365/ac17f3}, 257, 61

\bibitem[\protect\citeauthoryear{Leslie, Kewley, Sanders  \& Lee}{Leslie
  et~al.}{2016}]{Leslie2016QuenchingSequence}
Leslie S.~K.,  Kewley L.~J.,  Sanders D.~B.,   Lee N.,  2016, \mn@doi [MNRAS]
  {10.1093/mnrasl/slv135}, 455, 82

\bibitem[\protect\citeauthoryear{Lin et~al.,}{Lin
  et~al.}{2007}]{Lin2007AEGIS:1}
Lin L.,  et~al., 2007, The Astrophysical Journal, 660, 51

\bibitem[\protect\citeauthoryear{Lin et~al.,}{Lin
  et~al.}{2017}]{Lin2017SDSS-IVMaNGA}
Lin L.,  et~al., 2017, \mn@doi [The Astrophysical Journal]
  {10.3847/1538-4357/aa96ae}, 851, 18

\bibitem[\protect\citeauthoryear{Lin et~al.,}{Lin
  et~al.}{2019}]{Lin2019TheSequence}
Lin L.,  et~al., 2019, \mn@doi [Astrophysical Journal Letters]
  {10.3847/2041-8213/ab4815}, 884

\bibitem[\protect\citeauthoryear{Lin et~al.,}{Lin
  et~al.}{2020}]{Lin2020ALMAQUEST:SURVEY}
Lin L.,  et~al., 2020, \mn@doi [The Astrophysical Journal]
  {10.3847/1538-4357/abba3a}, 903, 19

\bibitem[\protect\citeauthoryear{Lotz, Primack  \& Madau}{Lotz
  et~al.}{2004}]{Lotz2004AClassification}
Lotz J.~M.,  Primack J.,   Madau P.,  2004, \mn@doi [The Astronomical Journal]
  {10.1086/421849}, 128, 163

\bibitem[\protect\citeauthoryear{Lotz, Jonsson, Cox  \& Primack}{Lotz
  et~al.}{2008}]{Lotz2008GalaxyMergers}
Lotz J.~M.,  Jonsson P.,  Cox T.,   Primack J.~R.,  2008, \mn@doi [Monthly
  Notices of the Royal Astronomical Society]
  {10.1111/j.1365-2966.2008.14004.x}, 391, 1137

\bibitem[\protect\citeauthoryear{McMullin, Waters, Schiebel, Young  \&
  Golap}{McMullin et~al.}{2007}]{McMullin2007CASAApplications}
McMullin J.~P.,  Waters B.,  Schiebel D.,  Young W.,   Golap K.,  2007, in
  Astronomical Data Analysis Software and Systems XVI. p.~127, \url
  {http://casa.nrao.edu}

\bibitem[\protect\citeauthoryear{McPartland, Sanders, Kewley  \&
  Leslie}{McPartland et~al.}{2019}]{Mcpartland2019DissectingUniverse}
McPartland C.,  Sanders D.~B.,  Kewley L.~J.,   Leslie S.~K.,  2019, \mn@doi
  [Monthly Notices of the Royal Astronomical Society: Letters]
  {10.1093/mnrasl/sly202}, 482, L129

\bibitem[\protect\citeauthoryear{Mihos \& Hernquist}{Mihos \&
  Hernquist}{1996}]{Mihos1996GasdynamicsMergers}
Mihos J.~C.,  Hernquist L.,  1996, \mn@doi [The Astrophysical Journal]
  {10.1086/177353}, 464, 641

\bibitem[\protect\citeauthoryear{Moreno, Torrey, Ellison, Patton, Bluck, Bansal
   \& Hernquist}{Moreno et~al.}{2015}]{Moreno2015MappingFormation}
Moreno J.,  Torrey P.,  Ellison S.~L.,  Patton D.~R.,  Bluck A.~F.,  Bansal G.,
    Hernquist L.,  2015, \mn@doi [Monthly Notices of the Royal Astronomical
  Society] {10.1093/mnras/stv094}, 448, 1107

\bibitem[\protect\citeauthoryear{Moreno et~al.,}{Moreno
  et~al.}{2019}]{Moreno2019InteractingContent}
Moreno J.,  et~al., 2019, \mn@doi [Monthly Notices of the Royal Astronomical
  Society] {10.1093/mnras/stz417}, 485, 1320

\bibitem[\protect\citeauthoryear{Moreno et~al.,}{Moreno
  et~al.}{2021}]{Moreno2021SpatiallyInteractions}
Moreno J.,  et~al., 2021, \mn@doi [Monthly Notices of the Royal Astronomical
  Society] {10.1093/mnras/staa2952}, 503, 3113

\bibitem[\protect\citeauthoryear{Narayanan, Krumholz, Ostriker  \&
  Hernquist}{Narayanan et~al.}{2012}]{Narayanan2012ALaw}
Narayanan D.,  Krumholz M.~R.,  Ostriker E.~C.,   Hernquist L.,  2012, \mn@doi
  [Monthly Notices of the Royal Astronomical Society]
  {10.1111/J.1365-2966.2012.20536.X}, 421, 3127

\bibitem[\protect\citeauthoryear{Nikolic, Cullen  \& Alexander}{Nikolic
  et~al.}{2004}]{Nikolic2004StarSurvey}
Nikolic B.,  Cullen H.,   Alexander P.,  2004, \mn@doi [Monthly Notices of the
  Royal Astronomical Society] {10.1111/j.1365-2966.2004.08366.x}, 355, 874

\bibitem[\protect\citeauthoryear{Pan et~al.,}{Pan
  et~al.}{2018}]{Pan2018TheProperties}
Pan H.-A.,  et~al., 2018, \mn@doi [The Astrophysical Journal]
  {10.3847/1538-4357/aaeb92}, 868, 132

\bibitem[\protect\citeauthoryear{Pan et~al.,}{Pan
  et~al.}{2019}]{Pan2019SDSS-IVInteractions}
Pan H.-A.,  et~al., 2019, \mn@doi [The Astrophysical Journal]
  {10.3847/1538-4357/ab311c}, 881, 119

\bibitem[\protect\citeauthoryear{Patton, Torrey, Ellison, Mendel  \&
  Scudder}{Patton et~al.}{2013}]{Patton2013GalaxyGalaxies}
Patton D.~R.,  Torrey P.,  Ellison S.~L.,  Mendel J.~T.,   Scudder J.~M.,
  2013, \mn@doi [Monthly Notices of the Royal Astronomical Society: Letters]
  {10.1093/mnrasl/slt058}, 433, 59

\bibitem[\protect\citeauthoryear{Patton, Qamar, Ellison, Bluck, Simard, Mendel,
  Moreno  \& Torrey}{Patton et~al.}{2016}]{Patton2016GalaxySeparations}
Patton D.~R.,  Qamar F.~D.,  Ellison S.~L.,  Bluck A. F.~L.,  Simard L.,
  Mendel J.~T.,  Moreno J.,   Torrey P.,  2016, \mn@doi [Monthly Notices of the
  Royal Astronomical Society] {10.1093/mnras/stw1494}, 461, 2589

\bibitem[\protect\citeauthoryear{Patton et~al.,}{Patton
  et~al.}{2020}]{Patton2020InteractingContext}
Patton D.~R.,  et~al., 2020, Monthly Notices of the Royal Astronomical Society,
  494, 4969

\bibitem[\protect\citeauthoryear{Perez, Michel-Dansac  \& Tissera}{Perez
  et~al.}{2011}]{Perez2011ChemicalInteractions}
Perez J.,  Michel-Dansac L.,   Tissera P.,  2011, \mn@doi [Monthly Notices of
  the Royal Astronomical Society] {10.1111/j.1365-2966.2011.19300.x}, 417, 580

\bibitem[\protect\citeauthoryear{Pessa et~al.,}{Pessa
  et~al.}{2021}]{Pessa2021AstronomyScale}
Pessa I.,  et~al., 2021, \mn@doi [A{\&}A] {10.1051/0004-6361/202140733}, 650,
  A134

\bibitem[\protect\citeauthoryear{Pettini \& Pagel}{Pettini \&
  Pagel}{2004}]{Pettini2004Oiii/NiiRedshift}
Pettini M.,  Pagel B. E.~J.,  2004, \mn@doi [Monthly Notices of the Royal
  Astronomical Society] {10.1111/j.1365-2966.2004.07591.x}, 348, L59

\bibitem[\protect\citeauthoryear{Querejeta et~al.,}{Querejeta
  et~al.}{2021}]{Querejeta2021StellarGalaxies}
Querejeta M.,  et~al., 2021, A{\&}A, 656, A133

\bibitem[\protect\citeauthoryear{Salpeter}{Salpeter}{1955}]{Salpeter1955THEEVOLUTION}
Salpeter E.~E.,  1955, \mn@doi [The Astrophysical Journal] {10.1086/175334},
  121, 161

\bibitem[\protect\citeauthoryear{S{\'{a}}nchez et~al.,}{S{\'{a}}nchez
  et~al.}{2012}]{Sanchez2012CALIFASurvey}
S{\'{a}}nchez S.~F.,  et~al., 2012, \mn@doi [Astronomy {\&} Astrophysics]
  {10.1051/0004-6361/201117353}, 538, A8

\bibitem[\protect\citeauthoryear{S{\'{a}}nchez et~al.,}{S{\'{a}}nchez
  et~al.}{2013}]{Sanchez2013Mass-metallicityRate}
S{\'{a}}nchez S.~F.,  et~al., 2013, \mn@doi [Astronomy {\&} Astrophysics]
  {10.1051/0004-6361/201220669}, 554, A58

\bibitem[\protect\citeauthoryear{S{\'{a}}nchez et~al.,}{S{\'{a}}nchez
  et~al.}{2016a}]{Sanchez2016Pipe3DFIT3D}
S{\'{a}}nchez S.~F.,  et~al., 2016a, \mn@doi [Revista Mexicana de Astronomia y
  Astrofisica] {2016RMxAA..52...21S}, 52, 21

\bibitem[\protect\citeauthoryear{S{\'{a}}nchez et~al.,}{S{\'{a}}nchez
  et~al.}{2016b}]{Sanchez2016Pipe3DDataproducts}
S{\'{a}}nchez S.~F.,  et~al., 2016b, \mn@doi [Revista Mexicana de Astronomia y
  Astrofisica] {2016RMxAA..52..171S}, 52, 171

\bibitem[\protect\citeauthoryear{S{\'{a}}nchez et~al.,}{S{\'{a}}nchez
  et~al.}{2018}]{Sanchez2018SDSSGalaxies}
S{\'{a}}nchez S.~F.,  et~al., 2018, Revista Mexicana de Astronomia y
  Astrofisica, 54, 217

\bibitem[\protect\citeauthoryear{Schmidt}{Schmidt}{1959}]{Schmidt1959TheFormation}
Schmidt M.,  1959, \mn@doi [The Astrophysical Journal] {10.1086/146614}, 129,
  243

\bibitem[\protect\citeauthoryear{Schruba et~al.,}{Schruba
  et~al.}{2011}]{Schruba2011AGALAXIES}
Schruba A.,  et~al., 2011, \mn@doi [The Astronomical Journal]
  {10.1088/0004-6256/142/2/37}, 142, 37

\bibitem[\protect\citeauthoryear{Scudder, Ellison, Torrey, Patton  \&
  Mendel}{Scudder et~al.}{2012}]{Scudder2012GalaxyKpc}
Scudder J.~M.,  Ellison S.~L.,  Torrey P.,  Patton D.~R.,   Mendel J.~T.,
  2012, \mn@doi [Monthly Notices of the Royal Astronomical Society]
  {10.1111/j.1365-2966.2012.21749.x}, 426, 549

\bibitem[\protect\citeauthoryear{Scudder, Ellison, Momjian, Rosenberg, Torrey,
  Patton, Fertig  \& Mendel}{Scudder
  et~al.}{2015}]{Scudder2015GalaxyInteractions}
Scudder J.~M.,  Ellison S.~L.,  Momjian E.,  Rosenberg J.~L.,  Torrey P.,
  Patton D.~R.,  Fertig D.,   Mendel J.~T.,  2015, \mn@doi [Monthly Notices of
  the Royal Astronomical Society] {10.1093/mnras/stv588}, 449, 3719

\bibitem[\protect\citeauthoryear{Spindler et~al.,}{Spindler
  et~al.}{2018}]{Spindler2018SDSS-IVEnvironment}
Spindler A.,  et~al., 2018, \mn@doi [Monthly Notices of the Royal Astronomical
  Society] {10.1093/mnras/sty247}, 476, 580

\bibitem[\protect\citeauthoryear{Steffen, Fu, Comerford, Dai, Feng, Gross  \&
  Xue}{Steffen et~al.}{2021}]{Steffen2021SDSS-IVPairs}
Steffen J.~L.,  Fu H.,  Comerford J.~M.,  Dai Y.~S.,  Feng S.,  Gross A.~C.,
  Xue R.,  2021, \mn@doi [The Astrophysical Journal]
  {10.3847/1538-4357/abe2a5}, 909, 13

\bibitem[\protect\citeauthoryear{Strauss, Weinberg, Lupton  \&
  Narayanan}{Strauss et~al.}{2002}]{Strauss2002SpectroscopicSample}
Strauss M.~A.,  Weinberg D.~H.,  Lupton R.~H.,   Narayanan V.~K.,  2002,
  \mn@doi [The Astronomical Journal] {10.1086/342343}, 124, 1810

\bibitem[\protect\citeauthoryear{Sun et~al.,}{Sun
  et~al.}{2020}]{Sun2020DynamicalGalaxies}
Sun J.,  et~al., 2020, \mn@doi [The Astrophysical Journal]
  {10.3847/1538-4357/ab781c}, 892, 28

\bibitem[\protect\citeauthoryear{Thorp, Ellison, Simard, S{\'{a}}nchez  \&
  Antonio}{Thorp et~al.}{2019}]{Thorp2019SpatiallyMaNGA}
Thorp M.,  Ellison S.,  Simard L.,  S{\'{a}}nchez S.,   Antonio B.,  2019,
  \mn@doi [Monthly Notices of the Royal Astronomical Society: Letters]
  {10.1093/mnrasl/sly185}, 482, L55

\bibitem[\protect\citeauthoryear{Tomi{\v{c}}i{\'{c}}
  et~al.,}{Tomi{\v{c}}i{\'{c}} et~al.}{2018}]{Tomicic2018Two2276}
Tomi{\v{c}}i{\'{c}} N.,  et~al., 2018, \mn@doi [The Astrophysical Journal
  Letters] {10.3847/2041-8213/aaf810}, 869, L38

\bibitem[\protect\citeauthoryear{Torrey, Cox, Kewley  \& Hernquist}{Torrey
  et~al.}{2012}]{Torrey2012THEGALAXIES}
Torrey P.,  Cox T.~J.,  Kewley L.,   Hernquist L.,  2012, \mn@doi [The
  Astrophysical Journal] {10.1088/0004-637X/746/1/108}, 746, 108

\bibitem[\protect\citeauthoryear{Usero et~al.,}{Usero
  et~al.}{2015}]{Usero2015VARIATIONSGALAXIES}
Usero A.,  et~al., 2015, \mn@doi [The Astronomical Journal]
  {10.1088/0004-6256/150/4/115}, 150, 115

\bibitem[\protect\citeauthoryear{Utomo et~al.,}{Utomo
  et~al.}{2017}]{Utomo2017THEGALAXIES}
Utomo D.,  et~al., 2017, \mn@doi [The Astrophysical Journal]
  {10.3847/1538-4357/aa88c0}, 849, 16

\bibitem[\protect\citeauthoryear{Violino, Ellison, Sargent, Coppin, Scudder,
  Mendel  \& Saintonge}{Violino et~al.}{2018}]{Violino2018GalaxyMergers}
Violino G.,  Ellison S.~L.,  Sargent M.,  Coppin K. E.~K.,  Scudder J.~M.,
  Mendel T.~J.,   Saintonge A.,  2018, \mn@doi [Monthly Notices of the Royal
  Astronomical Society] {10.1093/mnras/sty345}, 476, 2591

\bibitem[\protect\citeauthoryear{Vulcani et~al.,}{Vulcani
  et~al.}{2019}]{Vulcani2019GASP.Galaxies}
Vulcani B.,  et~al., 2019, \mn@doi [Monthly Notices of the Royal Astronomical
  Society] {10.1093/mnras/stz1829}, 488, 1597

\bibitem[\protect\citeauthoryear{Wang, Lilly, Pezzulli  \& Matthee}{Wang
  et~al.}{2019}]{Wang2019OnGalaxies}
Wang E.,  Lilly S.~J.,  Pezzulli G.,   Matthee J.,  2019, \mn@doi [The
  Astrophysical Journal] {10.3847/1538-4357/ab1c5b}, 877, 132

\bibitem[\protect\citeauthoryear{Wong et~al.,}{Wong
  et~al.}{2013}]{Wong2013CARMAPROPERTIES}
Wong T.,  et~al., 2013, \mn@doi [The Astrophysical Journal Letters]
  {10.1088/2041-8205/777/1/L4}, 777

\bibitem[\protect\citeauthoryear{Woods, Geller, Kurtz, Westra, Fabricant  \&
  Dell'Antonio}{Woods et~al.}{2010}]{Woods2010TRIGGERED0.08-0.38}
Woods D.~F.,  Geller M.~J.,  Kurtz M.~J.,  Westra E.,  Fabricant D.~G.,
  Dell'Antonio I.,  2010, \mn@doi [The Astronomical Journal]
  {10.1088/0004-6256/139/5/1857}, 139, 1857

\makeatother
\end{thebibliography}



\appendix

\section{ALMaQUEST - Merger Set}
\label{app:DP}

Included in figure \ref{fig:ALMaQUEST_DP_1}-\ref{fig:ALMaQUEST_DP_5} are data product maps for the ALMaQUEST merger set. Post-mergers are shown first, followed by pairs in order of increasing pair separation. All 31 mergers are included, even the 6 galaxies with insufficient overlap between CO $S/N>3$ and H$\alpha$+D4000-\SigSFR measurements which are excluded from individual galaxy analysis. Data products include, from left to right: the SDSS \emph{gri}-image, inclination corrected stellar mass surface density (from {\sc pipe3d}), inclination corrected molecular gas surface density (computed from CO luminosity), inclination corrected star formation rate surface density (computed from H$\alpha$ luminosity), inclination corrected star formation rate surface density (computed from H$\alpha$ luminosity and sSFR-D4000 fit), molecular gas fraction (\Siggasend/\Sigstarend), and star formation efficiency (\SigSFRend/\Siggasend).

\begin{figure*}
	\includegraphics[width=0.9\textwidth]{Figures/ALMaQUEST_DP_1.pdf}
	\centering
    \caption{}
    \label{fig:ALMaQUEST_DP_1}
\end{figure*}

\begin{figure*}
	\includegraphics[width=0.9\textwidth]{Figures/ALMaQUEST_DP_2.pdf}
	\centering
    \caption{}
    \label{fig:ALMaQUEST_DP_2}
\end{figure*}

\begin{figure*}
	\includegraphics[width=0.9\textwidth]{Figures/ALMaQUEST_DP_3.pdf}
	\centering
    \caption{}
    \label{fig:ALMaQUEST_DP_3}
\end{figure*}

\begin{figure*}
	\includegraphics[width=0.9\textwidth]{Figures/ALMaQUEST_DP_4.pdf}
	\centering
    \caption{}
    \label{fig:ALMaQUEST_DP_4}
\end{figure*}

\begin{figure*}
	\includegraphics[width=0.9\textwidth]{Figures/ALMaQUEST_DP_5.pdf}
	\centering
    \caption{}
    \label{fig:ALMaQUEST_DP_5}
\end{figure*}



\section{Metallicity dependent CO conversion factor}

\label{app:alpha_CO}

\begin{figure*}
	\includegraphics[width=0.8\textwidth]{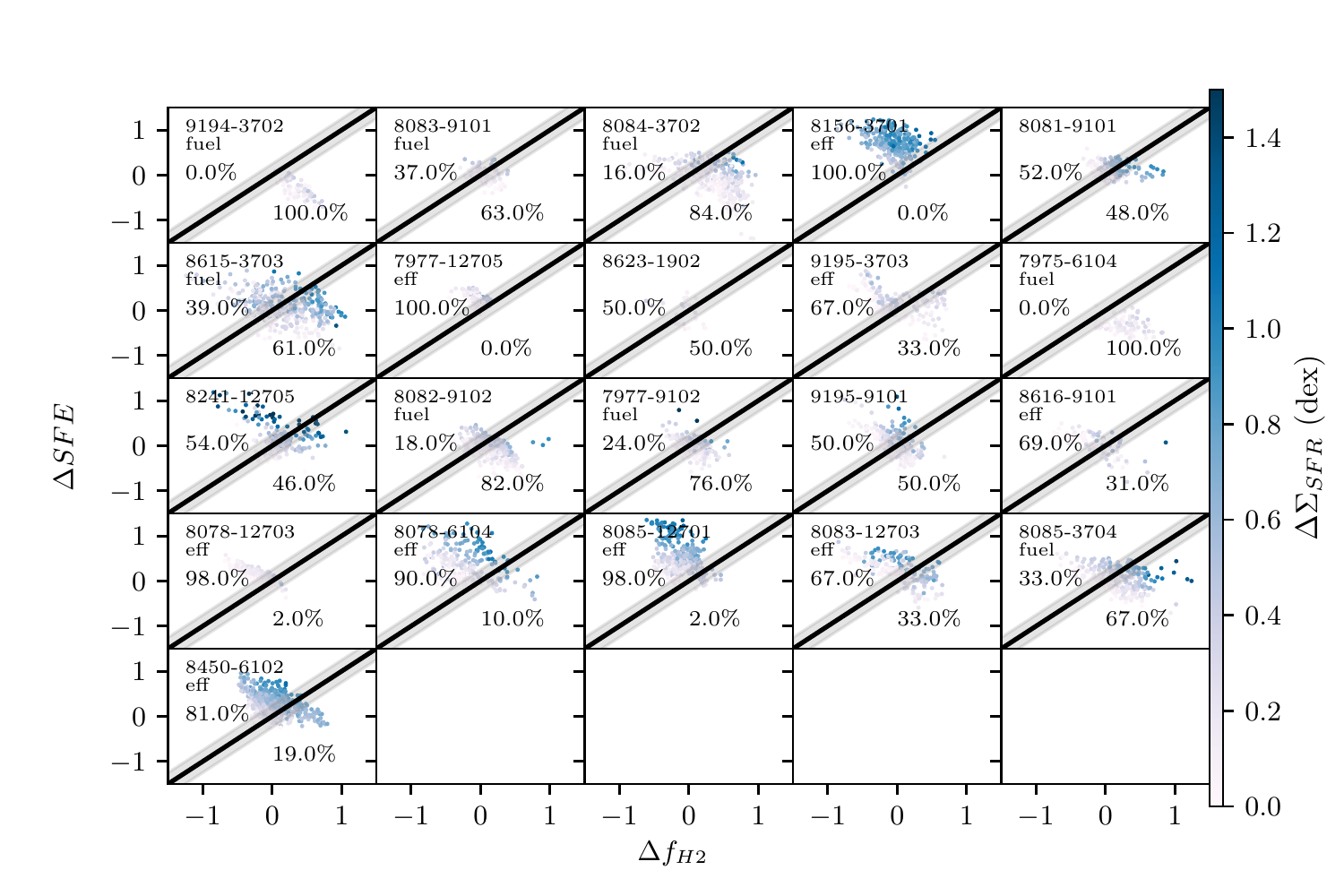}
	\centering
    \caption{A replication of Figure \ref{fig:delta_plots_all}, but now with a grey line indicating the 0.248 dex error on both offset parameters from using a constant $\alpha_{\textrm{CO}}$. Pixels within this grey region are not considering when calculating the fraction of fuel driven spaxels. This biases the distribution of fuel fractions to the extreme ends (very high or very low), as spaxels further away from the line become more important in the calculation.}
    \label{fig:delta_plot_buffer}
\end{figure*}

Throughout this work we elect to use a constant $\alpha_{\textrm{CO}}$ rather than a metallicity dependent one. This is highly beneficial to our analysis in that we can include spaxels which do not meet the requirements to measure accurate gas-phase metallicities (requiring high signal-to-noise in multiple emission lines). As a result our \Siggas measurements will be imperfect, leading to the possibility that variations \dSFE and \dfgas (both of which require \Siggasend) could be the result of inaccuracies in $\alpha_{\textrm{CO}}$. Depending on the degree of that inaccuracy this could sully our final conclusions concerning the role enhanced efficiency and fuel play in increasing star formation activity.

Though we cannot compute a metallicity dependent $\alpha_{\textrm{CO}}$ ($\alpha_{\textrm{CO, met}}$ from here on) for every spaxel, we can characterize the change in \Siggas for spaxels where $\alpha_{\textrm{CO, met}}$ is measurable. We compute a $\alpha_{\textrm{CO, met}}$ for all spaxels that are star-forming based on the criteria described in Subsection \ref{subsec:SFR-D4000}, for which we can also measure the gas phase metallicity using the O3N2 calibration of \cite{Pettini2004Oiii/NiiRedshift}. We set $\alpha_{\textrm{CO, met}}=4.35 \times Z'^{-1.6} (\textrm{M}_{\odot} (\textrm{K } \textrm{km/s } \textrm{pc}^2)^{-1})$ where $Z'$ is the metallicity normalized to solar metallicity ($Z'=10^{\textrm{PP04}_{\textrm{O3N2}}-8.69}$) \citep{Sun2020DynamicalGalaxies}. We find the median difference between \Siggas and \Siggasmet to be $-$0.068 dex, with a standard deviation of 0.124 dex. If we want to be conservative, we can say that \dfgas and \dSFE both have an uncertainty of 2$\sigma$ on the difference in \Siggasend, or 0.248 dex. Figure \ref{fig:delta_plot_buffer} is a replication of Figure \ref{fig:delta_plots_all}, with a grey bar representing 0.248 dex of uncertainty; points within this range could change from fuel to efficiency driven, or vice versa, if one switched from a constant to a metallicity dependent $\alpha_{\textrm{CO}}$. We can thus exclude spaxels within this grey bar from our calculation of the fuel fraction to achieve a value that is independent of variations in $\alpha_{\textrm{CO}}$. By excluding borderline spaxels, most galaxies tend to be more ``fuel'' or ``efficiency'' driven.

\begin{figure}
	\includegraphics[width=0.95\columnwidth]{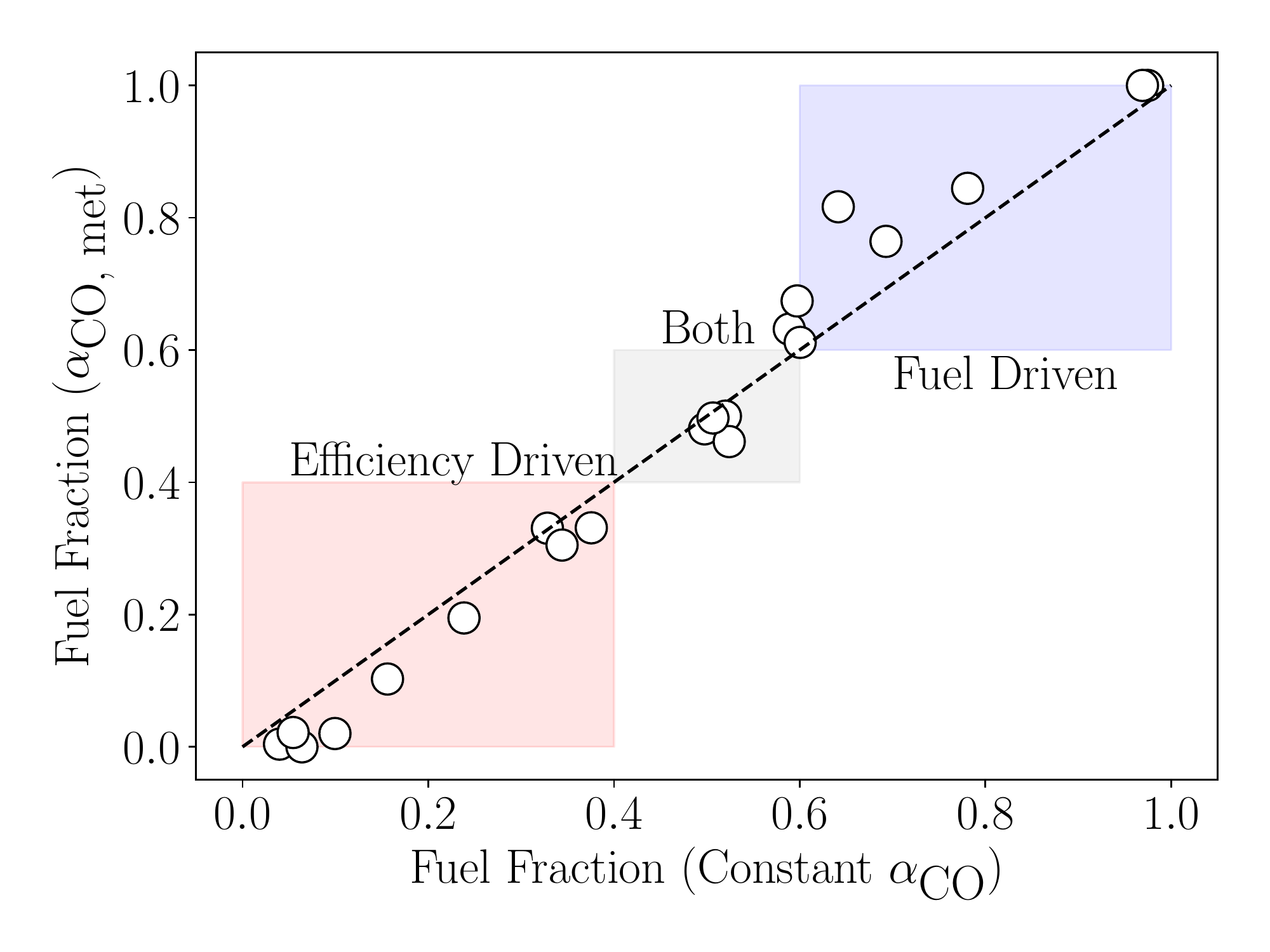}
	\centering
    \caption{The fuel fraction determined from Figure \ref{fig:delta_plot_buffer}, considering uncertainties from not using $\alpha_{\textrm{CO, met}}$, plot against the original fuel fraction determined from Figure \ref{fig:delta_plots_all}. Note that efficiency driven mergers move below the line of equality (black dashed line), and fuel driven mergers move above the line; thus the dominant star-forming mechanism only appears more dominant when uncertainties in $\alpha_{\textrm{CO}}$ are considered. We have highlighted regions that indicate the dominant mechanism has not changed (ideally where all points would lie): where both fractions imply star formation is efficiency driven (red), where both fractions imply star formation is fuel driven (blue), and where both fractions imply star formation is driven by both (grey). Only two mergers change category (from ``both'' to ``fuel driven''), though both galaxies were already at the border between these two categories.}
    \label{fig:FF_change_met}
\end{figure}

Figure \ref{fig:FF_change_met} directly compares the original fuel fraction, using a constant $\alpha_{\textrm{CO}}$, to this new one which incorporates error from not using $\alpha_{\textrm{CO, met}}$. Ideally all points would be along the line of equality, confirming that the fraction of fuel driven spaxels does not change. Instead galaxies which are fuel driven become more fuel driven, rising above the line of equality, and galaxies which are efficiency driven become more efficiency driven. As a result very few galaxies change whether we would consider them ``efficiency'' or ``fuel'' driven. We highlight regions where both the new and old fuel fractions agree on the dominant mechanism: where efficiency dominates (red), fuel fraction dominates (blue), or both equally influece star formation(grey). Only two mergers change which mechanism dominates: both are mergers where efficiency and fuel are equally important with a constant $\alpha_{\textrm{CO}}$, but then change to fuel driven. This shift is not the result of a significant change in fuel fraction, rather the two mergers were already at the border between categories. As discussed in the main analysis, these categories of ``efficiency'' and ``fuel'' driven are less robust than the continuously changing fuel fraction measurement.

In summary, our measured fuel fraction changes little when uncertainties in $\alpha_{\textrm{CO}}$ are considered. Moreover, our results concerning \dSFE being more important to star formation than \dfgas (or vice versa) are unlikely to stem from local variations in conversion factor. Rather they characterize the true relationship between star-formation and molecular gas in a merger event.

\section{sSFR-D4000}
\label{app:sSFR-d4000}

\begin{figure}
	\includegraphics[width=0.95\columnwidth]{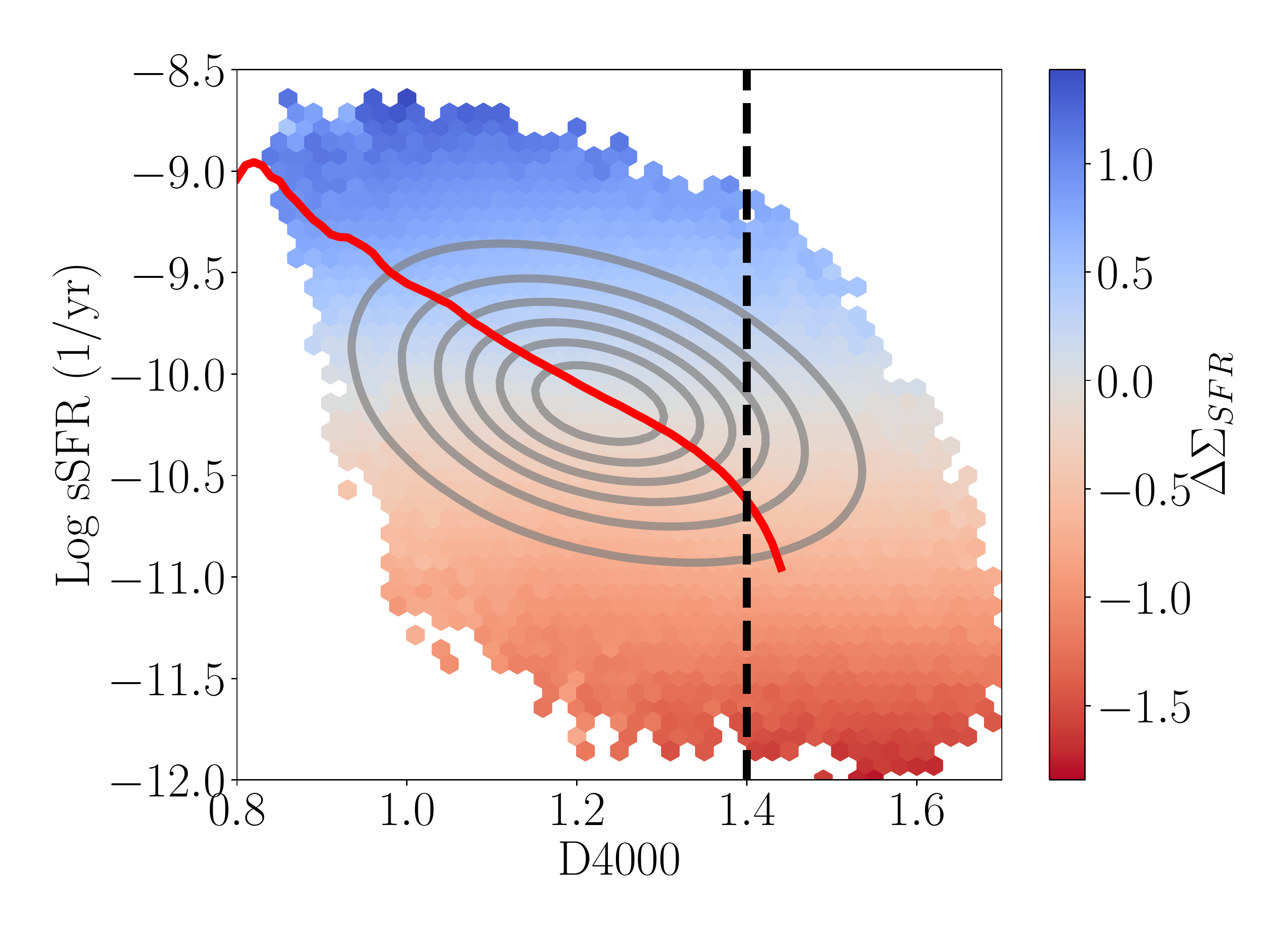}
	\centering
    \caption{sSFR versus D4000 for all star forming spaxels in MaNGA. Density contours of this distribution are shown in grey, over a hexbin distribution colour-coded by the median \dsigsfr value in each bin. The median sSFR value in a D4000 bin is shown in red: these points are used to approximate \SigSFR for spaxels that do not meet our star-forming criteria. We exclude spaxels with D4000>1.4, given the asymptotic nature of the sSFR-D4000 at high D4000. Note that there is significant scatter in this relation, which correlates with \dsigsfrend.}
    \label{fig:sSFR_D4000_dSFR}
\end{figure}

The SFRs presented herein include those that have been calibrated from D4000.  Here, we demonstrate that our conclusions are not significantly affected by the inclusion of SFRs computed in this way. Figure \ref{fig:sSFR_D4000_dSFR} shows the distribution of sSFR versus D4000 for all star forming spaxels in MaNGA, based on our star forming criteria described in Subsection \ref{subsec:SFR-D4000}, with contours representing the density of the distribution of hexbins colour-coded by the median \dsigsfr value in that bin. The asymptotic nature of the sSFR-D4000 relation as D4000 approaches 1.45 can lead to drastic difference in approximated \SigSFR for spaxels with very similar D4000 values. Therefore, only spaxels with D4000>1.4 are used for SFR calculations.

There is a non-negligible scatter in the sSFR-D4000 relation which is strongly correlated with \dsigsfrend. This is to be expected; by definition a spaxel with high sSFR given the D4000 index would have a higher \SigSFR given its \Sigstarend. The red line is a median sSFR in a bin so it makes sense that scatter significantly above the line would corresponds to large, positive \dsigsfrend. Thus by approximating \SigSFR from the median sSFR, we will always underestimate the offset of the star formation from regular behaviour (both enhancements and deficits). That bias works in favour of this analysis, where we are looking for strong offsets in \SigSFR. Thus for D4000-\SigSFR values, any enhancement or suppression in star formation is a lower limit on the true offset; the same can be said for enhancements and deficits in SFE.

When D4000-\SigSFR is used the absolute value of \dsigsfr and \dSFE will be underestimated, but \dfgas (which does not depend on a \SigSFR measurement) will not be changed. The dependence of \dSFE on \SigSFR is crucial when comparing \dSFE and \dfgas to determine which mechanism drives enhanced star formation. When using a D4000-\SigSFR the value of \dSFE is a lower limit, so \dSFEend$>$\dfgas will be consistent even if D4000-\SigSFR is less accurate than H$\alpha$-\SigSFRend. However if \dSFEend$<$\dfgasend, \dSFE could be underestimated and the statement might not be true if we could measure a H$\alpha$-\SigSFRend. We therefore have to consider that galaxies with a large fraction of fuel driven spaxels (i.e., \dSFEend$<$\dfgasend) might have a smaller fraction if we were not dependent on D4000-\SigSFR measurements.

\begin{figure}
	\includegraphics[width=0.95\columnwidth]{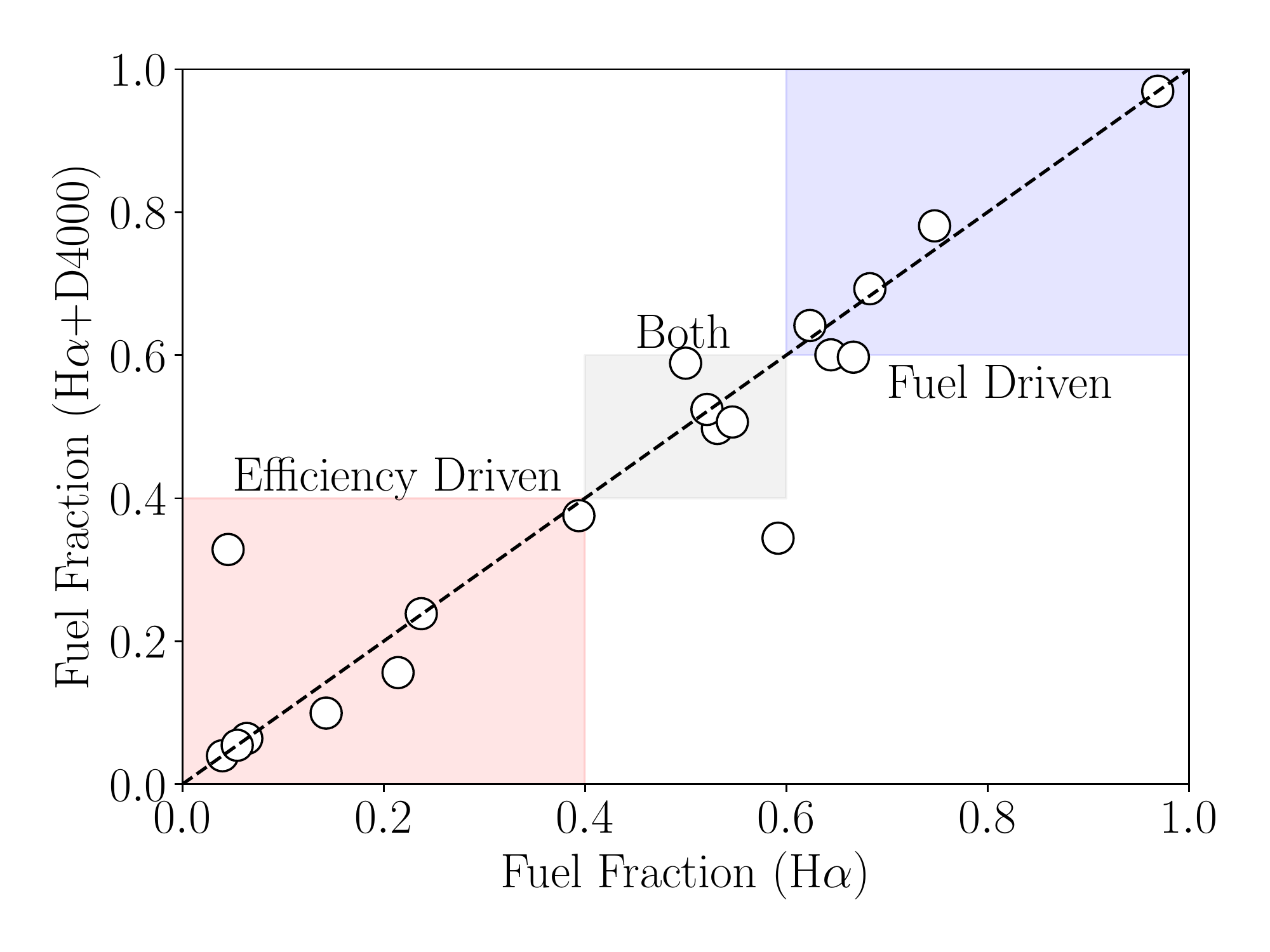}
	\centering
    \caption{The fuel fraction determined from only H$\alpha$ spaxels compared to the fuel fraction when both H$\alpha$ and D4000 spaxels are included. Ideally all galaxies would lie on the line of equality, implying that the inclusion of D4000 spaxels does not drastically change our results. But realistically there is scatter about the line of equality as more spaxels are included in the analysis. Difference between the two fuel fractions is alright, so long as it does not change which mechanism predominantly drives star formation. We have highlighted regions that ideally the points should lie within: where both fractions imply star formation is efficiency driven (red), where both fractions imply star formation is fuel driven (blue), and where both fractions imply star formation is driven by both (grey). Only one point lies outside these acceptable regions: 8616-9101, which is efficiency driven when both H$\alpha$ and D4000 spaxels are included, but fuel driven when only H$\alpha$ spaxels are considered.}
    \label{fig:sSFR_D4000_dSFR_change}
\end{figure}

To check how this bias might alter our results, we calculate the fraction of fuel driven spaxels both with our combined H$\alpha$+D4000 \SigSFR values, as well as those which only have H$\alpha$-\SigSFRend. Four galaxies do not have at least 20 spaxels with viable H$\alpha$-\SigSFRend, so can not be part of the test. Both fractions are included in Table \ref{tab:mechanism} and discussed in the text, though we provide additional comparison here. Figure \ref{fig:sSFR_D4000_dSFR_change} directly compares the fuel fraction determined from only H$\alpha$ spaxels to the fuel fraction when both H$\alpha$ and D4000 spaxels are included. Ideally these two values would be equal and all galaxies would lie on the line of equality, but as expected there is some difference between the two. Difference between the two fuel fractions can be acceptable so long as it does not change which mechanism predominantly drives star formation. To better demonstrate this distinction we highlight regions where both fractions agree: that star formation is efficiency drive (red), fuel driven (blue), or both (grey). Only one galaxy is outside these acceptable regions: 8616-9101, which is efficiency driven when both H$\alpha$ and D4000 spaxels are included, but fuel driven when only H$\alpha$ spaxels are considered. 8616-9101 is an interesting case given we expect spaxels that use D4000-\SigSFR to underestimate \dSFE and bias towards large fuel fractions, but in this case including D4000 spaxels leads to a smaller fuel fraction. Considering only one galaxy changes dominant mechanism, we are assured that our results are not dependent on the inclusion of D4000-\SigSFR values.

\section{Offset Maps for Entire Merger Sample}
\label{app:maps}

Figures \ref{offset_cat1}-\ref{offset_cat4} are a complete catalogue of the \dsigsfrend, \dSFEend, and \dfgas distributions for the merger galaxy sample examined in this work. Diverging colourbars are used such that blue spaxels represent enhancements in \SigSFRend, SFE, or \fgasend. Red spaxels represent a supression in either variable. The SDSS \emph{gri}-image is included as well, to provide context for the galaxy and the MaNGA IFU coverage. One will note that some galaxies have \dfgas measurements where no \dsigsfr or \dSFE is measured. This results from the stricter cuts required to measure \SigSFRend, a limitation not imposed on the calculation of \dfgasend. The fourth column provides a map of \dSFEend$-$\dfgas for all spaxels where \dsigsfrend$>0$, as described in Section \ref{sec:Discussion}. Note that most mergers have negative \dSFEend$-$\dfgas in the centre of the galaxy, implying gas fractions are more centrally enhanced than SFE.

\begin{figure*}
	\includegraphics[width=0.8\textwidth]{Figures/offsetmaps_1.pdf}
	\centering
    \caption{}
    \label{offset_cat1}
\end{figure*}

\begin{figure*}
	\includegraphics[width=0.8\textwidth]{Figures/offsetmaps_2.pdf}
	\centering
    \caption{}
    \label{offset_cat2}
\end{figure*}

\begin{figure*}
	\includegraphics[width=0.8\textwidth]{Figures/offsetmaps_3.pdf}
	\centering
    \caption{}
    \label{offset_cat3}
\end{figure*}

\begin{figure*}
	\includegraphics[width=0.8\textwidth]{Figures/offsetmaps_4.pdf}
	\centering
    \caption{}
    \label{offset_cat4}
\end{figure*}


\bsp	
\label{lastpage}
\end{document}